\documentclass[nomanuscript,mnsc,nonblindrev]{informs3}

\OneAndAHalfSpacedXI 

\usepackage[skip=10pt plus1pt]{parskip}

\usepackage{subfiles} 
\usepackage[colorlinks=true,linkcolor=blue,citecolor=blue,allcolors=blue
]{hyperref}%

\usepackage{cancel}    

\usepackage[section]{placeins}
\usepackage{tikz}
\usepackage{bbm}
\usepackage{dsfont}
\usepackage{algorithm2e}
\usepackage{enumitem}
\usepackage{fancyhdr}
\usepackage[font={footnotesize}]{subcaption} 
\usepackage{graphicx}
\usepackage{mathtools}
\usepackage{amsmath}
\usepackage{amssymb}
\usepackage{caption}
\usepackage{float}
\usepackage[normalem]{ulem}
\usetikzlibrary{arrows,positioning,shapes.geometric}
\usepackage{amsmath}
\DeclareMathOperator{\Tr}{Tr}
\DeclareMathOperator{\Cov}{Cov}

\usepackage{datetime}

\DeclareMathOperator{\diag}{diag}
\usepackage{natbib}
 \bibpunct[, ]{(}{)}{,}{a}{}{,}%
 \def\bibfont{\small}%

\usepackage{multibib}
\newcites{R}{References}

\TheoremsNumberedThrough     
\ECRepeatTheorems

\EquationsNumberedThrough    

\MANUSCRIPTNO{MS-1234-56}

\begin{document}


\RUNAUTHOR{Abdelhakmi and Lim}

\RUNTITLE{Dynamic Black-Litterman}

\TITLE{Dynamic Black-Litterman\footnote{\today}}

\ARTICLEAUTHORS{%

\AUTHOR{Anas Abdelhakmi}
\AFF{Institute of Operations Research and Analytics, \\ National University of Singapore, \EMAIL{a.anas@u.nus.edu}} 

\AUTHOR{Andrew E.B. Lim}
\AFF{Department of Analytics and Operations,   Department of Finance, and Institute of Operations Research and Analytics,\\ National University of Singapore, \EMAIL{andrewlim@nus.edu.sg}} 

} 

\ABSTRACT{%

The Black-Litterman model is a framework for incorporating forward-looking expert views in a portfolio optimization problem. Existing work focuses almost exclusively on single-period problems with the forecast horizon matching that of the investor. We consider a generalization where the investor trades dynamically and views can be over horizons that differ from the investor. By exploiting the underlying graphical structure relating the asset prices and views, we derive the conditional distribution of asset returns when the price process is geometric Brownian motion, and show that it can be written in terms  of a multi-dimensional Brownian bridge. The components of the Brownian bridge are dependent one-dimensional Brownian bridges with hitting times that are determined  by the statistics of the price process and views.  The new price process is an affine factor model with the conditional log-price process playing the role of a vector of factors. We derive an explicit expression for the optimal dynamic investment policy and analyze the hedging demand for changes in the new covariate.  More generally, the paper shows that Bayesian graphical models are a natural framework for incorporating complex information structures in the Black-Litterman model. The connection between Brownian motion conditional on noisy observations of its terminal value  and multi-dimensional Brownian bridge is novel and of independent interest.
}%



\maketitle


\section{Introduction}
\label{sec1}

 The Black-Litterman model \big(\cite{Black_1991, Black_1992}\big) is a framework for incorporating forward-looking expert views in a one-step portfolio optimization problem. The model uses a backward-looking equilibrium model like the Capital Asset Pricing Model (CAPM) as a preliminary (prior) forecast and Bayes' rule to combine this with forward-looking expert views. The updated return distribution is used in a single-period mean-variance optimization to obtain an asset allocation. It has been observed that 
 portfolios constructed using this approach tend to be less concentrated in a small number of assets and less sensitive to model inputs than those obtained without expert views. 

The classical Black-Litterman model has two limitations: Forecasts are assumed to match the investment horizon of the investor, and investors are restricted to making single-period decisions. 
When views coincide with an external event like the outcome of a firm's quarterly earnings report, they are unlikely to match the investment horizon of the investor. The  classical Black-Litterman model does not accommodate such an information structure as it does not specify how the predictions over a different horizon are related to the return distributions of interest to the investor.

\subsection*{Summary of contributions}
\begin{enumerate}
    \item We formulate a Bayesian graphical model of asset prices with forward looking views and derive conditional price dynamics for the assets.
    We show that the log-returns process, which was Brownian motion with drift under the prior model, is a  mean-reverting process after conditioning on views and plays the role of a covariate of the conditional price process.
 
 \item 
We show that a correlated multi-dimensional Brownian motion, conditional on noisy views of its terminal value, is the restriction to the view horizon of a multi-dimensional Brownian bridge. The components of the Brownian bridge are dependent, with hitting times determined by the correlation structure of the Brownian motions and the views that extend beyond the view horizon. Expert views in the dynamic Black-Litterman model transform the log-returns process into a linear combination of Brownian bridges with drift. We believe that this construction of multi-dimensional Brownian bridge, through noisy observations of a multi-dimensional Brownian motion, is novel and  of independent interest.

\item We generalize the classical Black-Litterman model to the dynamic setting by formulating a dynamic portfolio choice problem in terms of the conditional price process. 
We  derive an explicit expression for the optimal dynamic portfolio, and show that it  consists of a mean-variance term and a hedging demand for changes in the conditional log-returns (``views covariate"). The hedging demand depends on the solution of a nonlinear matrix Riccati equation which we solve in closed form. 

\item  We show that the hedging demand for the views covariate is increasing in the precision of the views. 
In an experiment where investors rebalance at discrete time points, the dynamic Black-Litterman investor outperforms an investor who solves a single-period Black-Litterman problem at each rebalancing epoch. Turnover for the dynamic investor is lower. Due to the hedging demand which offsets the risk of changes in the views covariate, investment performance is insensitive to the time between rebalancing epochs.

\item  We formulate and solve dynamic portfolio choice problems with alternative view models: when (a) views are revised during the investment horizon, and (b) views over shorter intervals are received during the investment horizon.

\end{enumerate}

\subsection*{Literature review}

The Black-Litterman model was introduced by \cite{Black_1991,Black_1992} using {\color{blue}Theil's mixed estimate approach}, then expanded and discussed with greater detail in \cite{Bevan1998} and \cite{He2002}. The Bayesian interpretation of the Black-Litterman model was introduced by \cite{Qian2001} and expanded by \cite{Cheung2009}; for a  survey we refer to \cite{Walters2011}. Much of the literature on the Black-Litterman model stays close to the classical setting where investments occur over a single-period and the horizon of the views matches that of the investor.  \cite{Andrew2020} show how complex information structures and uncertainty about the equilibrium model can be modeled using a Bayesian graphical model for a single-period problem.   We formulate a graphical model that accommodates continuous time asset price dynamics and expert views over multiple horizons. Analogous to the blending of the prior model and views in the classical Black-Litterman model, we derive recursive update equations for the price process conditional on views

While most of the papers on the Black-Litterman model consider single-period problems, dynamic extensions have been explored in \cite{DavisBL,DavisLleoJBF,DavisLleoMF,DavisLleoAOR,Frey2012,Sass2017,TimeDependent_BL}. A key distinction in this literature is the nature of expert views. Experts in \cite{Frey2012} and \cite{Sass2017} provide noisy information about the current value of the drift of the price process, which is treated as a latent variable and estimated using filtering methods. Experts in \cite{DavisBL,DavisLleoJBF,DavisLleoMF,DavisLleoAOR} provide forward-looking forecasts of the trajectory of risk factors that influence expected returns. \cite{DavisLleoJBF} also discuss calibration and debiasing of the view model in this setting. Our paper differs in that experts provide noisy views about the realization of future returns, aligning more closely with the information structure of the classical Black-Litterman model. Because of its forward-looking nature, our conditional price process is closer to a smoothed estimate than a filtered one.

A contribution we wish to highlight is to the literature on multi-dimensional Brownian bridge. The classical Brownian bridge (\cite{TheBBAlabama} and \cite{BrownianBridgesPINSKY}) is the stochastic process $B(t) = \{W(t)|W(T)=y\}$ obtained after conditioning on the terminal value of a one-dimensional Brownian motion $W(t)$. The classical literature characterizes its distributional properties and shows that it is the solution of a linear stochastic differential equation. One-dimensional Brownian bridge has been used in the finance literature to model price dynamics of an insider trader (\cite{JeanBlanc} and \cite{OverViewEnlargement}). The notion of multi-dimensional Brownian bridge is more intricate because the terminal value and time of {\it each component}, as well as the correlation between them, needs to be defined and there does not appear to be a canonical definition of this process. Applications of multi-dimensional versions of Brownian bridge in finance are quite scarce; one example we know of is \cite{mbb} who use it to model futures prices. We also mention \cite{mbb2} who consider a multi-dimensional Brownian bridge where the starting and ending values of each element  is known in advance and the hitting times are derived from the correlation structure.  

One contribution of this paper is to generalize the classical definition of Brownian bridge to the case where observations of the terminal value of a multi-dimensional Brownian motion is observed with noise. We derive a stochastic differential equation for generalized Brownian bridge and fully characterize its joint distribution. We show that its components are correlated one-dimensional Brownian bridges with different hitting times that are determined endogenously by the correlation structure of the original Brownian motion and the noisy terminal observation equation. Finally, we show that the conditional log-returns process from the Black-Litterman model can be written in terms of this generalized Brownian bridge. The connection between noisy views of the future value of a multi-dimensional Brownian motion and the stochastic process it defines is of independent interest.

\subsection*{Outline} 
We briefly review the classical Black-Litterman model in Section \ref{sec2} and  introduce a graphical model for the dynamic generalization in Section \ref{sec3}. We also derive the price dynamics conditioned on the views and show that it is an affine model with a new covariate $X^y(t)$ that dynamically blends  the views and real-time price information. We define the notion of generalized multi-dimensional Brownian bridge in Section \ref{sec4} and provide an interpretation of the covariate $X^y(t)$ in terms of this process. We formulate and solve the associated dynamic portfolio choice problem in Section \ref{sec5}. 
We discuss extensions to more complex information structures for the views (revisions, and multiple shorter-term views) in Section \ref{sec:Extensions}, and provide numerical results and simulations in Section \ref{sec7}. Section \ref{sec8} concludes.
To enhance the flow of the paper, many proofs have been relegated to the Appendix.

\section{The Black-Litterman Model}
\label{sec2}
To keep the paper self-contained, we now provide a brief review of the classical Black-Litterman model. Building on the single period model in \cite{Andrew2020} we adopt a graphical Bayesian representation of the joint distribution of views and asset prices.  

\subsection{Equilibrium Model}
Consider a financial market of $N$ risky assets with rate of return $r$ and one risk-free asset with rate $r_f$. The Black-Litterman model assumes that $r$ is normally distributed
\begin{equation}
\label{eq1}
    r \sim \mathcal{N}\big(\mu, \Sigma \big)
\end{equation}
and selects the mean return $\mu$ using a backward-looking equilibrium model such as the Capital Asset Pricing Model (CAPM) (see \cite{Sharp1964}), or an inverse optimization method (see \cite{Gupta2012,Sharp1974}). In a Bayesian setting, equation \eqref{eq1} can be considered a prior distribution on the unrealized returns, with views being noisy observations which are used to update the prior. For a comparison between the two approaches refer to \cite{Subekti2021}.
\subsection{Expert views}
The investor receives forward-looking views about risky asset returns. Conditional on the realized returns $r$, views are assumed to be normally distributed
\begin{equation}
\label{eq2}
    Y\, | \, r \sim \mathcal{N}\big(Pr, \Omega \big)
\end{equation}
where $P$ is a linear mapping from the set of returns to the set of views. The covariance matrix $\Omega$ models the accuracy of the views. Methods for estimating $\Omega$  are discussed in \cite{Omega2} and \cite{Omega1}.

To illustrate the idea,  consider a market with assets, A, B, and C, with $r = [r_A, r_B, r_C]^\top$ being the vector of returns. Before the realization of the returns, the investor receives expert views concerning the three assets. We distinguish here between two types of views:
\begin{enumerate}
    \item Absolute views: The expert gives a direct forecast about the return of one of the assets; for example, `The return of asset A will be $5\%$',
    \item Relative views: The expert compares the returns of two or more assets; For example, `Company C will outperform company B by $10\%$'.
\end{enumerate}
In this case, the expert is giving a noisy forecast of the realized return
\begin{equation*}
  Pr =   \begin{bmatrix}
1 & 0 & & 0\\
0 & -1 &  & 1\\
    \end{bmatrix}
    \begin{bmatrix}
        r_A \\ r_B \\ r_C
    \end{bmatrix}
     = \begin{bmatrix}
         r_A \\ r_C - r_B
     \end{bmatrix},
\end{equation*}
which by \eqref{eq2} is a sample of a two-dimensional normal random vector with mean $Pr$ and covariance matrix $\Omega$. In this example, the realization $y$ of $Y(0,T)$ is
\begin{equation*}
    y =  \begin{pmatrix}
        y_A \\
        y_{C-B}
    \end{pmatrix}
    = \begin{pmatrix}
        5\% \\ 10\%
    \end{pmatrix}.
\end{equation*}

\subsection{Graphical Representation}
Bayesian graphical models provide a clear and intuitive framework for capturing uncertainty, dependencies, and causal relationships among variables (see for example \cite{Andrew2020}). We represent random variables as nodes with 
unobserved random variables (the vector of unrealized returns) as circles, and observed random variables (experts views) as squares. Edges represent conditional dependencies.

In the classical single-period Black-Litterman model views can be interpreted as noisy observations of unrealized returns, which can be represented as shown in Figure \ref{fig:classical}.

\begin{figure}[!ht]
\centering
\begin{tikzpicture}[
roundnode/.style={circle, draw=black!80, fill=gray!5, very thick, minimum size=7mm},
squarednode/.style={rectangle, draw=black!80, fill=gray!5, very thick, minimum size=5mm},
]
\node[squarednode]      (maintopic)                              {\large Y};
\node[roundnode]        (uppercircle)       [above=of maintopic] {\large r};
\node[]      (rightsquare)       [right=of maintopic] {$Y \sim \mathcal{N}\big(Pr, \Omega \big) $};
\node[]        (lowercircle)       [right=of uppercircle] {$r \sim \mathcal{N}\big(\mu, \Sigma \big)$};

\draw[->] (uppercircle.south) -- (maintopic.north);

\end{tikzpicture}
\caption{Bayesian network of the classical Black-Litterman model}
\label{fig:classical}
\end{figure}
\subsection{Posterior Distribution of the Returns}
We can use the view \eqref{eq2} to update the equilibrium returns \eqref{eq1} using Bayes' rule. Specifically, the vector of returns $r$ given the view $Y(0,T) = y$ is still normal
\begin{equation}
\label{eq3}
    r \,|\, Y = y  \sim \mathcal{N}\big(\mu_{BL}, \Sigma_{BL} \big),
\end{equation}
with mean and covariance 
\begin{equation}
\label{eq4}
    \begin{split}
      &\mu_{BL} = \mathbb{E}\bigl[r \,| \,Y=y\bigr] = \big(\Sigma^{-1} + P^\top\Omega^{-1}P\big)^{-1} \bigl(\Sigma^{-1}\mu + P^\top\Omega^{-1}y\bigr),   \\
      &\Sigma_{BL} =  \mathbb{V}\bigl[r\,|\,Y=y\bigr] = \big(\Sigma^{-1} + P^\top\Omega^{-1}P\big)^{-1}.\\
    \end{split}
\end{equation}
Note that the posterior mean $\mu_{BL}$ is a combination of the prior mean  $\mu$ and the view $y$ weighted by their respective precision matrices $\Sigma^{-1}$ and $\Omega^{-1}$. The precision of the posterior return  increases from $\Sigma^{-1}$ to $\Sigma_{BL}^{-1} = \Sigma^{-1} + P^\top\Omega^{-1}P$ after the update. 
\subsection{Optimal Portfolio}
Using the updated return distribution in \eqref{eq4}, the optimal portfolio is obtained by solving a mean-variance optimization problem
\begin{equation}
    \label{eq5}
    \max_\pi \pi^\top \mathbb{E}\bigl[r\,|\,Y=y\bigr] + (1 - \pi^\top \mathbf{1}_N) r_f - \frac{\gamma}{2} \pi^\top \mathbb{V}\bigl[r\,|\,Y=y\bigr] \pi
\end{equation}
where $\gamma \in (0 ,\infty)$ is the risk-aversion parameter. The optimal portfolio is
\begin{equation}
\label{eq6}
\begin{split}
    \pi^* &= \frac{1}{\gamma}\,\mathbb{V}\bigl[r \,|\, Y = y\bigr]^{-1}
    \Bigl(\mathbb{E}\bigl[r \,|\, Y = y\bigr] - r_f \mathbf{1}_N\Bigr)\\[1mm]
         &= \frac{1}{\gamma}\,(\Sigma_{BL})^{-1}\Bigl(\mu_{BL} - r_f \mathbf{1}_N\Bigr).
\end{split}
\end{equation}
where $\mathbf{1}_N = (1, \dots, 1)^\top \in \mathbb{R}^N$ vector of ones.
\section{The Dynamic Black-Litterman Model}
\label{sec3}
We formulate a continuous time version of the Black-Litterman model with forward-looking expert views. 
We derive the price dynamics conditional on these views and show that it is closely related to the Kalman smoothing equations.

Throughout the paper, we assume that all random variables and stochastic processes are defined on the common probability space $(\Omega, \mathcal{F}, \mathbb{P})$. All vectors are column vectors. For a vector $S \in \mathbb{R}^N$,  we denote its $i^{th}$ element by $S_i$, $i \in [N]$, and the diagonal matrix of stock prices by $D(S) = \diag(S_1, \dots, S_N) \in \mathbb{R}^{N \times N}$. For a matrix $L$, we use $L_i$ to denote its $i^{th}$ column, and $I_N$ to denote the $N$ by $N$ identity matrix. $\top$ is used for the transpose operator.
\subsection{Financial Market}
\label{3.1}
Consider a financial market of $N$ risky assets and one risk-free asset. The interest rate $r_f$ for the risk-free asset is assumed to be constant and its price $S_0(t)$ satisfies
\begin{equation}
    \frac{dS_0(t)}{S_0(t)} = r_f dt.
\end{equation}
For $i \in [N]$, the price $S_i(t)$ of the risky asset $i$ evolves as a geometric Brownian motion
\begin{equation}
\label{eq8}
        \frac{dS_i(t)}{S_i(t)} =  \mu_i dt + dW_i(t)
\end{equation}
 where $W(t)$ is a vector of $N$ correlated Brownian motions with
\begin{equation*}
           W(t) \sim \mathcal{N} \big(0, t \Sigma \big).
\end{equation*}
As in the classical Black-Litterman model, the drift $\mu_i$ of stock $i$ is set to equal the expected return obtained from a backward-looking equilibrium model such as CAPM, and $\sigma_i$ is the associated volatility. If $\rho_{ij}$ is the correlation between $W_i(t)$ and $W_j(t)$, we can write the covariance matrix
\begin{equation*}
    \Sigma = \begin{pmatrix} 
    \sigma_{1}^2 & \rho_{12}\sigma_1 \sigma_2  & \cdots & \rho_{1N}\sigma_{1}\sigma_N \\
    \rho_{12}\sigma_1 \sigma_2 & \sigma_{2}^2 & \cdots & \rho_{2N}\sigma_{2}\sigma_N \\
    \vdots & \vdots & \ddots & \vdots \\
    \rho_{1N}\sigma_{1}\sigma_N & \rho_{2N}\sigma_{2}\sigma_N & \cdots & \sigma_{N}^2 \\
\end{pmatrix}.
\end{equation*}
There are various methods for specifying the covariance matrix $\Sigma$ (see, e.g., \cite{He2002}, \cite{walters2013}). 

Let $X(t) = (X_1(t), \dots, X_N(t))^\top \in \mathbb{R}^N$ denote the vector of the log-returns 
\begin{equation*}
    X_i(t) = \log \bigg(\frac{S_i(t)}{S_i(0)} \bigg).
\end{equation*}
Then 
\begin{equation}
\label{eq9}
    X(t) = t\mu^x + W(t) = t \mu^x + L V(t) \sim \mathcal{N} \big(t \mu^x, t\Sigma\big)
\end{equation}
is a multivariate Brownian motion with drift where $\mu^x = (\mu^x_1, \dots, \mu^x_N)^\top \in \mathbb{R}^N$ and  $\mu^x_i = \mu_i - \sigma_i^2/2$, $L \in \mathbb{R}^{N \times N}$ is a lower triangular matrix such that $LL^\top  = \Sigma$, the so-called Cholesky decomposition of $\Sigma$
(\cite{Moler1978} and \cite{Cholesky2000}), and $V(t) = L^{-1}W(t)$ is a $N$-dimensional standard Brownian motion. We adopt the parameterization \eqref{eq9} of the log-return throughout the paper. We use $\mathcal{F}_t := \sigma(W(s) ; s \leq t)$ to denote the natural filtration generated by the Brownian motion $\{W(t), t \in [0,T]\}$. It is easy to see that $V(t)$ generates the same filtration as $W(t)$. 

Equation \eqref{eq9} is analogous to the equilibrium model in the classical single period problem. We now introduce the model for forward-looking views. Analogous to the classical model \eqref{eq3}--\eqref{eq4}, we then  show how they can be used to update the stochastic model of returns.

\subsection{Expert Views}
\label{sec32}
A key difference between our model and \cite{Frey2012}, \cite{Sass2017}, and \cite{TimeDependent_BL}  is that our expert gives forward-looking views about future returns, for instance, a view at time $t_1$ about the return between $t_1$ and $t_2$. As in the classical Black-Litterman model, this is modeled as a noisy observation of a linear mapping of the vector of log-returns $X(t_2) -X(t_1)$. Note that views about the returns can be transformed to views about the log-returns, and vice versa.

An important component of the views model is the specification of the noise in the prediction. We assume that the noise is increasing in the return horizon $t_2 - t_1$.  Let $Y(t_1,t_2) \in \mathbb{R}^K$ be the vector of $K$ views given at time $t_1$, about the value of the log-returns vector at time $t_2$. Conditioned on the true log-returns being $X(t_2) -X(t_1)$, we assume that $Y(t_1,t_2)$ is normally distributed
\begin{equation*}
    Y(t_1,t_2) \,|\, X(t_2) -X(t_1) = P (X(t_2)-X(t_1)) + \sqrt{t_2 - t_1} \, \epsilon \sim \mathcal{N} \big(  P (X(t_2)-X(t_1)), (t_2 - t_1)\Omega \big)
\end{equation*}
where $\epsilon \sim \mathcal{N}\big(0, \Omega)$ and $P \in \mathbb{R}^{K \times N}$ is a linear mapping from returns to views. We assume for simplicity that all views are given at the start of the investment horizon and the horizon of the views match the investment problem, so $t_1 = 0$, $X(t_1)=0$ and $t_2 = T$. That is 
\begin{equation}
\label{eq10}
    Y(0,T) \,|\, X(T) = P X(T) + \sqrt{T} \epsilon \sim \mathcal{N}\big( PX(T), T \Omega\big).
\end{equation}
We show in Section \ref{sec:Extensions} how the model can be extended to accommodate more complex information structures like when views over shorter time periods are given during the investment horizon or old views are revised. 
We make the following assumptions about the covariance matrices for the log-returns ($\Sigma$) and the views ($\Omega$).
\begin{assumption}\label{ass-covariance}
    The covariance matrices $\Sigma$ for the log-returns and $\Omega$ for the views are invertible.
\end{assumption}
This assumption implies that the market is arbitrage free\footnote{It is a sufficient (but not necessary) condition for the market to be arbitrage free.} (see \cite{DHAENE2020112310}) and that expert views are not redundant\footnote{In practice, even if an expert is redundant (his view can be written as a linear combination of other views), we can add a small noise to the view to make $\Omega$ positive-definite.}.   

Figure \ref{fig:figure2} shows a Bayesian network representation of the discrete time version of the problem. An investor at time $\tau \in \{0 ,1,\dots, T\}$  knows the  history of log-returns $\{X(0), \dots, X(\tau - 1)\}$ and has the forward-looking view $Y(0,T)$ that was given at $t = 0$. The noisy forward-looking view changes the distribution of future returns $\{X(\tau), \dots, X(T)\}$.


\begin{figure}
\centering
\begin{tikzpicture}
[
roundnode/.style={circle, draw=black!80, fill=gray!5, very thick, minimum size=7mm},
squarednode/.style={rectangle, draw=black!80, fill=gray!5, very thick, minimum size=5mm},
]
\node[squarednode]      (observation)                              {$Y(0,T)$};
\node[roundnode]        (rT)       [above=of observation] { X(T)};
\node[]        (dots)       [left=of rT] {\dots};
\node[roundnode]        (rtau)       [left=of dots] { $X(\tau)$};
\node[squarednode]        (rtau1)       [left=of rtau] { $X(\tau-1)$};
\node[]        (dots2)       [left=of rtau1] {\dots};
\node[squarednode]        (r1)       [left=of dots2] { $X(1)$};
\node[squarednode]        (r0)       [left=of r1] { $X(0)$};

\draw[->] (rT.south) -- (observation.north);
\draw[<-] (rT.west) -- (dots.east);
\draw[<-] (dots.west) -- (rtau.east);
\draw[<-] (dots2.west) -- (r1.east);
\draw[<-] (r1.west) -- (r0.east);
\draw[<-] (rtau.west) -- (rtau1.east);
\draw[<-] (rtau1.west) -- (dots2.east);

\end{tikzpicture}
\captionsetup{format=hang, singlelinecheck=false, justification=raggedright}
\caption{Bayesian network of the Dynamic Black-Litterman model. \textmd{The figure shows a discrete time version of the problem where $t = 0, \dots, T$, and the noisy view $Y(0,T)$ of the log-return $X(T)$ is revealed at $t = 0$.  Although our model is in continuous time, this figure illustrates the idea that  noisy information about $X(T)$ changes the distribution and dynamics of future values of $X(s)$, $s=\tau+1, \cdots T$ for a investor at time $\tau$.}}
\label{fig:figure2}
\end{figure}
\subsection{Conditional Dynamics of the Asset Price}
\label{sec33}

Absent views, the  log-returns \eqref{eq9} satisfy the stochastic differential equation (SDE)
\begin{eqnarray*}
    dX(t) & = & \mu^x dt  + dW(t) \\
    X(0) & = & 0.
\end{eqnarray*}
The view $Y(0,T) = y$ 
changes the probability measure for the investor from $\mathbb{P}$ to $\mathbb{Q} = \mathbb{P}(\; \cdot \; | \, Y(0,T) = y)$. The following result gives the dynamics of the log-returns after conditioning on the view, which we denote by $X^y(t)$. $\{\mathcal{F}^Y_t, 0\leq t \leq T\}$ where $\mathcal{F}^Y_t := \sigma (\mathcal{F}_t \vee \sigma(Y(0,T)))$ is the filtration generated by the views and the history of prices. The proof of the following result can be found in the Appendix.


\begin{proposition}
\label{proposition1}
Suppose that the price process satisfies \eqref{eq8} and expert views $Y(0,T)$ satisfy \eqref{eq10}. Assume that $PL_j \neq 0$ for $j \in [N]$. Conditional on $Y(0,T) = y$, the log-returns $X(t)$ satisfy
\begin{equation}
\label{eq11}
\begin{split}
        dX^y(t) =&   \bigg(\mu^x + \beta_1 \big(y - T P \mu^x \big) + \beta_2(t) \big(\mathbb{E}[X^y(t)] - X^y(t)\big)\bigg) dt + dW^y(t), \\
        X^y(0) = & 0
\end{split}
\end{equation}
where
\begin{eqnarray*}
    \beta_1 & = & \frac{1}{T}\Sigma P^\top (P \Sigma P^\top+ \Omega)^{-1} \in  \mathbb{R}^{N \times K},\\ [8pt]
    \beta_2(t) & = & \Sigma P^\top \left((T-t) P\Sigma P^\top + T \Omega\right)^{-1} P \in \mathbb{R}^{N \times N},
\end{eqnarray*}
and
\begin{equation*}
   \mathbb{E}[X^y(t)]  =t \left( \mu^x +\beta_1 (y -T P \mu^x ) \right).
\end{equation*}
$W^y(t) \sim \mathcal{N}\big(0,t\Sigma\big)$ is a $N-$dimensional Brownian motion adapted to the filtration $\mathcal{F}_t^Y$. Conditional on the views, the stock price $S^y(t) = S(t) \,|\, (Y(0,T) = y)$ has dynamics
\begin{equation}
\label{eq12}
    dS^y(t) = D(S^y(t)) \big( \tilde{\mu}(t, X^y(t)) dt + dW^y(t)\big)
\end{equation}
where $D(S^y(t))$ a diagonal matrix with elements $\{S^y_i(t), i \in [N]\}$ and 
\begin{equation}
\label{eq:drift-base}
    \tilde{\mu}(t, x) =  \mu  +  \beta_1 \big(y - T P \mu^x\big) + \beta_2(t) \big(\mathbb{E}[X^y(t)] -x \big)
\end{equation}    
is the new drift of the stock price.
\end{proposition}


\begin{remark}
    The condition $PL_j \neq 0$, for $j \in [N]$ ensures that the views give information about each element of the standard Brownian motion $V(t) = L^{-1}W(t)$. In the next section, we show how this condition can be dropped without affecting the results.
\end{remark}

\begin{remark}
\label{rem:Omega}
    Our model assumes that $\Omega$ is known and remains constant for the entire horizon.  An interesting though challenging extension is to allow $\Omega$ to be uncertain but learned over time. For example, by putting a prior on  $\Omega$, the investor's confidence in the expert can be updated as prices are observed.
\end{remark}


\begin{remark}
\label{rem:Girsanov}
The view $Y(0,T) = y$ changes the probability measure for the investor from $\mathbb{P}$ to $\mathbb{Q} = \mathbb{P}(\; \cdot \; | \; Y(0,T) = y)$.  By Girsanov’s Theorem (see \cite{GaussianBridges} for a discussion on the change of measure related to the restriction of a Brownian bridge), it follows from  \eqref{eq11} that  the Radon-Nikodym derivative of $\mathbb Q$ with respect to $\mathbb P$ is
\begin{equation}
\label{eq:CoM}
  \frac{d\mathbb{Q}}{d\mathbb{P}} = \exp\Big(\int_0^{T} k(s)^\top   dW(s) - \frac{1}{2}\int_0^{T} ||k(s)||^2 ds  \Big), \quad t \in [0,T),
\end{equation}
where
\begin{equation*}
    k(t) =  \beta_1 \big(y - T P \mu^x \big) + \beta_2(t) \big(\mathbb{E}[X^y(t)] - X^y(t)\big),\quad t \in [0,T).
\end{equation*}
We show in Appendix \ref{app:VF} that ${\mathbb E}_{\mathbb P}\left[\frac{d\mathbb Q}{d \mathbb P}\right]=1$. 
Under the new measure $\mathbb Q$, 
$W(t)$ is Brownian motion $W^y(t)$ with drift $k(t)$
\begin{equation*}
\begin{split}
           W(t) = W^y(t) + \int_0^t k(s)ds. 
\end{split}
\end{equation*}
\end{remark}



For the information structure in Figure \ref{fig:figure2}, the investor at time $t$ has the forward-looking views $Y(0,T) = y$ provided at the beginning of the investment period and the history of log returns on $[0, t]$ including $X^y(t) = x$. Using this information
\begin{equation*}
    X^y(t+dt) - x 
    =\bigg(\mu^x + \beta_1 \big(y - T P \mu^x \big) + \beta_2(t) \big(\mathbb{E}[X^y(t)] - x \big)\bigg)dt + dW^y(t) 
\end{equation*}
predicts log-returns over the interval $[t, t+dt]$
where $\beta_1$ and $\beta_2(t)$ are coefficients of a linear regression model with covariates $y - TP \mu^x$ and $\mathbb{E}[X^y(t)] - x$, and $dW^y(t)$ is the uncertainty in the prediction.
The conditional price process \eqref{eq12} is no longer Geometric Brownian motion but has a drift which is a function of time, the vector of views $y$, and  the conditional log-return $X^y(t) = x$ as a predictor. 


\subsubsection*{Kalman smoothing}
In the problem of Kalman smoothing, we have a hidden linear stochastic process $\{X(s), 0\leq s \leq T\}$  driven by Gaussian noise, and noisy observations $\{Y(\tau_j)=P X(\tau_j) + \epsilon_j, 0 < \tau_1<\cdots < \tau_m\}$ of the hidden process where the $\epsilon_j$'s are zero-mean Gaussian. Given a (normal) prior on the initial state $X(0)$ and the noisy observations, the objective is to find the joint distribution of all the states ${\mathbb P}[X(s), 0 \leq s \leq T\,|\, Y(\tau_j), j=1, \cdots, m]$.

For the Black-Litterman investor at $t=0$ the yet-to-be-realized log-returns are analogous to the hidden states in the Kalman smoother; there is a (degenerate) prior ($X(0)=0$) and a single observation $Y(0, T)$ of $X(T)$. The distribution of log-returns given the view, $\{X(s),0\leq s \leq T \,|\, X(0)=0, Y(0, T)=y\}$, is analogous to the conditional distribution of the latent process in the smoothing problem and is given by the distribution of $\{X^y(s), 0\leq s \leq T\}$ in \eqref{eq11}. One important difference is that the log-returns $X(t)$  is now being revealed over time and the Black-Litterman investor must account for this information. 

Given the analogy to smoothing, it is natural to allow views of returns over multiple horizons, all expressed at $t=0$, instead of a single view  $Y(0, T)$. This extension is discussed in the Appendix (Section \ref{Sec:EC_ViewsDifferentHorizons}).  We also consider extensions in which views arrive during the investment horizon (Section \ref{sec:Extensions}). All build on the case of a single view of returns expressed at $t=0$. The problem of filtering, where the decision maker receives information at $t$ about the {\it current} value of the latent state, $X(t)$, is more naturally associated with stochastic control problems (\cite{Frey2012,Sass2017}). Forward looking expert views are an example where smoothing equations define state dynamics.

\medskip

\begin{example}
\label{example1}
Suppose we have a single asset with price $S(t) \in \mathbb{R}$ that is geometric Brownian motion with drift $\mu \in \mathbb{R}$ and volatility $\sigma \in \mathbb{R}$. Let $X(t) \in \mathbb{R}$ be its log-return. Given a noisy forward-looking view $y \sim X(T) + \epsilon$ where $\epsilon \sim \mathcal{N}\big(0, T\omega^2\big)$ is independent of $W(t)$, the conditional log returns satisfies
\begin{equation}
\label{eq13}
     dX^y(t) = \left( \mu^x+ \frac{1}{\tilde{T}} ( y - \mu^x T) + \frac{1}{\tilde{T} - t}(\mathbb{E}[X^y(t)] - X^y(t))   \right) dt + \sigma dW^y(t)
\end{equation}
where $\tilde{T} = T(1 + \dfrac{\omega^2}{\sigma^2})$ and $\mu^x = \mu - \sigma^2 / 2$. The stochastic differential equation (SDE) \eqref{eq13} has an explicit solution
\begin{equation}
\label{eq14}
    X^y(t) =  \mu^x t + \frac{t}{\tilde{T}} (y - \mu^x T) + \sigma  (\tilde{T} - t) \int_0^t \frac{1}{\tilde{T} - s} dW^y(s),  \, \, \, \text{for} \, \, \, t \in [0,T].
\end{equation}
\end{example}

We see that the view induces an adjustment in the posterior mean that is proportional to the difference between the view $y$ and $\mu^x T$ so prices  drift upwards if the forecast $y$ exceeds prior expectations $\mu^xT = \mathbb{E}[Y(0,T)]$. For an alternative interpretation  note that 
\begin{equation*}
    X^y(t) = \mu^x t + \sigma B(t),  \, \, \, \text{for} \, \, \, t \in [0,T]
\end{equation*}
where 
\begin{equation*}
    B(t) = \frac{t}{\tilde{T}}\frac{1}{\sigma}(y - \mu^x T) +  (\tilde{T} - t) \int_0^t \frac{1}{\tilde{T} - s} dW^y(s),  \, \, \, \text{for} \, \, \, t \in [0,T]
\end{equation*}
is the restriction to $[0, T]$ of a Brownian bridge from $0$ to $\dfrac{1}{\sigma}(y - \mu^x T)$ with a hitting time  $\tilde{T}\geq T$. 
The view changes the Brownian motion $W(t)$ in \eqref{eq9} to the restriction of a Brownian motion with drift $B(t)$.

\section{Forward-looking Views and Brownian Bridge}
\label{sec4}
Example \ref{example1} shows that the conditional log-returns process for a single asset problem  can be written in terms of a Brownian bridge. We now explore this connection in the multi-asset case. This leads to the notion of generalized multi-dimensional Brownian bridge and an alternative derivation of the conditional log-returns process.

\subsection{One-dimensional case}
\label{sec41}

The classical Brownian bridge (\cite{TheBBAlabama} and \cite{BrownianBridgesPINSKY}) is the stochastic process that is obtained after conditioning on the terminal value of a one-dimensional Brownian motion $W(t)$ on the closed interval $[0, T]$.

\begin{definition}
\label{definition1}
     Let $W(t) \in \mathbb{R}$ be a Brownian motion with initial value $W(0) = a$. Then the process  $\{B(t) = (W(t) \,|\, W(T) = y) , t \in [0, T]\}$ is called a Brownian bridge from $a$ to $y$ with hitting time $T$.
\end{definition}


The following result characterizes properties of one-dimensional Brownian bridge (\cite{JeanBlanc} and \cite{GaussianBridges}).

\begin{proposition}
\label{SDE:BB}
    A stochastic process $B(t) \in \mathbb{R}$ is a Brownian bridge from $a$ to $y$ with hitting time $T$ if and only if it satisfies the following 5 conditions:
    \begin{enumerate}
        \item $B(0) = a$ and $B(T) = y$ (with probability $1$),
        \item $\{B(t), t \in [0,T]\}$ is a Gaussian process,
        \item $\mathbb{E}[B(t)] = a + \frac{t}{T}(y-a)$ for $t \in [0,T]$,
        \item $cov\big(B(t), B(s) \big) = \min\{s,t \} -\dfrac{st}{T}$, for $s,t \in [0,T]$,
        \item With probability $1$, $t \to B(t)$ is continuous in $[0,T]$.
    \end{enumerate}
    The Brownian bridge $B(t)$ is the solution to the SDE
     \begin{equation}
    \begin{cases}
                dB(t) &= \dfrac{y-B(t)}{T - t} dt + dW^y(t)\\
B(0) &= a,
    \end{cases}
    \label{eq:1dBbSDE}
    \end{equation}
        where $W^y(t)$ is a Brownian motion. The explicit solution of this equation is 
    \begin{equation}
    \label{eq:1dBb}
        B(t) = a + \frac{t}{T}(y - a) + (T - t) \int_0^t \frac{1}{T - s} dW^y(s).
    \end{equation}
\end{proposition}

\begin{figure}[ht]
 \begin{center}
       \includegraphics[width= 0.6\textwidth]{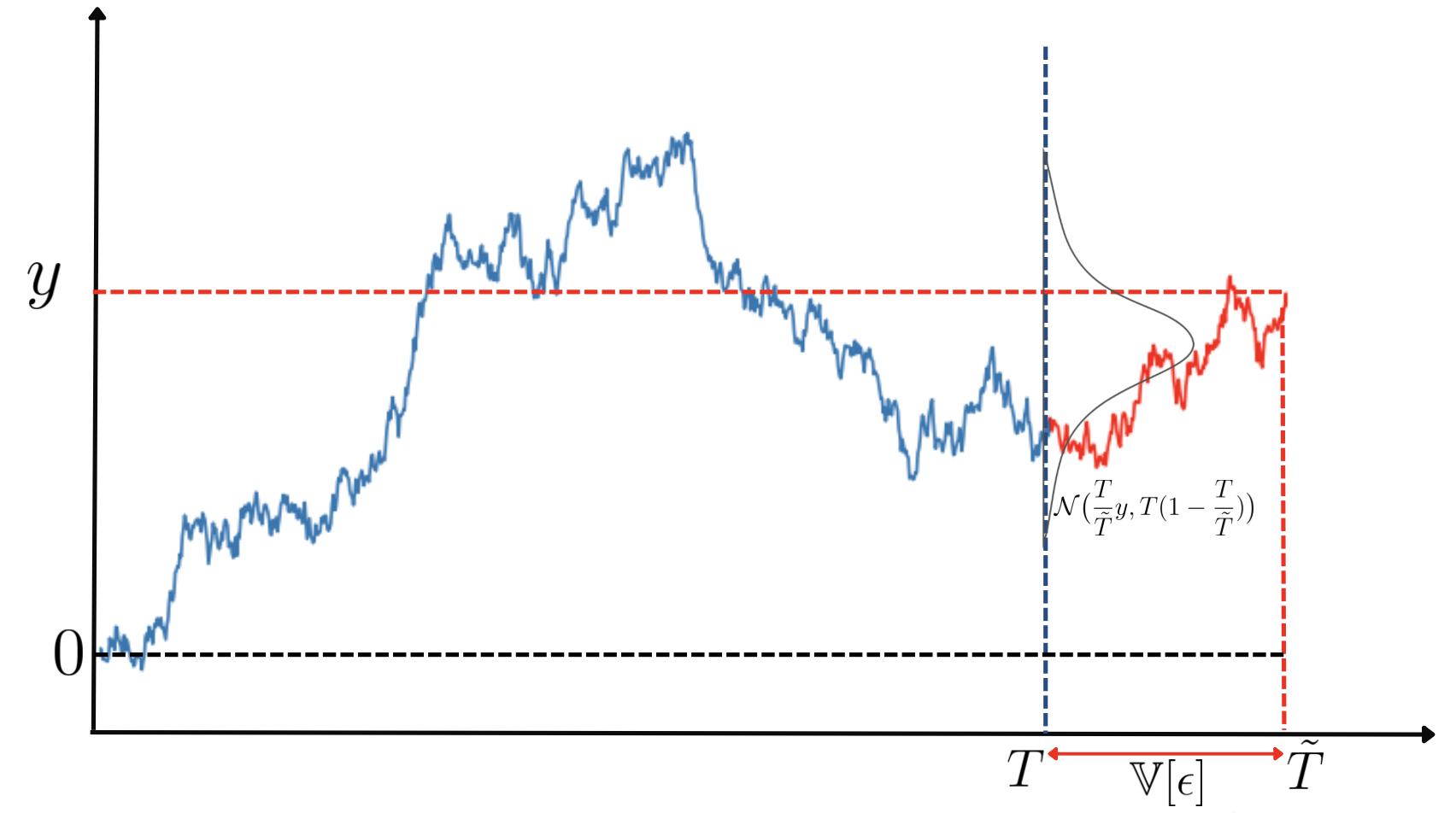}
     \caption{Relationship between the forward-looking view and the Bb. \textmd{The view $Y(0,T) = y$ changes the distribution of the Brownian motion $W(t)$, for $t \in [0,T]$, to a Brownian bridge hitting $y$ at a time $\tilde{T} = T + \mathbb{V}[\epsilon]$}.}
      \label{fig:Figure3}
 \end{center}
\end{figure}

We now generalize Definition \ref{definition1} to the case when we have a noisy observation of the terminal value of a Brownian motion of the form $Y(0, T) = W(T) + \epsilon$. Conditional on $Y(0, T)$, the Brownian motion is the restriction to $[0, T]$ of a Brownian bridge with a hitting time $\tilde{T} >T$ (Figure \ref{fig:Figure3}). This enables us to connect the conditional log returns process from the Black-Litterman model and Brownian bridge.
\begin{proposition}
\label{proposition2}    
Let $W(t) \in \mathbb{R}$ be a standard Brownian motion such that $W(0) = a$. Let $T > 0$ and suppose we observe a sample from $Y(0,T) = W(T) + \epsilon$ at $t = 0$ where $\epsilon \sim \mathcal{N}\big(0, T \omega^2\big)$ is independent of $W(T)$. Then the stochastic process $\{B(t) = (W(t) \,|\, Y(0,T) = y), t \in [0,T]\}$ is a restriction of a Brownian bridge from $a$ to $y$ with hitting time $\tilde{T} = T(1 + \omega^2)$ to the interval $[0,T]$. Additionally, $B(t)$ is the solution the SDE
 \begin{equation*}
    \begin{cases}
                dB(t) &= \dfrac{y-B(t)}{\tilde{T} - t} dt + dW^y(t), \; t \in [0, T] \\
B(0) &= a,
    \end{cases}
    \end{equation*}
which is given by
    \begin{equation*}
        B(t) = a + \frac{t}{\tilde{T}}(y - a) + (\tilde{T} - t) \int_0^t \frac{1}{\tilde{T} - s} dW^y(s), \; t \in [0, T].
    \end{equation*}
\end{proposition}

Returning to Example \ref{example1}, observe that
\begin{equation*}
\begin{split}
    X^y(t) &=  (X(t)\,|\, Y(0,T) = y) \\
    &= \mu^x t + \sigma \big( W(t)\,|\,Y(0,T) = y \big).
    \end{split}
\end{equation*}
It now follows from Proposition \ref{proposition2} that  $B(t) = (W(t) \,|\, Y(0,T) = y)$ is the restriction to the interval $[0,T]$ of a Brownian bridge from $0$ to $\frac{1}{\sigma}(y - \mu^x T)$ with hitting time $\tilde{T} = T(1 + \frac{\omega^2}{\sigma^2})$, and hence
\begin{equation*}
    X^y(t) = \mu^x t + \sigma B(t).
\end{equation*}

\subsection{Multidimensional Case}
\label{sec42}
In the multi-asset case, observe that the conditional log-returns process \eqref{eq11} can be written
\begin{eqnarray*}
    X^y(t) & = & \Big(X(t) \,| \, Y(0, T) = y\Big) \\
    & = & \mu^xt + \Big( W(t) \, | \, Y(0, T) = y - T P \mu^x\Big) \\ & = & \mu^x t + B(t)
\end{eqnarray*}
where  $Y(0, T) = P W(T) + \epsilon$ 
is a vector of noisy  observations of the terminal value $W(T)$ of an $N$-dimensional correlated Brownian motion with view matrix $P$ and normal random vector $\epsilon$. This suggests we consider  
\begin{eqnarray}
\label{eq:noisy_multi_Bb}
    \Big\{B(t) = \big(W(t) \,|\, Y(0, T)= y\big), \; t \in [0, T] \Big\}
\end{eqnarray} 
which generalizes the setting considered in Proposition \ref{proposition2} to the multi-dimensional case.
The multi-dimensional version is more nuanced because information about each component  of $W(t)$ is obtained not only  through $Y(0, T)$ but  the noisy observations of other components of $W(t)$ which are correlated. We illustrate this in the following example.


\medskip

\begin{example}
\label{example2}
Let $W_1(t)$ and $W_2(t)$ be standard Brownian motions with correlation $\rho \in (0,1]$
\begin{equation*}
    dW_1(t) \, dW_2(t) = \rho \, dt.
\end{equation*}
Let $\epsilon \sim \mathcal{N}\big(0, \omega^2\big)$ and suppose we have a noisy observation $Y(0,T) = W_2(T) + \epsilon$ of $W_2(T)$. Let  
$\big\{B(t) \equiv \big(B_1(t), B_2(t)\big)^\top = (W(t) \,|\, Y(0, T) = y), t \in [0,T]\big\}$
be the distribution of $W(t)$ conditioned on the vector of observations $Y(0, T)=y$.
It follows from Proposition \ref{proposition2} that $B_2(t)$ is the restriction to $[0,T]$ of a Brownian bridge $(W_2(t) \,|\, W_2(\tilde{T}_2) = y)$ with terminal value $W_2(\tilde{T}_2) = y$ at $\tilde{T}_2 = T + \omega^2$. Additionally, the correlation between the two Brownian motions transforms the view about $W_2(T)$ to a noisy observation of $W_1(T)$ and it can also be shown (see Appendix \ref{App:example2}) that $B_1(t)$ is the restriction of a Brownian bridge $(W_1(t) \,|\, W_1(\tilde{T}_1) = \frac{y}{\rho})$ with terminal value $W_1(\tilde{T}_1) =  \frac{y}{\rho}$ at $\tilde{T}_1 = \frac{T+\omega^2}{\rho^2}$ to $[0,T]$. Note that $\tilde{T}_1 \geq \tilde{T}_2$ with equality if and only if $\rho = 1$; there is less information about $W_1$ than $W_2$. If $\rho = 0$, then $B_1(t)$ is unaffected by the observation of $W_2(T)$ and remains a standard Brownian motion. Plots showing the impact of correlation and view noise on hitting times can be found in Appendix \ref{sec:BBexample}.
\end{example}


\subsubsection{Main Results.}


The following result gives properties of $B(t)$. In contrast to the one-dimensional setting in Proposition \ref{proposition2}, $B(t)$ is not a Brownian bridge but a linear combination of (dependent) one-dimensional Brownian bridges from $0$ to $0$, each with its own hitting time.

\begin{proposition}
    \label{proposition3}
    Let $W(t)$ be an $N$-dimensional Brownian motion such that $W(t) \sim \mathcal{N} \big(a, t \Sigma \big)$, where $\Sigma \in \mathbb{R}^{N \times N}$ is symmetric positive definite with Cholesky decomposition $L$ $(\Sigma = LL^\top)$. Let  $Y(0,T) = P W(T) + \epsilon$ where $P \in \mathbb{R}^{K \times N}$ is such that $PL_j \neq 0$ for $j \in  [N]$ and $\epsilon \sim \mathcal{N}\big(0, T \Omega \big)$ for some symmetric positive definite covariance matrix $\Omega \in \mathbb{R}^{K \times K}$. Given $Y(0,T) = y$, the stochastic process $\{B(t) = (W(t) \,|\, Y(0,T) = y) , t \in [0, T]\}$ satisfies
     \begin{enumerate}
        \item $B(0) = a$ (with probability $1$). 
        \item $B$ is a Gaussian process.
      \item For $t \in [0,T]$, $$\mathbb{E}[B(t)] = a + t \beta_1\big(y - Pa\big)$$  where $$\beta_1 = \frac{1}{T}\Sigma P^\top (P\Sigma P^\top + \Omega)^{-1}.$$
        \item $\Cov\big(B(t), B(s) \big) = L\big(\min\{s,t \} I_N - st H\big)L^\top$  for $s,t \in [0,T]$
        where 
    \begin{equation*}
        H = \frac{1}{T} (PL)^\top(P \Sigma P^\top+ \Omega)^{-1} PL \in \mathbb{R}^{N \times N}
    \end{equation*}
    is a symmetric, positive-definite $N \times N$  matrix.
        \item With probability $1$, $t \to B_i(t)$ is continuous in $[0,T]$ for $i \in [N]$.
    \end{enumerate}
\end{proposition}

Let
\begin{eqnarray*}
\bar{B}(t) = L^{-1}(B(t) - \mathbb{E}[B(t)]).
\end{eqnarray*}
It follows from Theorem \ref{proposition5} that $\bar{B}(0) = 0$, $\mathbb{E}[\bar{B}(t)] = 0$,  $\bar{B}(t)$ is a Gaussian process with continuous sample paths on $[0,T]$ and $\Cov\big(\bar{B}(t), \bar{B}(s) \big) = \big(\min\{s,t \} I_N - st H\big)$ (property 4). In particular,  
\begin{equation}
\label{eq:covBbar}
Cov\big(\bar{B}_i(t), \bar{B}_j(s) \big) =  \begin{cases}
        \min\{s,t \} - \dfrac{st}{\tilde{T}_i}, \, \, \, \text{if} \, \, \, i = j,\\
        - \dfrac{st}{H_{i,j}}, \, \, \, \text{if} \, \, \, i \neq j,\\
    \end{cases}
    \end{equation}
so 
$var(\bar{B}_i(\tilde{T}_i))=0$ and  $\bar{B}_i(\tilde{T}_i)=0$ at the hitting time
\begin{equation}
\label{eq15}
    \tilde{T}_{i} = \frac{1}{H_{i,i}} =  T \big((PL_i)^\top(P \Sigma P^\top+ \Omega)^{-1} PL_i\big)^{-1}.
\end{equation} 
It follows from Proposition \ref{SDE:BB} that each component $\bar{B}_i(t)$ of $\bar{B}(t)$ is the restriction to $[0, T]$ of a one-dimensional Brownian bridge from $0$ to $0$ with hitting time $\tilde{T}_{i}$. Note that the diagonal elements of the matrix $H$, and hence the hitting times $\tilde{T} = [\tilde{T}_1,\cdots,\,\tilde{T}_N]^\top \in \mathbb{R}^N$, are strictly positive because the covariance matrices $\Sigma$ and $\Omega$ are positive definite and  $PL_i \neq 0$ (by assumption). It can also be shown (Proposition \ref{proposition4} below) that $\tilde{T}_i \geq T$.

Since $\bar{B}_i(t)$ is the restriction to $[0,\, T]$ of a one-dimensional Brownian bridge, it follows from Proposition \ref{proposition2}  that it satisfies the SDE
\begin{eqnarray*}
\left\{\begin{array}{l}
   \begin{displaystyle} d\bar{B}_i(t) = \frac{y - \bar{B}_i(t)}{\tilde{T}_i - t} dt + dW_i^y(t),\; t\in[0, T]\end{displaystyle} \\ \vspace{-0.15cm}\\
 \bar{B}_i(0) = 0.  
 \end{array}\right.
\end{eqnarray*}
The following result gives a characterization of $\{B(t) = (W(t) \,|\, Y(0, T)= y), \, t \in [0,T]\}$ in terms of the multi-dimensional Brownian bridge $\bar{B}(t)$ and the SDE for $\bar{B}(t)$. Since the components of $\bar{B}(t)$ are correlated, the SDE for $\bar{B}(t)$ is not a simple ``stacking together" of the SDEs for its components. The proof is in Appendix \ref{App:thm1}.

\begin{theorem}
    \label{theorem1}
    Consider the stochastic process $\{B(t), \, t \in [0,T]\}$ defined in Proposition \ref{proposition3}. Then
\begin{equation}
    \label{eq16}
        B(t) = a + t \beta_1\big(y - Pa\big) + L \bar{B}(t)
    \end{equation}
    where
    \begin{equation*}
        \beta_1 = \frac{1}{T}\Sigma P^\top (P\Sigma P^\top + \Omega)^{-1}
    \end{equation*}
    and $\bar{B}(t)$ is a solution of the SDE
\begin{equation}
\label{eq17}
    d\bar{B}(t) 
   = - dt \cdot \bar{\beta}_2(t) \bar{B}(t) + dV^y(t).\\
\end{equation}
Here, $V^y(t)$ is a standard $N-$dimensional Brownian motion adapted to the enlarged filtration $\mathcal{F}_t^Y$ and
\begin{equation*}
       \bar{\beta}_2(t) =  \frac{1}{T} (PL)^\top \big( (1 - \frac{t}{T}) P \Sigma P^\top + \Omega)^{-1}PL \in \mathbb{R}^{N \times N}.
\end{equation*}
$\{\bar{B}(t) \in \mathbb{R}^N, \, t \in  [0,T]\}$ is a $0$ mean stochastic process and each element 
$\{\bar{B}_i(t), \, t \in [0,T]\}$ $(i \in [N])$ is a restriction to the interval $[0,T]$ of a Brownian bridge from $0$ to $0$ with hitting time $\tilde{T}_i$ defined in \eqref{eq15}. The covariance matrix of $\bar{B}(t)$ is \eqref{eq:covBbar}.
\end{theorem}

\medskip

\begin{remark}
To understand the condition $PL_j \neq 0$, notice that when $PL_j = 0$, the view $Y(0,T) = y$ gives no additional information about the Brownian motion $\{V_j(t), t \in [0,T]\}$, where $V(t) = L^{-1}W(t)$ is the vector of $N-$independent Brownian motions derived from $W(t)$. Therefore, $Y(0,T)$ and $V_j(t)$ are independent and the Brownian bridge $\{V_j(t) \,|\, Y(0,T) = y, \, t \in [0,T]\}$ remains a Brownian motion (in this case, its hitting time $\tilde{T}_{j}$ is infinite). The condition can be dropped by separating the elements of $B(t)$ into a set $\mathcal{I}$ where $PL_i \neq 0$ for $i \in \mathcal{I}$, and a set $\mathcal{J}$ where $PL_j = 0$ for $j \in \mathcal{J}$, with $\mathcal{I} \cup \mathcal{J} = [N]$, and $\mathcal{I} \cap \mathcal{J} = \emptyset$. The process $B_\mathcal{I}(t) = \{ B_i(t), \, i \in \mathcal{I}\}$ is then a vector of Brownian bridges and Theorem \ref{theorem1} applies, whereas $B_\mathcal{J}(t) = \{ B_j(t), \, j \in \mathcal{J}\}$ is a vector of Brownian motions where $B_j(t) = V_j(t)$ for $j \in \mathcal{J}$. Without loss of generality, we assume in the rest of the paper that $PL_j \neq 0$ for all $j \in [N]$.
\end{remark}

\subsubsection{Hitting Times of the Brownian Bridges.} 
\label{sec422}

We saw in Example \ref{example1}  that the hitting time for a 1-dimensional Brownian bridge induced by a noisy forward-looking view is monotone in the noise in the view. The following result shows that this is also true in the multi-dimensional setting.
Here we say that a covariance matrix $\Omega^1$ is greater than or equal another covariance matrix $\Omega^2$ ($\Omega^1 \succeq \Omega^2$) if their difference is positive semi-definite ($\Omega^1 - \Omega^2 \succeq 0$)
\begin{proposition}
\label{proposition4}
    Consider the process $\{\bar{B}(t), t \in [0,T]\}$ satisfying \eqref{eq17}. Then for each $i \in [N]$, the hitting time $\tilde{T}_i$ satisfying \eqref{eq15} of the Brownian bridge $\{\bar{B}_i(t), \, t \in [0,T]\}$, is strictly larger than the views horizon $T$ and is increasing in the views covariance matrix $\Omega$. 
\end{proposition}
The proof is given in Appendix \ref{App:PropHittingTimes}.

\subsection{Application: Black-Litterman model}
\label{Application_BL}

Recall the price process \eqref{eq8}, log-returns process $X(t)$ given by \eqref{eq9} and the vector of views $Y(0, T)$ in \eqref{eq10}. Given $Y(0,T) =y$, the conditional log-returns
    \begin{equation}
    \label{eq:X^y_B(t)}
    X^y(t) = t \mu^x + B(t)
\end{equation}
where 
\begin{eqnarray*}
    B(t) = \big(W(t) \,|\, Y(0, T) = y\big) = \big(W(t) \,|\, P W(T) + \epsilon = y - T P \mu^x\big)
\end{eqnarray*}
is a generalized Brownian bridge. By Theorem \ref{theorem1},  
\begin{equation*}
 B(t) =   t \beta_1 (y - T P \mu^x) + L \bar{B}(t)
\end{equation*}
where
\begin{eqnarray*}
    \beta_1 = \frac{1}{T}\Sigma P^\top (P\Sigma P^\top + \Omega)^{-1} \in \mathbb{R}^{N \times K}
\end{eqnarray*}
and $\bar{B}(t)$ is a vector of dependent Brownian bridges
that solve the SDE \eqref{eq17} with 
\begin{equation*}
\bar{\beta}_2(t) =  \frac{1}{T} (PL)^\top \big( (1 - \frac{t}{T}) P \Sigma P^\top + \Omega)^{-1}PL.
\end{equation*}

To recover the SDE \eqref{eq11} for $X^y(t)$, define
\begin{equation*}
\begin{split}
            \beta_2(t) = L \bar{\beta}_2(t) L^{-1}
        =   \Sigma P^\top \left((T-t) P\Sigma P^\top + T \Omega\right)^{-1} P \in \mathbb{R}^{N \times N}.
        \end{split}
\end{equation*}

Observing that $L \bar{B}(t)= X^y(t) - \mathbb{E}[X^y(t)]$, it follows from \eqref{eq17} that
\begin{eqnarray}
\label{eq:LB}
L \bar{B}(t) = - \int_0^t \beta_2(s) (X^y(s) - \mathbb{E}[X^y(s)])ds + L V^y(t)
\end{eqnarray}
where $V^y(t)$ is an $N$-dimensional standard Brownian motion and
\begin{equation*}
   B(t) = t \beta_1 (y - T P \mu^x) - \int_0^t \beta_2(s) (X^y(s) - \mathbb{E}[X^y(s)])ds + W^y(t)
\end{equation*}
where $W^y(t) = LV^y(t)$
is an $N$-dimensional Brownian motion with $W^y(t) \sim N(0, t \Sigma)$.  It follows that $X^y(t) = \mu^x t  + B(t)$ solves \eqref{eq11}. This derivation also shows that  $\mu^x + \beta_1(y - T P \mu^x)$ is the drift of the conditional log-return \eqref{eq11} and \eqref{eq:LB} is a linear combination of $N$ dependent Brownian bridges $\bar{B}(t)$, which generalizes the observation from Example \ref{example1} to the multi-asset case.

\section{Optimal Portfolio Choice}
\label{sec5}
The investor maximizes the expected utility of terminal wealth over dynamic portfolios conditional on views $Y(0,T) = y$ satisfying \eqref{eq10}. By Proposition \ref{proposition1}, risky asset prices are given by \eqref{eq12} where the factor $X^y(t)$ in the drift  is the conditional log-returns process \eqref{eq11}.
The investor dynamically chooses the proportion $\pi(t)$ of her wealth to be invested in the risky assets. 
The class of admissible policies is
\begin{equation*}
    \mathcal{A} = \left\{ \pi: [0,T] \to \mathbb{R}^N, \pi \text{ is adapted to } \{\mathcal{F}_t^Y\}_{t \in [0,T]}, \int_{0}^T |\pi(t)|^2 dt \leq \infty\right\}.
\end{equation*}
Under the assumption that the portfolio is self-financing, the investor's wealth satisfies
\begin{equation}
\label{eq18}
    dZ(t) = Z(t) \bigg( r_f dt + \pi(t)^\top \big(\tilde{\mu}(t,X^y(t)) - r_f \mathbf{1}_N\big) dt + \pi(t)^\top dW^y(t) \bigg)
\end{equation}
where $r_f$ is the risk-free rate.  We assume that the investor has an isoelastic utility function
\begin{equation*}
    U(Z) = \frac{Z^{1 - \gamma}}{1 - \gamma}
\end{equation*}
with relative risk aversion $\gamma$ and maximizes the expected utility of her terminal wealth at the end of the horizon $T$. We assume throughout that $\gamma > 1$. Results are possible when $0<\gamma<1$ though investment strategies in this regime can be excessively risky and the value function might not be finite if the investment horizon is too long. For a  discussion of why strategies in this regime are best avoided, see \cite{Fleming2001} and \cite{ziemba2003stochastic}. The limit $\gamma \to 1$ corresponds to the log-utility.
The investor's value function is
\begin{equation*}
    V(t,z,x) = \max_{\pi \in \mathcal{A}} \mathbb{E}\big[ U(Z(T)) \,|\, X^y(t) = x, Z(t) = z  \big]
\end{equation*}
where $X^y(t)$ and $Z(t)$ satisfy \eqref{eq11} and \eqref{eq18}. 

\subsection{Value Function and Optimal Policy}
\label{sec51}
The Hamilton-Jacobi-Bellman (HJB) partial differential equation is
\begin{equation}
\label{eq:HJB-base}
\begin{split}
    \max_{\pi} \Big\{&\frac{\partial V}{\partial t} +  z \big( r_f + \pi(t)^\top(\tilde{\mu}(t,x) - r_f \mathbf{1}_N \big) \nabla_z V
  +  \big(\tilde{\mu}(t,x) - \frac{1}{2}\diag(\Sigma)\big)^\top \nabla_x V    + \frac{1}{2} z^2 \pi(t)^\top \Sigma \pi(t)  \nabla^2_z V \\ & + \frac{1}{2} \Tr(\Sigma \nabla_x^2 V)  + z \pi^\top(t) \Sigma \nabla^2_{x,z} V\Big\} = 0
\end{split}
\end{equation}
with terminal condition
\begin{equation*}
    V(T,z,x) = \frac{1}{1-\gamma}z^{1-\gamma}.
\end{equation*}
The optimal investment policy is
\begin{equation}
\label{eq:opt-policy}
    \pi^*(t, x) = \underbrace{- \frac{\nabla_z V}{ z \nabla_z^2 V} \Sigma^{-1}(\tilde{\mu}(t,x) - r_f \mathbf{1}_N \big)}_{\text{Mean-Variance Holding}} \underbrace{- \frac{ \nabla^2_{x,z}V}{ z \nabla_z^2 V}}_{\text{Hedging}}.
\end{equation}
The first component of the optimal portfolio is the mean-variance holding while the second hedges changes in the value function that are driven by changes in the factor $X^y(t)$ (\cite{ChackoViceira2005}). 
The following result shows that we can compute the value function and optimal policy by solving a system of ordinary differential equations. (ODEs). We derive explicit expressions for the solutions of these equations, including the nonlinear Ricccati equation, in the next section. 
\begin{proposition}
\label{prop:VF1}
Suppose Assumption \ref{ass-covariance} holds. The solution of the HJB equation \eqref{eq:HJB-base} is  
\begin{equation}
        V(t,z,x) = \frac{z^{1 - \gamma}}{1 - \gamma} \exp(g(t,x))
        \label{eq:V-base}
    \end{equation}
    where
    \begin{equation*}
        g(t,x) = \frac{1}{2}x^\top A(t)x + x^\top b(t) + c(t)
    \end{equation*}
    and $A:[0, T]\rightarrow {\mathbb R}^{N\times N}$, $b:[0, T]\rightarrow {\mathbb R}^{N}$ and $c:[0, T]\rightarrow {\mathbb R}$ are unique solutions of a system of ordinary differential equations (ODEs) 
\begin{eqnarray}
    & & \left\{\begin{array}{l}
    A'(t) + \dfrac{1 - \gamma}{\gamma} \eta_t \Sigma \eta_t + \dfrac{1}{\gamma} (A(t) \Sigma \eta_t + \eta_t \Sigma A(t)) + \dfrac{1}{\gamma} A(t) \Sigma A(t) = 0,   
    \label{eq19}
    \\\vspace{-0.2cm}\\
         A(T) = 0, 
         \end{array}\right. \\ [5pt]
  & &      \left\{\begin{array}{l}
         b'(t) + \dfrac{1}{\gamma}\big(\eta_t + A(t)  \big) \Sigma b(t) + \dfrac{1 - \gamma}{\gamma}\big( \eta_t +A(t) \big)(\alpha_t - r_f \mathbf{1}_N) + A(t) \big(\alpha_t - \dfrac{1}{2}\diag(\Sigma) \big) = 0, \\ \vspace{-0.2cm}\\
        b(T) = 0, 
            \end{array}\right. 
            \label{eq20} \\[5pt]
 & &   \left\{
    \begin{array}{l}
    c'(t) + (1 - \gamma) r_f + \frac{1}{2} \Tr\big(A(t) \Sigma \big) + \frac{1 - \gamma}{2 \gamma}\big(\alpha_t - r_f \mathbf{1}_N \big)^\top \Sigma^{-1} \big(\alpha_t - r_f \mathbf{1}_N\big)  \\ \vspace{-0.2cm}\\
         + \big(\alpha_t - \frac{1}{2}\diag(\Sigma)\big)^\top b(t) + \frac{1 - \gamma}{\gamma} \big(\alpha_t - r_f \mathbf{1}_N\big)^\top b(t) + \frac{1}{2\gamma} b^\top (t) \Sigma b(t) = 0, \\ \vspace{-0.2cm} \\
          c(T) = 0,
          \end{array}\right.
          \label{eq20b}
\end{eqnarray}
where
\begin{equation}
    \begin{split}
    \eta_t & =  -P^\top ((T - t) P \Sigma P^\top + T \Omega)^{-1} P,\\
 \alpha_t &= \mu + \beta_1 (y - TP \mu^x) - \Sigma \eta_t \mathbb{E}[X^y(t)].
 \end{split}
 \label{eq:alpha_and_eta}
    \end{equation}
$A(t)$ is strictly negative definite when $t<T$. There is a unique optimal  policy
        \begin{equation}
        \label{eq21a}
        \pi^*(t, x) = \frac{1}{\gamma} \Sigma^{-1} \big(\tilde{\mu}(t,x) - r_f \mathbf{1}_N \big) +\frac{1}{\gamma}\frac{\partial g}{\partial x}(t,x)
    \end{equation}
    where $\tilde{\mu}(t, x)$ is given by \eqref{eq:drift-base} and
\begin{eqnarray}
\label{eq22a}
    \frac{1}{\gamma}\frac{\partial g}{\partial x}(t,x)   =  \frac{1}{\gamma} \big( A(t) x + b(t)\big).
\end{eqnarray}
\end{proposition}

\proof{Proof}
The ODEs for $A(t)$, $b(t)$ and $c(t)$ can be derived by substituting \eqref{eq:V-base} into the HJB equation \eqref{eq:HJB-base}.
The optimal policy \eqref{eq21} is the maximizer in the HJB equation. Details can be found in Section \ref{app:policy} of the Appendix. 
    \Halmos
\endproof

Aside from being more economically reasonable, the assumption $\gamma > 1$ ensures the solution of the Riccati equation remains bounded. If $0<\gamma<1$, it may be possible for $A(t)$ to become unbounded when solving \eqref{eq19} backwards from $T$, the so-called finite escape time phenomenon.

\subsection{Explicit solutions}
\label{sec52}
 
The following result gives explicit expressions for the solutions of the ordinary differential equations \eqref{eq19}--\eqref{eq20} and the optimal dynamic portfolio. 
\begin{theorem}
\label{proposition5}
    Suppose $\gamma > 1$. The solution to the HJB equation \eqref{eq:HJB-base} is given by \eqref{eq:V-base} with coefficients $A:[0, T]\rightarrow {\mathbb R}^{N\times N}$, $b:[0, T]\rightarrow {\mathbb R}^{N}$ and $c:[0, T]\rightarrow {\mathbb R}$ where
    \begin{eqnarray}
    \label{eq:Riccati-soln}
        A(t) & = &      
        \frac{1}{2}\left\{M(t)  \eta_t + \eta_t^\top M(t)^\top\right\},\\ [5pt]
    \label{eq:b}
        b(t) & = & M(t) \Sigma^{-1} (\alpha_t - r_f \mathbf{1}_N),
\end{eqnarray}
with $\alpha_t$ and $\eta_t$ are given by \eqref{eq:alpha_and_eta} and
    \begin{eqnarray*}
    M(t) & = & (\gamma - 1) (1 - \frac{t}{T}) P^\top \Omega^{-1}P \Big(\gamma\Sigma^{-1} + (1 - \frac{t}{T}) P^\top \Omega^{-1} P\Big)^{-1}, 
    \end{eqnarray*}
and $c:[0, T]\rightarrow{\mathbb R}$ is the solution of \eqref{eq20b}. $A(t)$ is strictly negative definite when $t<T$. The unique optimal allocation policy is
        \begin{equation}
        \label{eq21}
        \pi^*(t, x) = \frac{1}{\gamma} \Sigma^{-1} \big(\tilde{\mu}(t,x) - r_f \mathbf{1}_N \big) +\frac{1}{\gamma}\frac{\partial g}{\partial x}(t,x)
    \end{equation}
    where $\tilde{\mu}(t, x)$ is given by \eqref{eq:drift-base} and the intertemporal hedging demand
\begin{eqnarray}
\label{eq22}
    \frac{1}{\gamma}\frac{\partial g}{\partial x}(t,x) 
    = \frac{1}{\gamma} M(t) \Sigma^{-1} \big(\tilde{\mu}(t,x) - r_f \mathbf{1}_N \big). 
    \nonumber
\end{eqnarray}
\end{theorem}

\proof{Proof}
The derivation of the explicit expressions for $A(t)$ and $b(t)$ is given in Appendix \ref{App:thm2}.
    \Halmos
\endproof

It is interesting that the matrix Riccati equation \eqref{eq19} has an explicit solution \eqref{eq:Riccati-soln}, which is typically not the case. In fact, it can also be shown (see Appendix \ref{App:thm2})
\begin{eqnarray*}
    A(t) =  - \eta_t \Big(\frac{\gamma}{(\gamma - 1)(T-t)}  \Sigma^{-1} + \eta_t \Big)^{-1}\eta_t\, , \; t \in [0,T).
\end{eqnarray*}

As observed in \eqref{eq:opt-policy} the optimal policy \eqref{eq21} consists of a mean-variance term\footnote{\label{footnote:kelly} The first term in  \eqref{eq21} can also be interpreted as a fractional Kelly strategy (\cite{DavisLleoJBF}).}
$\pi_{\text{MV}}^*(t) =\frac{1}{\gamma}\Sigma^{-1} \big(\tilde{\mu}(t,x) - r_f \mathbf{1}_N \big)$
and a hedge $\frac{1}{\gamma} \frac{\partial g}{\partial x}(t,x)$ (\cite{ChackoViceira2005}).
Returns from the hedge offset changes in the value function that occur when the predictor $X^y(t)$ changes, which reduces the risk for the investor.
Intuitively, suppose there is a single asset and $X^y(t) = x$ is the current value of the predictor. Since  $A(t)$ is strictly negative when $t < T$,  $g(t, x)$ is strictly concave in $x$ with a global maximum at $x_0(t) = - A(t)^{-1}b(t)$
; equivalently  $V(t,z,x)$ has a global minimum at $x_0(t)$ (recall that $\gamma > 1$ in \eqref{eq:V-base} so $V(t, z,x)$ is negative). 
If $X^y(t) = x < x_0(t)$, an  increase (decrease) in the predictor $X^y(t)$ results in a decrease (increase) in the value function (i.e., $\nabla_x V(t, z, x)<0$). 
Since the change in the predictor $dX^y(t)$  and the asset price $dS^y(t)$  are positively correlated, a long position in the risky asset   generates a return that offsets the decrease in the value function; that is, the hedging demand is positive when $x<x_0(t)$.
A similar argument explains why the hedging demand is negative 
when $x > x_0(t)$.

The magnitude of the hedge is determined by the matrix $M(t)$. $M(t)$ is the ratio of the precision of the views $(1 - \frac{t}{T}) P^\top \Omega^{-1}P$ to the precision of the return 
$\gamma\Sigma^{-1} + (1 - \frac{t}{T}) P^\top \Omega^{-1} P$
(with adjustments for risk-aversion) and measures the information content of the views. 
The value function is more sensitive to changes in the predictor $X^y(t)$ when views are informative ($M(t)$ is large), requiring a larger hedge. Consequently,  the hedge is increasing in $M(t)$. $M(t)$ and hence the hedging demand vanish as we approach the  maturity date $T$ since the value function becomes less sensitive to changes in the predictor $X^y(t)$ as the time remaining in the market diminishes.  Finally,  $M(t)=0$ when $\gamma = 1$ so there is no hedging demand for an investor with log-utility; this is not too surprising.

The following result gives an alternative expression for the optimal holdings.
\begin{corollary}
\label{theorem2}
Let $\gamma>1$. The optimal holding for the dynamic Black-Litterman investor is
    \begin{equation}
    \label{eq:MPBL_final}
    \begin{split}
             \pi^*(t) 
             &= \frac{1}{\gamma} \left(\Sigma_{\text{DBL}
             }(t)\right)^{-1} \big(\tilde{\mu}(t,x) - r_f \mathbf{1}_N \big)
    \end{split}
    \end{equation}
    where 
    \begin{equation*}
    \Sigma_{\text{DBL}}(t) = \Big( \Sigma^{-1} + (1 - \frac{t}{T}) P^\top \Omega^{-1} P\Big)^{-1} + \frac{1}{\gamma} \Big\{\Sigma - \Big( \Sigma^{-1} + (1 - \frac{t}{T}) P^\top \Omega^{-1} P\Big)^{-1}\Big\}.
    \end{equation*}
\end{corollary}

Though our focus is the dynamic problem, our framework allows a single-period investor at $t$ to use an ``old" view $Y(0, T)$  and the history of prices on $[0, t]$, an information structure that differs from the classical single-period Black-Litterman model, to adjust her prior forecast of returns. 

\begin{example}[Single-period investor with an ``old" view]
\label{example:Single_Period_Policy}
Given the view $Y(0, T)=y$ and the history of returns until $t\in(0, T)$,  it can be shown (see Appendix \ref{App:SinglePeriod_BL}) that the distribution of log-returns over the remainder of the investment period
\begin{eqnarray*}
\big[X(T)-X(t) \, \big| \, X(t)=x, Y(0, T)=y \big]\sim N\left((T-t)\tilde{\mu}^x(t,x), (T-t)\,\Sigma_{BL|t}\right)
\end{eqnarray*}
where
\begin{align*}
\tilde{\mu}^x(t,x) &= \tilde{\mu}(t,x) - \frac{1}{2}\diag(\Sigma),\\[1mm]
\Sigma_{BL|t}^{-1} &=  \Sigma^{-1} + \Bigl(1 - \frac{t}{T}\Bigr) P^\top \Omega^{-1} P.
\end{align*} 
It follows that a single-period investor over $[t, T]$ with an old view $Y(0, T)=y$ 
has the holding
\begin{equation}
\label{eq:Dynamic_CBL}
\pi^*_{BL|t}  =  \frac{1}{\gamma} \Sigma_{BL|t}^{-1}\big(\tilde{\mu}^x(t,x) - r_f \mathbf{1}_N \big).
\end{equation}
Note that the impact of the view $Y(0, T)$ on the precision update from $\Sigma^{-1}$ to $\Sigma_{BL|t}^{-1}$
diminishes as the view ages. 

\end{example}

\section{Other View Structures}
\label{sec:Extensions}
The model we have studied assumes that views are provided at $t=0$. We now consider two extensions, when views are revised during the investment horizon in Section \ref{sec:revisions}, and
 short term views are provided during the investment horizon (e.g., quarterly forecasts when the investor's horizon spans one year) in Section \ref{sec:short-term_views}. Proofs can be found in Appendix \ref{Sec:EC_ViewsRevision} and Appendix \ref{Sec:EC_QuarterlyViews}, respectively. (Section \ref{Sec:EC_ViewsDifferentHorizons} in the Appendix considers the case when multiple views over views over different horizons are given at $t=0$). There are certainly other information structures that can be modeled and solved by adapting the methods used to solve our base problem.


\subsection{Revisions} \label{sec:revisions}
We first consider the case where experts revise their views during the investment horizon.

\subsubsection*{Views Model}
Suppose that the view provided at time $t_j$ ($0=t_0 < t_1 < \cdots < t_M<T$) is a noisy signal of returns over the remainder of the investment horizon 
\begin{equation}
\label{eq:UpdatedViews}
    Y^j(t_j, T) \, | \, (X(t_j), X(T)) = P \big(X(T) - X(t_j)\big) + \epsilon^j \sim \mathcal{N}\big( P \big(X(T) - X(t_j)\big), \Omega^j \big)
\end{equation}
where $\epsilon^{j} \sim \mathcal{N}\big(0, \Omega^{j}\big)$ and the covariance matrix $\Omega^j$ is positive definite.  We denote by $\{\mathcal{F}^Y_t, 0\leq t \leq T\}$ the filtration generated by the history of prices and views where for  $t \in [t_j, t_{j+1})$, \begin{eqnarray*}\mathcal{F}^Y_t = \sigma\left(\mathcal{F}_t \vee \sigma(Y^0(0,T), \dots, Y^j(t_j,T))\right).\end{eqnarray*}

A feature of a revision is that existing views become obsolete after they are revised. this requires additional conditions  on the noise terms in \eqref{eq:UpdatedViews}.
Suppose that
\begin{equation}
    \epsilon^j = \epsilon^{j,0} + \epsilon^{j+1}
    \label{eq:noise-redundant-views}
\end{equation}
where $\epsilon^{j,0} \sim \mathcal{N}\big(0, \Omega^{j,0}\big)$. It follows that the view at $t_j$ can be decomposed into into a view over $(t_j, t_{j+1})$ and another over $(t_{j+1}, T)$:
\begin{equation*}
\begin{split}
        Y^j(t_j, T)  &=  P \big(X(T) - X(t_j)\big) + \epsilon^j\\
        &= P (X(t_{j+1}) - X(t_j)) + \epsilon^{j,0} + Y^{j+1}(t_{j+1},T).
\end{split}
\end{equation*}
The investor knows $Y^j(t_j, T)$ at $t_j$ and the return  $X(t_{j+1}) - X(t_j)$ and the revised view $Y^{j+1}(t_{j+1},T)$ at $t_{j+1}$ (but not the return $X(T)-X(t_{j+1})$ or the value of $\epsilon^{j+1}$). It follows that the noise term 
\begin{eqnarray*}
 \epsilon^{j, 0} = Y^j(t_j, T) - P (X(t_{j+1}) - X(t_j))-Y^{j+1}(t_{j+1},T)
\end{eqnarray*}
is known at time $t_{j+1}$. 
If $\epsilon^{j, 0}$ and $\epsilon^{j+1}$ are dependent, $\epsilon^{j,0}$ can be used to estimate $\epsilon^{j+1}$ which reduces the noise in the view $Y^{j+1}(t_{j+1}, T)$ and leads to a refined estimate of $X(T)-X(t_{j+1})$. 
In this case, the past view $Y^j(t_j, T)$ and the log-return $X(t_{j+1}) - X(t_j)$, both of which are needed to compute $\epsilon^{j, 0}$, are part of the state. The following assumption ensures that past views become obsolete.
\begin{assumption}\label{ass1}
    $\epsilon^j$ satisfies \eqref{eq:noise-redundant-views},  $\{\epsilon^{j, 0}, j=1, \cdots, M\}$ are mutually independent and $\epsilon^{j, 0}$ is independent of $\epsilon^{j+1}, \cdots, \epsilon^{M}$.
\end{assumption}  
Observe that $\Omega^j = \Omega^{j, 0}+\Omega^{j+1}$.

\subsubsection*{Market Dynamics}

In Section \ref{sec3}, views are provided at the beginning of the investment horizon and the  log-returns $X^y(t)$ and wealth $Z(t)$ are  state variables with dynamics adjusted using the views.
When there are revisions during the investment horizon, the most recent view  
\begin{eqnarray}
\label{eq:revision_i(t)}
    I(t) = Y^j(t_j,T), \quad t \in [t_j,t_{j+1}),
\end{eqnarray}
is also a state variable because it changes the dynamics of the log-returns and wealth. Since revisions are made during the investment horizon, it is convenient to reset the reference point of the cumulative log-returns to its price at the time of the new view:
\begin{eqnarray*}
    \bar{X_i}(t) & = & \log\Big(\frac{S_i(t)}{S_i(t_j)}\Big) =X_i(t) - X_i(t_j), i = 1, \cdots, N,  \\ [5pt]
    \bar{X}(t) &  :=  & [\bar{X}_1(t), \cdots, \bar{X}_N(t)] \, \, \, \text{for} \, \, t \in [t_j, t_{j+1}),
\end{eqnarray*}
We can now use the approach from Section \ref{sec3} to derive the dynamics of $\bar{X}(t)$ and $Z(t)$ until the next revision.

\begin{proposition}
\label{cor:UpdatedViews}
    Let $t\in[t_j, t_{j+1})$, given the view $I(t) =  y^j$,
    the conditional log-returns $\bar{X}^y(t) = \bar{X}(t) \, | \, (I(t) = y^j)$ 
    is given by
    \begin{eqnarray}
        d\bar{X}^y(t) & = &   
        \bigg(\mu^x + \beta_1^j \big(y^j - (T-t_j) P \mu^x \big) + \beta_2^j(t) \big(\mathbb{E}[\bar{X}^y(t)] - \bar{X}^y(t)\big)\bigg) dt 
        + dW^y(t),  \nonumber
        \\ [5pt]
        \bar{X}^y(t_j) & = & 0, \label{eq:dX_updatedViews}    \end{eqnarray}
    where
    \begin{eqnarray*}
    \beta_1^j & = & \Sigma P^\top \big((T-t_j) P \Sigma P^\top+\Omega^j\big)^{-1} \in  \mathbb{R}^{N \times K},\\ [5pt]
    \beta_2^j(t) & = & \Sigma P^\top \big((T-t) P \Sigma P^\top + \Omega^j\big)^{-1} P  \in \mathbb{R}^{N \times N},\\ [5pt]
   \mathbb{E}[\bar{X}^y(t)]  & =& (t - t_j) \big(\mu^x +  \beta_1^j \big(y^j -(T-t_j) P \mu^x \big)\big).
\end{eqnarray*}
The investor's wealth satisfies
\begin{eqnarray}
\label{eq:wealth-revision}
    dZ(t) & = & Z(t) \bigg( r_f dt + \pi(t)^\top \big(\tilde{\mu}^j(t,\bar{X}^y(t), y^j) - r_f \mathbf{1}_N\big) dt + \pi(t)^\top dW^y(t) \bigg), \; t \in [t_j, t_{j+1}) \\ [5pt]
    Z(0) & = & z.\nonumber
\end{eqnarray}
where
\begin{eqnarray}
\label{eq:mu-revise}
    \tilde{\mu}^j(t,\bar{x}, y^j) = \mu + \beta_1^j (y^j - (T-t_j)P \mu^x) + \beta_2^j(t) (\mathbb{E}[\bar{X}^y(t)] - \bar{x}).
\end{eqnarray}
$I(t)=Y^j(t_j, T)$ remains constant for $t\in[t_j, t_{j+1})$ and changes to $I(t_{j+1}) = Y^{j+1}(t_{j+1},T)$ according to \eqref{eq:UpdatedViews} at $t_{j+1}$.
\end{proposition}
Note that  $\bar{X}^y(t)$  resets to $0$ after every revision. The drift of $Z(t)$ and $\bar{X}^y(t)$ also change.

\subsubsection*{Optimal Policy}
The investor maximizes expected utility of terminal wealth 
\begin{equation*}
    V(t,z,\bar{x}, y) = \max_{\pi \in \mathcal{A}} \mathbb{E}\big[ U(Z(T)) \,|\, \bar{X}^y(t) = \bar{x}, Z(t) = z, I(t) = y  \big]
\end{equation*}
where the dynamics of $(Z(t), \bar{X}^y(t), I(t))$ are described in Proposition \ref{cor:UpdatedViews}. 
We use dynamic programming to derive the value function and optimal policy, which we now summarize.
A detailed proof is provided in Appendix \ref{EC_Views_Revision_Control}. 

In the last interval $[t_M, T]$, the problem is exactly the same as discussed in Section \ref{sec3}. 
From Section \ref{sec51}, the value function is of the form
\begin{eqnarray}
   V(t, z, \bar{x}, y^{j})  = \frac{z^{1-\gamma}}{1-\gamma} \exp(g^{j}(t,\bar{x},y^{j}))
   \label{eq:V-ext1}
\end{eqnarray}
where $g^{j}(t,\bar{x},y^{j})$  is quadratic in $x$ and $y^j$ ($j=M$). 

For the preceding epoch $[t_{M-1}, t_M)$, the Principle of Optimality implies that
\begin{eqnarray}
V(t_{M}^-,z,\bar{x},y^{M-1}) &= \mathbb{E}[V(t_{M}, z, 0, Y^{M}) \, | \, \bar{X}^y(t_{M}^-) = \bar{x}, Z(t) = z, I(t) = y^{M-1}]
\label{eq:continuation}
\end{eqnarray}
which gives the terminal condition for the value function on $t\in[t_{M-1}, t_M)$ in terms of the value function on $[t_M, T]$. For $t\in[t_{M-1}, t_M)$, the value function $V(t, z, \bar{x}, y^{M-1})$ satisfies a HJB equation. It can be shown that the value function on $t\in[t_{M-1}, t_M)$ has the form \eqref{eq:V-ext1} where $g^{M-1}(t, \bar{x}, y)$ is quadratic in $\bar{x}$ and $y$ so the problem of solving the HJB equation over $[t_{M-1}, t_{M})$ reduces to solving a system of ODEs for the coefficients of this quadratic function. This process can be repeated until we have the value function over $[0, T]$.

Since one of the ODEs in each epoch is a (nonlinear) matrix Riccati equation, we are usually content with numerical solutions of this ODE system.  
As was the case in Section \ref{sec5}, this can be avoided as the ODEs that characterize the optimal policy have explicit solutions. The proof can be found in Appendix \ref{EC_Views_Revision_Control}.

\begin{theorem}
\label{thm:updatedViews}
Suppose that views satisfy \eqref{eq:UpdatedViews} and \eqref{eq:noise-redundant-views} 
and Assumption \ref{ass1} holds. 
For $t\in[t_j, t_{j+1})$, $j=0, 1, \cdots, M$, the value function is of the form \eqref{eq:V-ext1} where
    \begin{eqnarray}
    \label{eq:g-view1}
            g^{j}(t,\bar{x},y^{j})&   = &  -\frac{1}{2} (P\bar{x} - y^{j})^\top C^{j}(t) (P \bar{x} - y^{j}) - (P\bar{x} - y^j)^\top \hat{c}^{j}(t) + \bar{c}^{j}(t).
    \end{eqnarray}
The coefficients
    \begin{equation*}
    \begin{split}
              &C^j(t) = \frac{1}{2}\big\{\bar{M}^j(t) P^\top \bar{\eta}_t^j + (\bar{M}^j(t) P^\top \bar{\eta}_t^j)^\top\big\} \in \mathbb{R}^{K \times K},\\
        &\hat{c}^j(t) = \bar{M}^j(t) \left( \Sigma^{-1} (\mu - r_f \mathbf{1}_N)  + (T-t) P^\top \bar{\eta}_t^j P \mu^x\right) \in \mathbb{R}^{K}
    \end{split}
    \end{equation*}
    where 
    \begin{equation*}
        \bar{M}^j(t) =- (\gamma - 1) (T - t) (\Omega^{j})^{-1}P \Big(\gamma\Sigma^{-1} + (T - t) P^\top (\Omega^{j})^{-1} P\Big)^{-1} \in \mathbb{R}^{K \times N}
    \end{equation*}
    \begin{eqnarray*}
        \bar{\eta}^j_t = - \left((T-t) P \Sigma P^\top + \Omega^j\right)^{-1} \in \mathbb{R}^{K \times K},
    \end{eqnarray*}
and  $\bar{c}^j(t)$ is a real-valued function defined by \eqref{eq:EC_barcj_update} and \eqref{terminal:C} in the Appendix.
The investor's optimal policy is the sum of the mean-variance holding and a hedging demand
    \begin{equation}
        \label{eq:Pi0*_revisions}
        \pi^{j *}(t, \bar{x}, y^j) = \frac{1}{\gamma} \Sigma^{-1} \big(\tilde{\mu}^j(t,\bar{x}, y^j) - r_f \mathbf{1}_N\big) + \frac{1}{\gamma} \frac{\partial g^j}{\partial \bar{x}}(t,\bar{x}, y^j), \quad \text{for} \quad t \in [t_j, t_{j+1})
    \end{equation}
 where $\tilde{\mu}^j(t, \bar{x}, y^j)$ is given by \eqref{eq:mu-revise}. The hedging demand is
    \begin{equation*}
    \begin{split}
               \frac{1}{\gamma} \frac{\partial g^j}{\partial \bar{x}}(t,\bar{x}, y^j) &= \frac{1}{\gamma} M^j(t) \Sigma^{-1}\big(\tilde{\mu}^j(t,\bar{x}, y^j) - r_f \mathbf{1}_N\big).
    \end{split}
    \end{equation*}
    where
    \begin{eqnarray*}
         M^j(t) = (\gamma - 1) (T-t) P^\top (\Omega^j)^{-1}P \Big(\gamma\Sigma^{-1} + (T-t) P^\top (\Omega^j)^{-1} P\Big)^{-1} \in \mathbb{R}^{N \times N}.
    \end{eqnarray*}
   Furthermore, the optimal policy \eqref{eq:Pi0*_revisions} can be expressed as 
    \begin{equation*}
    \begin{split}
             \pi^{j *}(t, \bar{x}, y^j) 
             &= \frac{1}{\gamma} \big(\Sigma_{\text{DBL}}^j(t) \big)^{-1} \big(\tilde{\mu}^j(t, \bar{x}, y^j) - r_f \mathbf{1}_N \big), \quad \text{for} \quad t \in [t_j, t_{j+1})
    \end{split}
    \end{equation*}
    where 
    \begin{equation*}
    \Sigma_{\text{DBL}}^j(t) = \Big( \Sigma^{-1} + (1 - \frac{t}{T}) P^\top (\Omega^j)^{-1} P\Big)^{-1} + \frac{1}{\gamma} \Big\{\Sigma - \Big( \Sigma^{-1} + (1 - \frac{t}{T}) P^\top (\Omega^j)^{-1} P\Big)^{-1}\Big\} \in \mathbb{R}^{N \times N}.
    \end{equation*}
\end{theorem}



Consider two dynamic investors, one who receives a view at $t_k$ and anticipates future revisions at $t_{k+1}, \cdots, t_M$, and another who receives the same view at $t_k$ but none of the revisions. Theorem \ref{thm:updatedViews} shows that both have the same dynamic policy from $t_k$ until the revision at $t_{k+1}$ is received by the first investor. The value function of the investor who receives future updates is larger because the revisions improve her forecast. Mathematically, the policy of the investor who does not receive revised views is admissible for the more informed investor.



\subsection{Short-term views}
\label{sec:short-term_views}
{\color{red}}
We consider now the case when the investor receives views over portions of the investment horizon (e.g., quarterly views received every quarter over an investment horizon of a year).

\subsubsection*{Views Model}

Suppose that times $0=T_0 < T_1 < \cdots <T_M < T_{M+1}=T$\footnote{We refer to the time points \(T_1, T_2, \dots\) in uppercase because, in this setting, they represent both the time at which the current view was given and the horizon of the previous view.} are known at the start of the investment horizon.
The investor receives the view $Y^j(T_j, T_{j+1})$  of return $X(T_{j+1})-X(T_j)$ at $t=T_j$. We assume
\begin{equation}
\label{eq:Quarterlyviews}
    Y^j(T_j, T_{j+1})\, | \, (X(T_j), X(T_{j+1})) = P \big(X(T_{j+1}) - X(T_j)\big) + \epsilon^j \sim \mathcal{N}\big( P \big(X(T_{j+1}) - X(T_j)\big), \Omega^j \big)
\end{equation}
where $P \in \mathbb{R}^{K \times N}$ and $\epsilon^j \sim \mathcal{N}\big(0, \Omega^j\big)$ is the noise term associated with the $j^{th}$ view. 

Although the intervals $[T_j, T_{j+1})$ are non-overlapping, views may be correlated due to dependence in the noise. We assume that the noise follows a VAR(p) process
\begin{equation}
\label{eq:quarterlyNoises}
    \epsilon^{j} = \Phi^1 \epsilon^{j-1} + \dots + \Phi^{\bar{p}(j)} \epsilon^{j-\bar{p}(j)} +  \epsilon^{j,0}, \, \, \, \text{for} \, \, j \in \{1, \dots, M\},
\end{equation}
where $\bar{p}(j) = \min\{p, j\}$ and $p$ is  some known non-negative integer. $\Phi^i \in \mathbb{R}^{K \times K}$ are matrices of autoregression coefficients and $\epsilon^{j,0} \sim \mathcal{N}\big(0,\Omega^{j,0}\big)$ is the idiosyncratic noise; the $\epsilon^{j, 0}$'s are assumed to be independent.
We assume that the investor knows $\{ \Phi^{i}, i \in \{1, \dots, p\}\}$,  and $\{\Omega^{j,0}, j \in \{1, \dots, M\}\}$.  For details on how to estimate the autoregression coefficients, see \cite{VAR_Estimation1} and \cite{VAR_Estimation2}.

\subsubsection*{Market dynamics}

In the baseline model described in Section \ref{sec3}, the investor's wealth $Z(t)$ and log-returns process $X(t)$ are state variables. Although the view $Y(0, T)$ can be thought of as a state variable, it  remains fixed over the investment horizon after being realized at $t=0$ and is treated like a parameter when solving the problem. As in Section \ref{sec:revisions} views are now part of the state. 

Since the noise
\begin{eqnarray}
\label{eq:eps_i}
    \epsilon^i = y^i - P[X(T_{i+1})-X(T_i)], \; i=0, 1, \cdots, j-1,
\end{eqnarray}
can be computed at $T_{i+1}$ once the view $Y^i(T_i,T_{i+1})$ and return $X(T_{i+1})-X(T_i)$ have been realized, the history of noises $(\epsilon^{j-1}, \epsilon^{j-2}, \cdots, \epsilon^{j-p})$ can be use to predict a portion of the error in the next view 
\begin{eqnarray*}
\hat{\epsilon}^j = \sum_{i=1}^{\bar{p}} \Phi^i \epsilon^{j-i}.
\end{eqnarray*}
We can refine the prediction $Y^{j}(T_{j}, T_{j+1})$
by stripping out the predictable component of the noise from \eqref{eq:eps_i}
    \begin{eqnarray}
            \label{eq:refined_view}
  \bar{Y}^j(T_j, T_{j+1}) & = & Y^j(T_j, T_{j+1}) - \sum_{i=1}^{\bar{p}} \Phi^i \epsilon^{j-i}, \,\, \, \text{for} \, \, j \in \{1, \dots, M\} \\
        \bar{Y}^j(0, T_1) & = & Y^0(0, T_1).\nonumber
    \end{eqnarray}
It follows that the past $p$ views affect the dynamics of the returns and one might expect that they should be part of the state. 
The following result shows that this is not the case and that the information in the past views is fully captured by the most recent adjusted view
\begin{eqnarray}
\label{eq:quarterlyViews_i(t)}
    I(t) = \bar{Y}^j(T_j,T_{j+1}), \quad t \in [T_j,T_{j+1}).
\end{eqnarray}
It also shows that the sequence of adjusted views are Gaussian, that $\bar{Y}^j(T_j,T_{j+1})$ is a noisy view of   $X(T_{j+1})-X(T_j)$, and that the collection of adjusted views are  independent. One important consequence is that $Z(t)$, $X(t)$, and the most recent adjusted view $I(t)$ are the state variables for this problem, which is a substantial simplification from having to keep track of the last $p$ views. As in Section \ref{sec:revisions}, the log-returns process resets to $0$ at the start of every new epoch $[T_j, T_{j+1})$.
\begin{proposition}
\label{prop:independence_quarterly_views}
    Suppose that the asset price satisfies \eqref{eq8} and views  \eqref{eq:Quarterlyviews}--\eqref{eq:quarterlyNoises}. Let $\bar{Y}^j(T_j, T_{j+1})$ be the  refinement of the view $Y^i(T_i, T_{i+1})$ defined by \eqref{eq:eps_i}--\eqref{eq:refined_view}. 
    Then $\{\bar{Y}^j(T_j, T_{j+1}), j \in \{0, \dots, M\}\}$ are jointly independent with
    \begin{eqnarray}
    \label{eq:Y-bar}
        \bar{Y}^j(T_j, T_{j+1}) \, | \, X(T_j), X(T_{j+1}) \sim \mathcal{N}\big( P \big(X(T_{j+1}) - X(T_j)\big), \Omega^{j,0} \big).
    \end{eqnarray}
Let
    \begin{eqnarray}
    \label{eq:barX_quarterlyViews}
        \bar{X}(t) := X(t) - X(T_j), \, \, \, \text{for} \, \, t \in [T_j, T_{j+1})
    \end{eqnarray}
    be the log-return between the time of the last view $T_j$ and the current time $t$. Then conditional log-returns satisfy
    \begin{equation*}
        \begin{split}
            \bar{X}^y(t) &= \bar{X}(t) \, | \, (Y^0(0,T_1) = y^0, \dots,Y^j(T_j,T_{j+1}) = y^j)\\
            &= \bar{X}(t) \, | \, (\bar{Y}^j(T_j,T_{j+1}) = \bar{y}^j).
        \end{split}
    \end{equation*}
\end{proposition}

The following result gives dynamics for the conditional log-returns process.
\begin{proposition}
    \label{cor:QuarterlyViews}
    For $t \in [T_j, T_{j+1})$, let  
    \begin{eqnarray*}
    \bar{X_i}(t) & = & \log\Big(\frac{S_i(t)}{S_i(T_j)}\Big) =X_i(t) - X_i(T_j),
\end{eqnarray*}
and $\bar{X}(t) = [\bar{X}_1(t), \cdots, \bar{X}_N(t)]$ be the vector of log-returns over $[T_j, t]$.
    Suppose that  views satisfy \eqref{eq:Quarterlyviews}--\eqref{eq:quarterlyNoises} and $I(t)$ be defined by \eqref{eq:quarterlyViews_i(t)}.
     The conditional log-returns $\bar{X}^y(t) = \bar{X}(t) \, | \, (I(t) = \bar{y}^j)$ satisfies
    \begin{eqnarray}
        d\bar{X}^y(t) & = &
        \bigg(\mu^x + \beta_1^j \big(\bar{y}^j - (T_{j+1}-T_j) P \mu^x \big) + \beta_2^j(t) \big(\mathbb{E}[\bar{X}^y(t)] - \bar{X}^y(t)\big)\bigg) dt 
    + dW^y(t), \; t \in [T_j, T_{j+1}) \nonumber \\
\bar{X}^y(T_j) & = & 0
\label{eq:dX_quarterlyViews}
    \end{eqnarray}
     where
\begin{eqnarray*}
    \beta_1^j & = & \Sigma P^\top \big((T_{j+1}-T_j) P \Sigma P^\top+\Omega^{j,0}\big)^{-1} \in  \mathbb{R}^{N \times K},\\ [8pt]
    \beta_2^j(t) & = & \Sigma P^\top \big((T_{j+1}-t) P \Sigma P^\top + \Omega^{j,0}\big)^{-1} P  \in \mathbb{R}^{N \times N},
\end{eqnarray*}
and
\begin{equation*}
   \mathbb{E}[\bar{X}^y(t)]  = (t - T_j) \big(\mu^x +  \beta_1^j \big(\bar{y}^j -(T_{j+1}-T_j) P \mu^x \big)\big).
\end{equation*}
The investor's wealth satisfies
\begin{eqnarray}
\label{eq:wealth-short}
    dZ(t) & = & Z(t) \bigg( r_f dt + \pi(t)^\top \big(\tilde{\mu}^j(t,\bar{X}^y(t), \bar{y}^j) - r_f \mathbf{1}_N\big) dt + \pi(t)^\top dW^y(t) \bigg), \; t \in [T_j, T_{j+1}) \\ [5pt]
    Z(0) & = & z.\nonumber
\end{eqnarray}
where
\begin{equation}
    \tilde{\mu}^j(t,\bar{x}, \bar{y}^j) = \mu + \beta_1^j (\bar{y}^j - (T_{j+1}-T_j)P \mu^x) + \beta_2^j(t) (\mathbb{E}[\bar{X}^y(t)] - \bar{x}), \; t \in [T_j, T_{j+1}).
    \label{eq:tilde_mu_j}
\end{equation}
$I(t)=\bar{Y}^j(T_j, T_{j+1})$ remains constant on $t\in[T_j, T_{j+1})$ and changes to $I(T_{j+1}) = \bar{Y}^{j+1}(T_{j+1}, T_{j+2})$ at $t_{j+1}$. $\{\bar{Y}^j(T_j, T_{j+1}), j \in \{0, \dots, M\}\}$ are mutually independent and satisfy \eqref{eq:Y-bar}.
\end{proposition}

\subsubsection*{Optimal Policy}

The investor maximizes expected utility of terminal wealth 
\begin{equation*}
    V(t,z,\bar{x}, \bar{y}) = \max_{\pi \in \mathcal{A}} \mathbb{E}\big[ U(Z(T)) \,|\, \bar{X}^y(t) = \bar{x}, Z(t) = z, I(t) = \bar{y}  \big]
\end{equation*}
where the dynamics of $(Z(t), \bar{X}^y(t), I(t))$ are described in Proposition \ref{cor:QuarterlyViews}. 
As in the case with revisions (Section \ref{sec:revisions}), this can be solved using dynamic programming, and the elegant properties of the value function and optimal policy described in Theorem \ref{thm:updatedViews} generalize to the present situation. A detailed proof is given in the Appendix.


\begin{theorem}
\label{thm:QuarterlyViews}
For $j \in \{0, \dots, M-1\}$ and $t \in [T_j,T_{j+1})$, the value function is
    \begin{equation*}
        V(t,z,\bar{x}, \bar{y}^j) = \frac{z^{1-\gamma}}{1-\gamma} \exp(g^j(t,\bar{x},\bar{y}^j))
    \end{equation*}
where
    \begin{equation*}
            g^j(t,\bar{x},\bar{y}^j) =  -\frac{1}{2} (P\bar{x} - \bar{y}^j)^\top C^j(t) (P \bar{x} - \bar{y}^j) - (P\bar{x} - \bar{y}^j)^\top \hat{c}^j(t) + \bar{c}^j(t).
    \end{equation*}
    For $t\in[T_j, T_{j+1})$, 
    \begin{equation}
    \begin{split}
              &C^j(t) = \frac{1}{2}\big\{\bar{M}^j(t) P^\top \bar{\eta}_t^j+ (\bar{M}^j(t) P^\top \bar{\eta}_t^j)^\top\big\} \in \mathbb{R}^{K \times K},\\
        &\hat{c}^j(t) = \bar{M}^j(t) \left( \Sigma^{-1} (\mu - r_f \mathbf{1}_N)  + (T_{j+1}-t) P^\top \bar{\eta}_t^j P \mu^x\right) \in \mathbb{R}^K,
    \end{split}
    \end{equation}
    where 
        \begin{equation*}
        \bar{\eta}^j_t = - \big((T_{j+1}-t) P \Sigma P^\top + \Omega^{j,0})^{-1} \in \mathbb{R}^{K\times K},
    \end{equation*}
    \begin{equation*}
        \bar{M}^j(t) =- (\gamma - 1) (T_{j+1} - t) (\Omega^{j,0})^{-1}P \Big(\gamma\Sigma^{-1} + (T_{j+1} - t) P^\top (\Omega^{j,0})^{-1} P\Big)^{-1} \in \mathbb{R}^{K \times N},
    \end{equation*}
and the real-valued function  $\bar{c}^j(t)$ is defined by \eqref{eq:EC_barcj_quarterly} and \eqref{eq:bar_c_terminal_quarterly}. 
The optimal policy 
    \begin{equation}
        \label{eq:Pi0*}
        \pi^{j *}(t,\bar{x},\bar{y}^j) = \frac{1}{\gamma} \Sigma^{-1} \big(\tilde{\mu}^j(t,\bar{x}, \bar{y}^j) - r_f \mathbf{1}_N\big) + \frac{1}{\gamma} \frac{\partial g^j}{\partial \bar{x}}(t,\bar{x}, \bar{y}^j), \; t \in [T_j, T_{j+1})
    \end{equation}
    is the sum of a mean-variance holding and the hedging demand
    \begin{equation*}
    \begin{split}
               \frac{1}{\gamma} \frac{\partial g^j}{\partial \bar{x}}(t,\bar{x}, \bar{y}^j) &= \frac{1}{\gamma} M^j(t) \Sigma^{-1}\big(\tilde{\mu}^j(t,\bar{x}, \bar{y}^j) - r_f \mathbf{1}_N\big)
    \end{split}
    \end{equation*}
    where $\tilde{\mu}^j(t,\bar{x}, \bar{y}^j)$ is given by \eqref{eq:tilde_mu_j} and 
    \begin{equation*}
         M^j(t) = (\gamma - 1) (T_{j+1} - t) P^\top (\Omega^{j,0})^{-1}P \Big(\gamma\Sigma^{-1} + (T_{j+1} - t) P^\top (\Omega^{j,0})^{-1} P\Big)^{-1} \in \mathbb{R}^{N \times N}.
    \end{equation*}    
    Finally, the optimal policy \eqref{eq:Pi0*} can be expressed as 
    \begin{equation*}
    \begin{split}
             \pi^{j *}(t, \bar{x}, \bar{y}^j) 
             &= \frac{1}{\gamma} \big(\Sigma_{\text{DBL}}^j(t) \big)^{-1} \big(\tilde{\mu}^j(t, \bar{x}, \bar{y}^j) - r_f \mathbf{1}_N \big), \; t \in [T_j, T_{j+1})
    \end{split}
    \end{equation*}
    where 
    \begin{equation*}
    \Sigma_{\text{DBL}}^j(t) = \Big( \Sigma^{-1} + (1 - \frac{t}{T_{j+1}}) P^\top (\Omega^{j,0})^{-1} P\Big)^{-1} + \frac{1}{\gamma} \Big\{\Sigma - \Big( \Sigma^{-1} + (1 - \frac{t}{T_{j+1}}) P^\top (\Omega^{j,0})^{-1} P\Big)^{-1}\Big\} \in \mathbb{R}^{N \times N}.
    \end{equation*}
\end{theorem}
It follows from Theorem \ref{proposition5} that for  $t \in [T_j, T_{j+1}]$ the portfolio weights \eqref{eq:Pi0*}  optimizes the expected utility of terminal wealth at $T_{j+1}$, namely \[{\mathbb E}[U(Z(T_{j+1}))\,|\,\bar{X}^y(t)=\bar{x}, Z(t)=z, \bar{Y}^j(T_j, T_{j+1})=\bar{y}^j],\] with the view  
\begin{eqnarray*}
        \bar{Y}^j(T_j, T_{j+1}) \, | \, X(T_j), X(T_{j+1}) \sim \mathcal{N}\big( P \big(X(T_{j+1}) - X(T_j)\big), \Omega^{j,0} \big).
    \end{eqnarray*}
    provided at $T_j$. That is, after receiving the adjusted view, it is optimal to solve dynamic portfolio choice problems over the horizons $[T_j, T_{j+1}]$.

\section{Experiments}
\label{sec7}

We conduct two experiments. In Section \ref{sec61}, we compare  dynamic Black-Litterman  to an investor who rebalances periodically to the solution of a single-period Black-Litterman model. We evaluate the impact of view revisions on investment performance in Section \ref{sec:experiments-revisions}. 

We consider a market of $N = 5$ risky assets ($A$, $B$, $C$, $D$, and $F$), one-risk free asset with fixed return $r_f = 3\%$ per year, and $K = 3$ forward-looking views about returns over a $T = 1$ year horizon. The views matrix $P = (p_1, p_2, p_3)^\top \in \mathbb{R}^{K \times N}$ with $p_1 = (1, -1, 0 ,0 ,0)$, $p_2 = (1, 0, 0 ,0, -1)$, and $p_3 = (0, 0, 1, 0 ,0)$. Forward-looking views are given about the difference in returns between asset $A$ and asset $B$, the difference in returns between asset $A$ and asset $F$, and the return of asset $C$. No view is expressed about the return of asset $D$. Expected return estimates 
are taken from \cite{CAPMdata} which uses the Capital Asset Pricing Model (CAPM) to estimate expected annual returns from daily stock price data for companies listed on the S\&P/ASX 50 Index. For the five assets we consider
\begin{eqnarray*}
    \mu = [0.0320, 0.0447, 0.0269, 0.0679, 0.0672]^\top.
\end{eqnarray*}
We set the covariance of the assets as
\[
\Sigma = 
\begin{bmatrix}
0.0641 & 0.0175 & 0.0086 & 0.0266 & 0.0363 \\ \vspace{-0.35cm}\\
0.0175 & 0.1191 & 0.0234 & 0.0303 & 0.0353 \\ \vspace{-0.35cm}\\
0.0086 & 0.0234 & 0.1154 & 0.0322 & 0.0278 \\\vspace{-0.35cm}\\
0.0266 & 0.0303 & 0.0322 & 0.1230 & 0.0431 \\ \vspace{-0.35cm}\\
0.0363 & 0.0353 & 0.0278 & 0.0431 & 0.1679
\end{bmatrix}.
\]

\subsection{Comparison of Dynamic and Classical Black-Litterman Models}
\label{sec61}

We use Monte-Carlo simulation to compare the performance of a Dynamic Black-Litterman (DBL) investor  \eqref{eq:MPBL_final} (equivalently \eqref{eq21}) and the Rebalanced Classical Black-Litterman investor (RBCL)  whose portfolio \eqref{eq:Dynamic_CBL} at $t$ is obtained by solving a single-period Markowitz problem over the horizon $[t, T]$ with the same view and price information as the DBL investor. We  compare investment performance and turnover of each approach with continuous, daily, weekly and (in some cases) monthly rebalancing epochs. For   weekly rebalancing (e.g.), we compute DBL and RCBL portfolio weights at the start of each week and rebalance our holding in each stock  to match these weights. These stocks until the next rebalancing epoch one week later. We assume a horizon of $T=1$ year.

In each simulation run, we 
generate  prices  \eqref{eq8} and views $Y(0,T)$  \eqref{eq10} with covariance matrix
\begin{eqnarray}
\label{eq:alpha}
    \Omega = \alpha P \Sigma P^\top,
\end{eqnarray}
and wealth under the DBL and RCBL policies. Efficient frontiers, certainty equivalents and portfolio turnovers are computed by averaging over all simulation runs. The parameter $\alpha$  captures the confidence in the equilibrium model, with lower values corresponding to higher confidence in the  views.
 Consistent with the literature, we choose $\alpha < 1$ (see \cite{Meucci2006, Walters2011}).



\begin{figure}[ht]
     \centering
     \begin{subfigure}[b]{0.49\textwidth}
         \centering
             \captionsetup{font=footnotesize} 
          \caption{$\alpha = 0.4$}
\includegraphics[width=\textwidth]{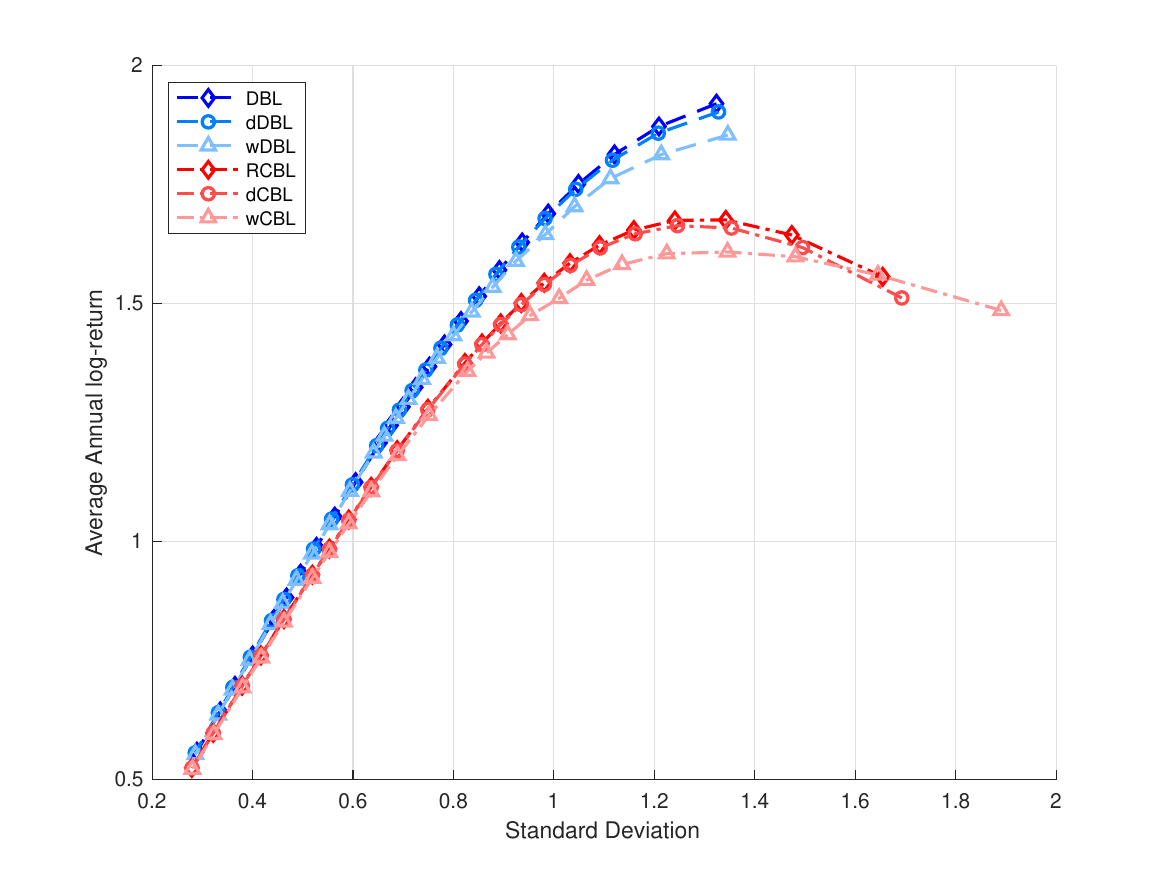}
         \label{fig:EFa}
     \end{subfigure}
     \begin{subfigure}[b]{0.49\textwidth}
         \centering
          \caption{$\alpha = 0.8$}
         \includegraphics[width=\textwidth]{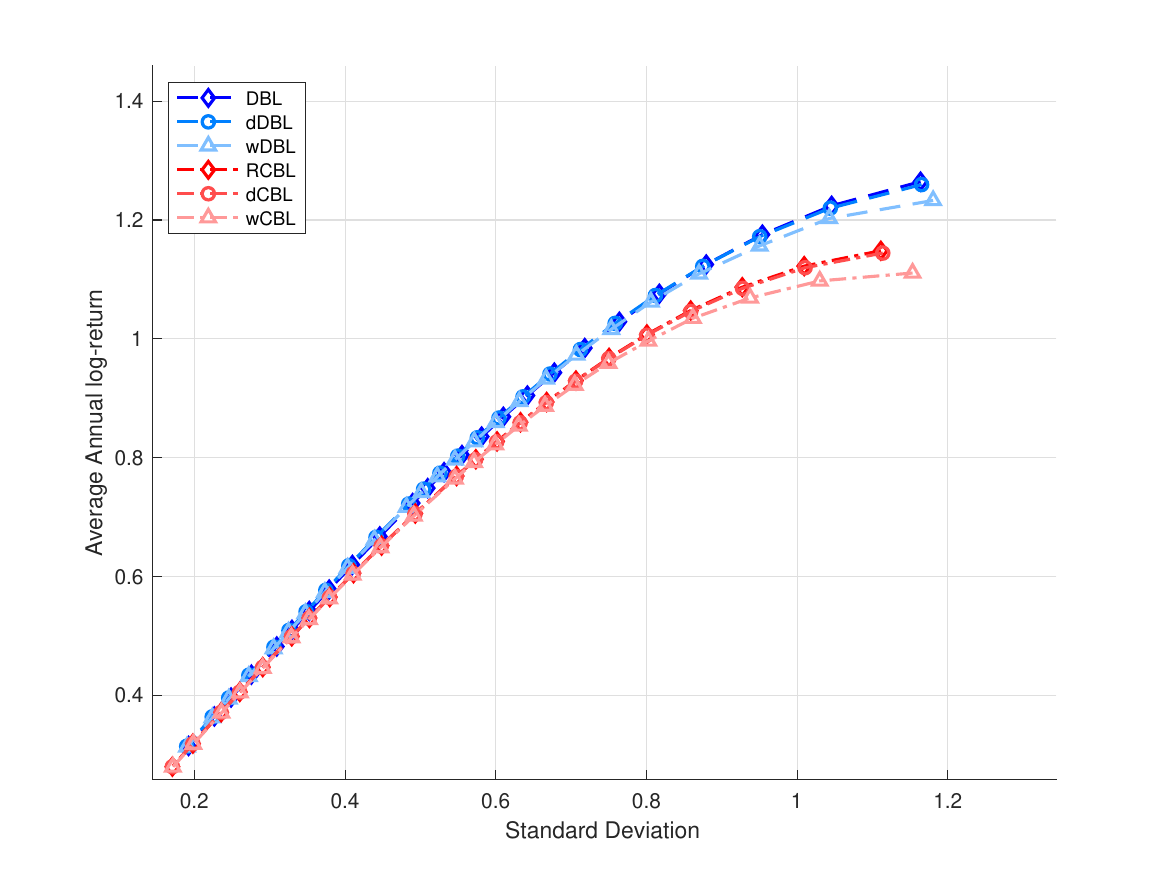}
         \label{fig:EFb}
     \end{subfigure}
        \caption{Efficient Frontier as a function of risk aversion. \textmd{The figure compares the Efficient Frontier of investors adopting the DBL and RCBL strategies for different levels of risk aversions under moderate and high noise in the expert views.}}
        \label{fig:EF_alpha0.4-0.8}
\end{figure}

Figure \ref{fig:EF_alpha0.4-0.8} shows efficient frontiers generated by DBL and RCBL when $\alpha=0.4$ and $0.8$. 
The  Efficient Frontier generated by DBL dominates  RCBL over all rebalancing frequencies, with the performance gap being larger when the expert views are more precise ($\alpha = 0.4$). 

Intuitively, DBL is a dynamic policy so every decision assumes future rebalancing opportunities whereas RCBL solves a single-period problem at each rebalancing epoch that assumes it is her last.  DBL also hedges the risk of  $X^y(t)$ changing over time using \eqref{eq22} which reduces the sensitivity of her value function to changes in $X^y(t)$ between trading epochs. As a result, investment performance of DBL is robust to the rebalancing interval.
There is no such hedge in RCBL so investment performance is more sensitive to changes in $X^y(t)$. This becomes more significant as views and hence $X^y(t)$ become more informative, which explains the faster breakdown in the performance of RCBL as the rebalancing interval increases when $\alpha=0.4$.

We now compare the Certainty Equivalent Return (CER) for DBL and RCBL investors. The CER is the constant rate of return $r_c$ that satisfies
\begin{eqnarray*}
    U\big(z_0 e^{T r_c}\big) =  \mathbb{E} \big[U\big(Z(T)\big) \, | \, X(0) = 0, \; Z(0) = z_0\big]
\end{eqnarray*}
with utility function $U(Z) = \dfrac{1}{1-\gamma}Z^{1-\gamma}$. In other words, the investor is indifferent between investing in the market and putting her money in a risk-free asset with a fixed return $r_c$. The preferred investment strategy has a higher CER. As in the mean-variance case, expected utilities in the calculation of CER are computed by averaging over prices and views.


\begin{figure}[ht]
     \centering
     \begin{subfigure}[b]{0.49\textwidth}
         \centering
             \captionsetup{font=footnotesize} 
          \caption{$\alpha = 0.4$}
          \includegraphics[width=\textwidth]{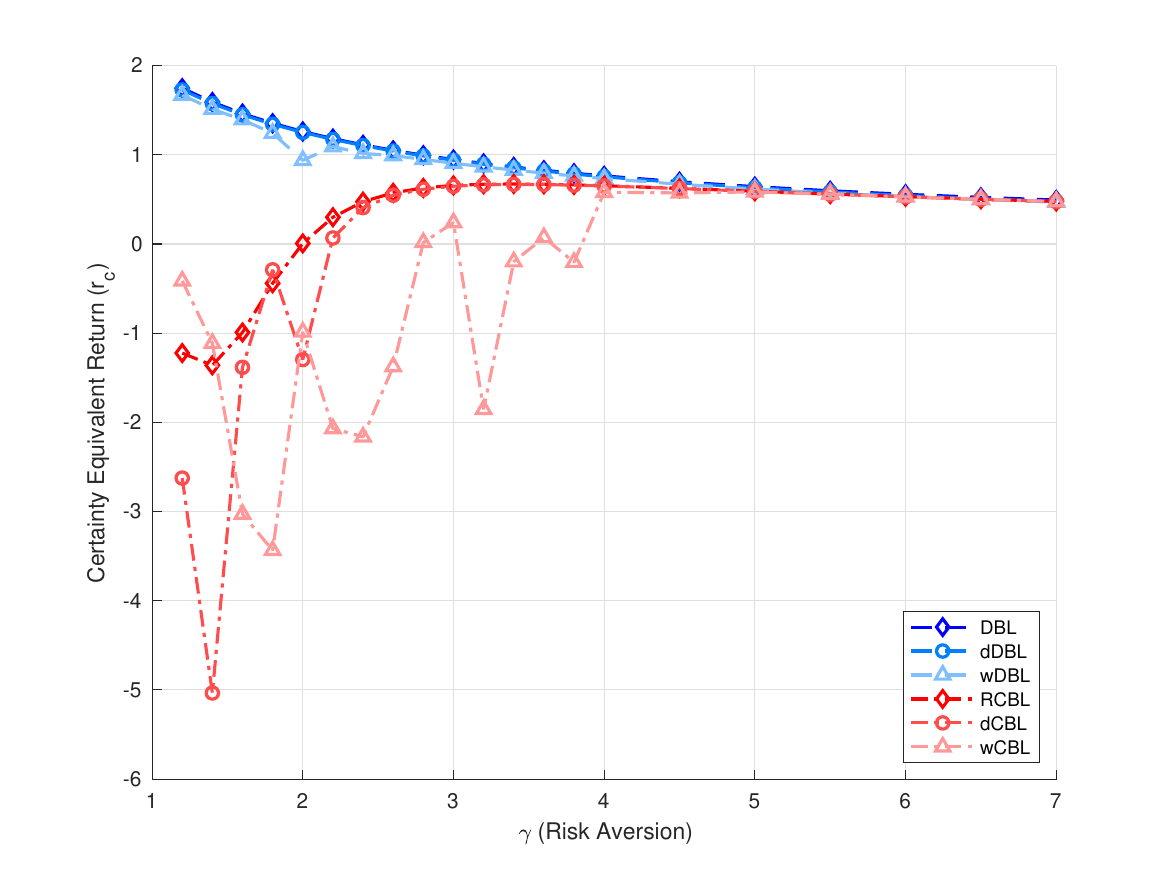}
         \label{fig:figure7a}
     \end{subfigure}
     \begin{subfigure}[b]{0.49\textwidth}
         \centering
          \caption{$\alpha = 0.8$}
         \includegraphics[width=\textwidth]{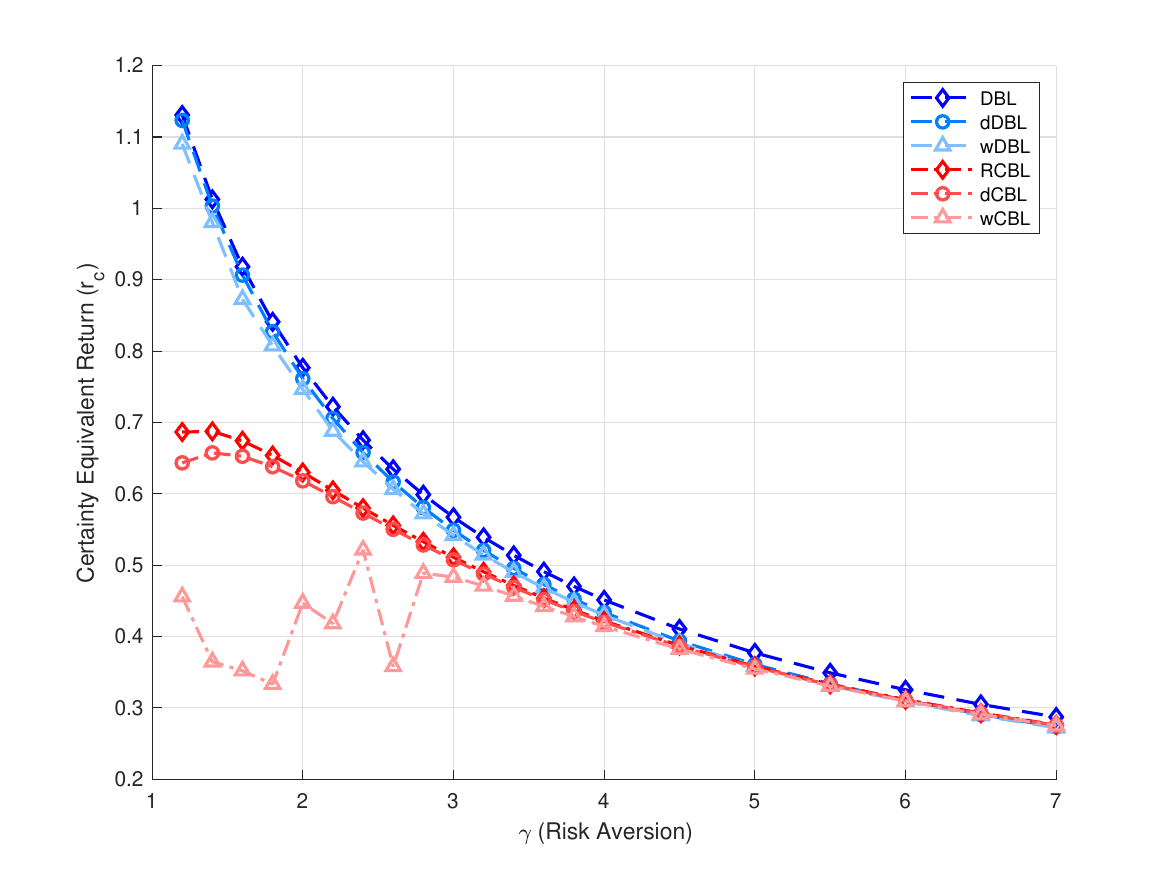}
         \label{fig:figure7c}
     \end{subfigure}
        \caption{Certainty Equivalent Return as a function of risk aversion. \textmd{The figure compares the Certainty Equivalent Return of investors adopting the DBL and RCBL strategies for different levels of risk aversions under moderate and high noise in the expert views.}}
        \label{fig:CE_alpha0.4-0.8}
\end{figure}

Figure \ref{fig:CE_alpha0.4-0.8} compares the Certainty Equivalent Return  for risk aversions $\gamma$ between $1$ and $7$.
We see the same patterns as Figure \ref{fig:EF_alpha0.4-0.8}: DBL consistently outperforms RCBL, performance of DBL is less sensitive to the rebalancing horizon, and the performance of RCBL degrades substantially when views become more important and the rebalancing interval increases.


We now compare the trading volumes associated with the DBL and RCBL. 
Let 
\begin{equation*}
    n_i(t) = \pi_i(t) \frac{Z(t)}{S_i(t)},
\end{equation*}
denote the number of assets $i \in \{A,B,C,D,E\}$ held at time $t \in [0,T]$
where $Z(t)$ is wealth at $t$, $\pi_i(t)$ the proportion of wealth invested in asset $i$, and $S_i(t)$ is the price of asset $i$ at time $t$. 
When trading is continuous, turnover is given by the total variation
\[
\text{Turnover}_{\text{cont}} = \sum_{i=A}^E \mathbb{E}\left[ \int_0^T \left| dn_i(t) \right| \right].
\]

When trading is discrete, 
\[
\text{Turnover}_{\text{disc}} = \sum_{i=A}^E \mathbb{E}\left[ \sum_{t \in \mathcal{T}} \left| n_i(t+\Delta t) - n_i(t) \right| \right]
\]
where \(\mathcal{T} \subset [0,T]\) is the set of rebalancing times, and \(\Delta t\) the time between them.





\begin{figure*}[ht]
     \centering
     \begin{subfigure}[t]{0.49\textwidth}  
         \centering
         \caption{CBL, $\alpha=0.4$}
         \includegraphics[width=\textwidth]{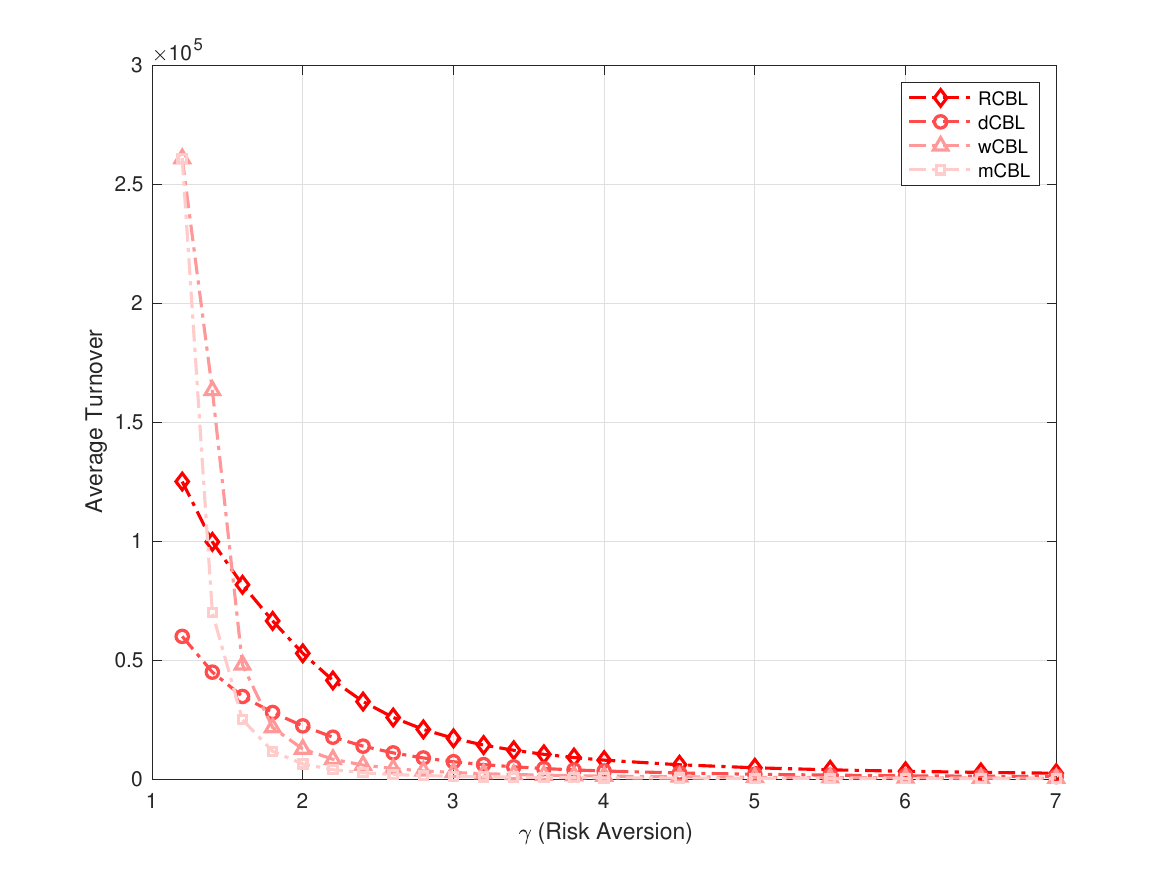}
     \end{subfigure}
     \begin{subfigure}[t]{0.49\textwidth}  
         \centering
         \caption{DBL, $\alpha=0.4$}
         \includegraphics[width=\textwidth]{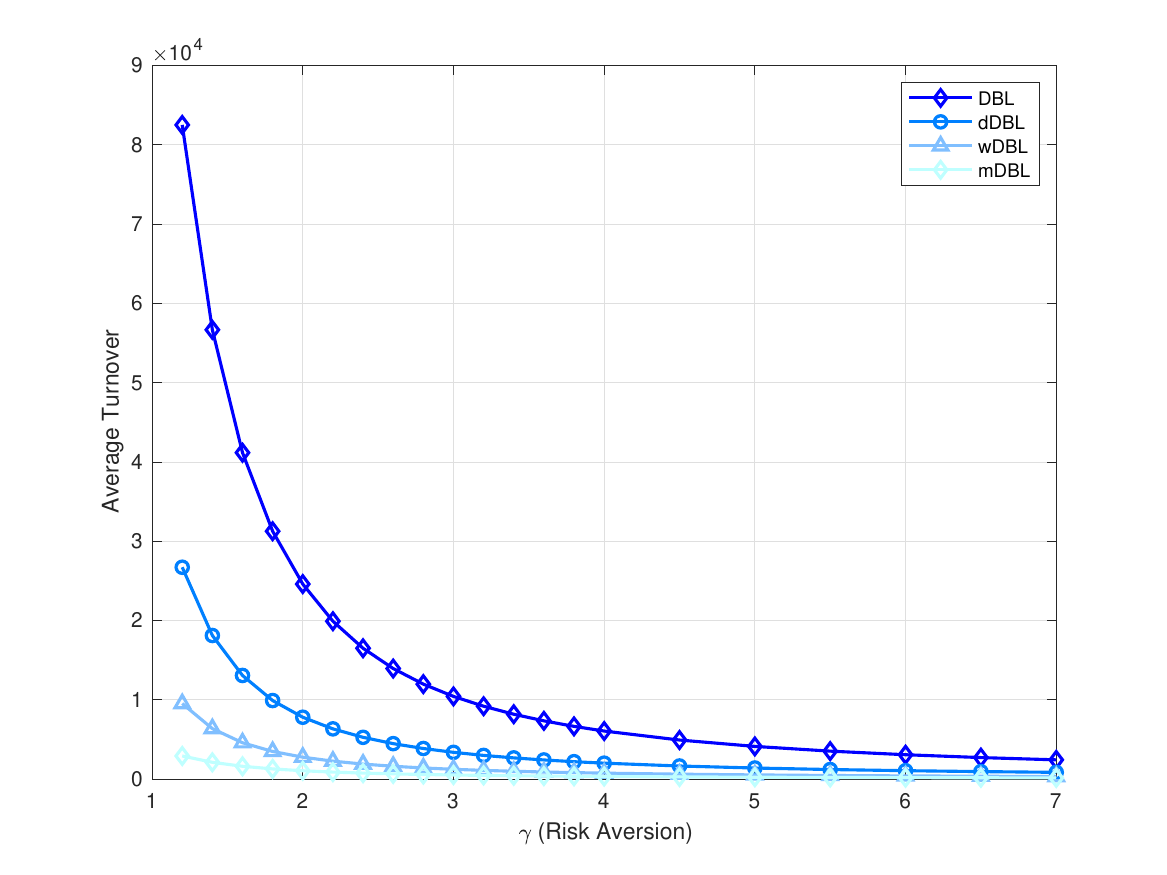}
     \end{subfigure}
\medskip
     \centering

     \begin{subfigure}[t]{0.49\textwidth}  
         \centering
         \caption{CBL, $\alpha=0.8$}
         \includegraphics[width=\textwidth]{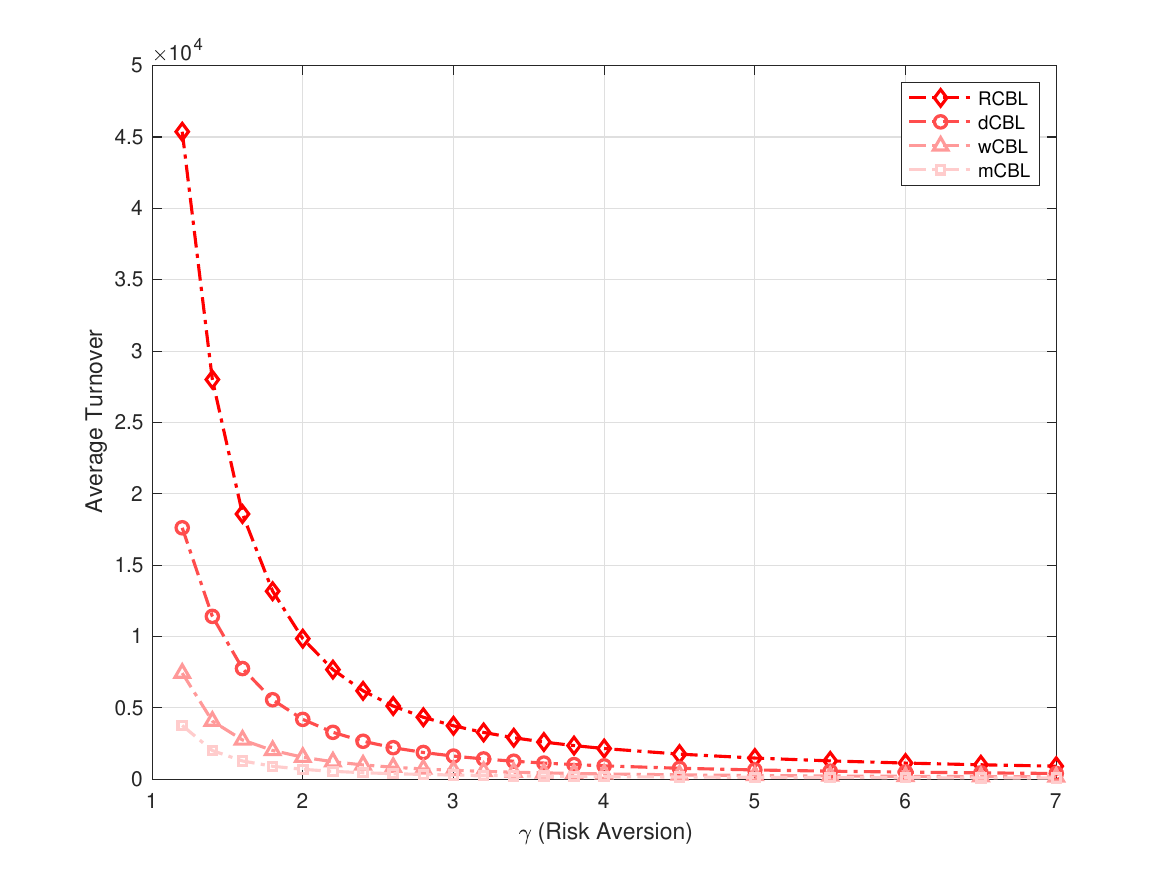}
     \end{subfigure}
     \begin{subfigure}[t]{0.49\textwidth}  
         \centering
         \caption{DBL, $\alpha=0.8$}
         \includegraphics[width=\textwidth]{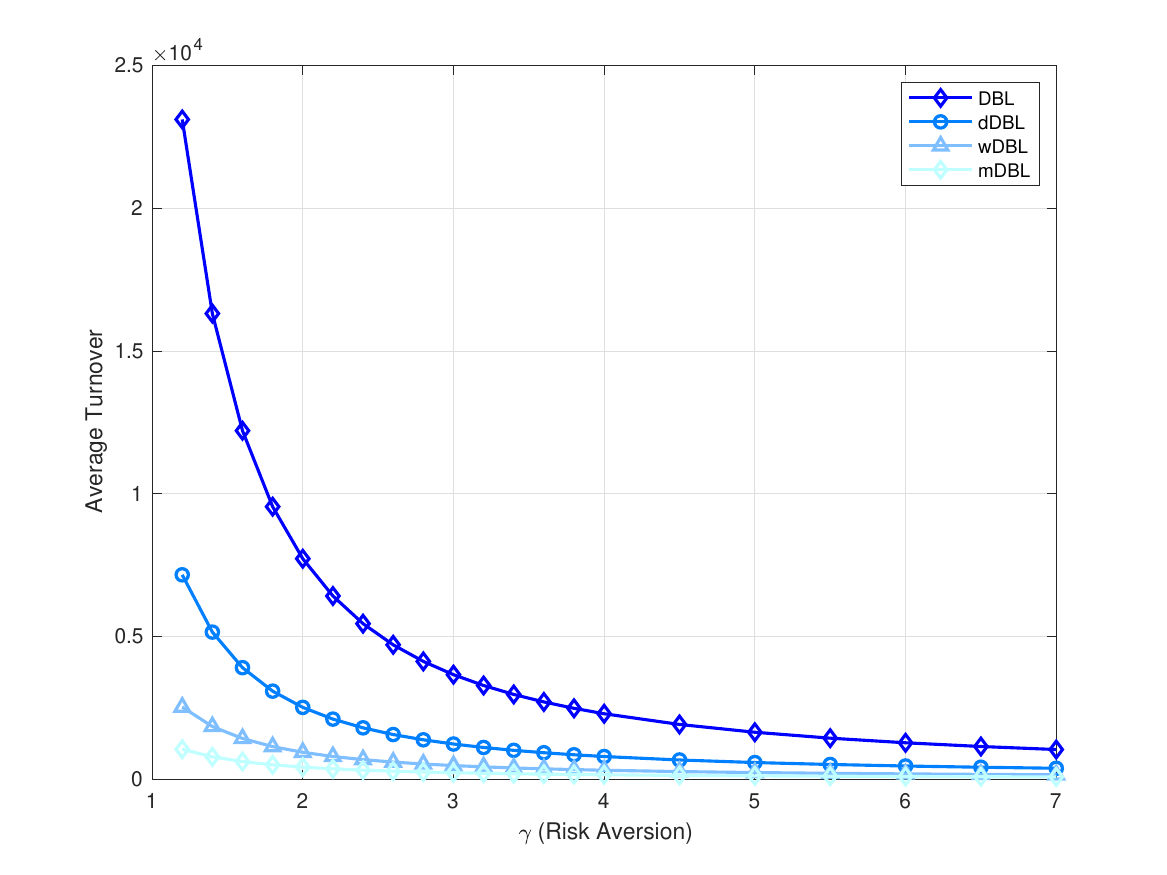}
     \end{subfigure}
     \caption{Average Turnover as a function of risk aversion. \textmd{The figure compares the average turnover for investors adopting the DBL and RCBL strategies for different levels of risk aversions under moderate and high noise in the expert views. Turnover is substantially lower for the DBL investor, particularly when view noise is low ($\alpha=0.4$). Note the difference in scale when comparing DBL and RCBL.}}
     \label{fig:TradingVolumes_alpha0.4-0.8}
\end{figure*}

Figure \ref{fig:TradingVolumes_alpha0.4-0.8} shows turnover for DBL and RCBL for risk-aversion $\gamma \in \{1.2, \dots, 7\}$ for low ($\alpha = 0.4$) and high ($\alpha = 0.8$)  view noise. DBL has lower trading volume across all levels of risk aversion $\gamma$. The difference in trading volume between DBL and RCBL is largest  views are more precise ($\alpha=0.4$). This again is a consequence of DBL being less sensitive to changes in $X^y(t)$ because it hedges changes in the predictor. RCBL makes larger adjustments at each investment epoch because it does not hedge. (Note that the annual trading volume for CBL with weekly and monthly rebalancing is much higher than shown in plot (a) of Figure \ref{fig:TradingVolumes_alpha0.4-0.8}, but has been capped to make the plot more readable.)



In summary, RCBL and DBL are different approaches to making investment decisions over time. DBL is a dynamic policy which includes a hedge for changes in the value function when the view covariate changes value. The payoff from this hedge reduces the sensitivity of the investment performance of DBL to misalignment between her portfolio holding and the current value of the covariate. As a result the performance of DBL is insensitive to being constrained to discrete trading. RCBL  solves a single-period problem at each rebalancing epoch  and does not hedge changes in $X^y(t)$. Her performance deteriorates more quickly when the rebalancing interval increases, especially when  the view covariate is informative (view noise is low), because she needs to wait longer before she can realign her position with  $X^y(t)$. Likewise, trading volume for RCBL is much higher.

\subsection{View Revisions}
\label{sec:experiments-revisions}

We now analyze the scenario where an investor anticipates revisions to her views during the investment horizon \(T = 1 \text{ year}\). We compare three types of investors:
\begin{enumerate}
    \item Baseline Investor: Receives the original view with no revisions.
    \item Single Revision Investor: Anticipates one revision at \(t_{1/2} = 6 \text{ months}\).
    \item Multi-Revision Investor: Anticipates a revision every quarter.
\end{enumerate}

The original view follows \eqref{eq:UpdatedViews} with a covariance matrix \(\Omega^0 = \alpha P \Sigma P^\top\), where \(\alpha \in (0, 2]\). Additionally, the noise in each update is proportional to the horizon of the update. For instance, an update at the midpoint of the investment period has a noise matrix \(\Omega^{1/2} = \frac{1}{2}\alpha P \Sigma P^\top\), while an update at the beginning of the second quarter has \(\Omega^{2/4} = \frac{3}{4} \alpha P \Sigma P^\top\).


\begin{figure}[ht]
     \centering
     \begin{subfigure}[b]{0.49\textwidth}
         \centering
             \captionsetup{font=footnotesize} 
          \caption{$\gamma = 5$}
         \includegraphics[width= \textwidth]{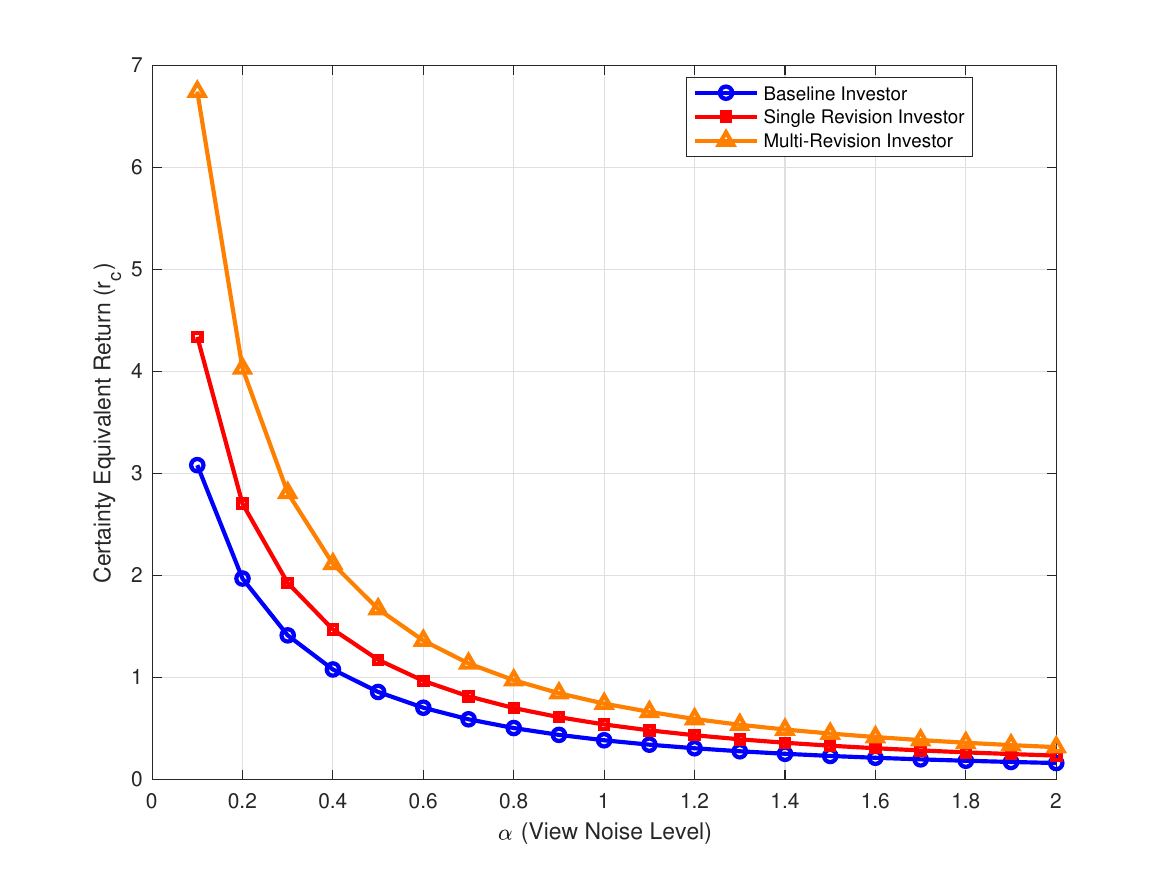}
         \label{fig:revisions_a}
     \end{subfigure}
     \begin{subfigure}[b]{0.49\textwidth}
         \centering
          \caption{$\alpha = 0.6$}
         \includegraphics[width=\textwidth]{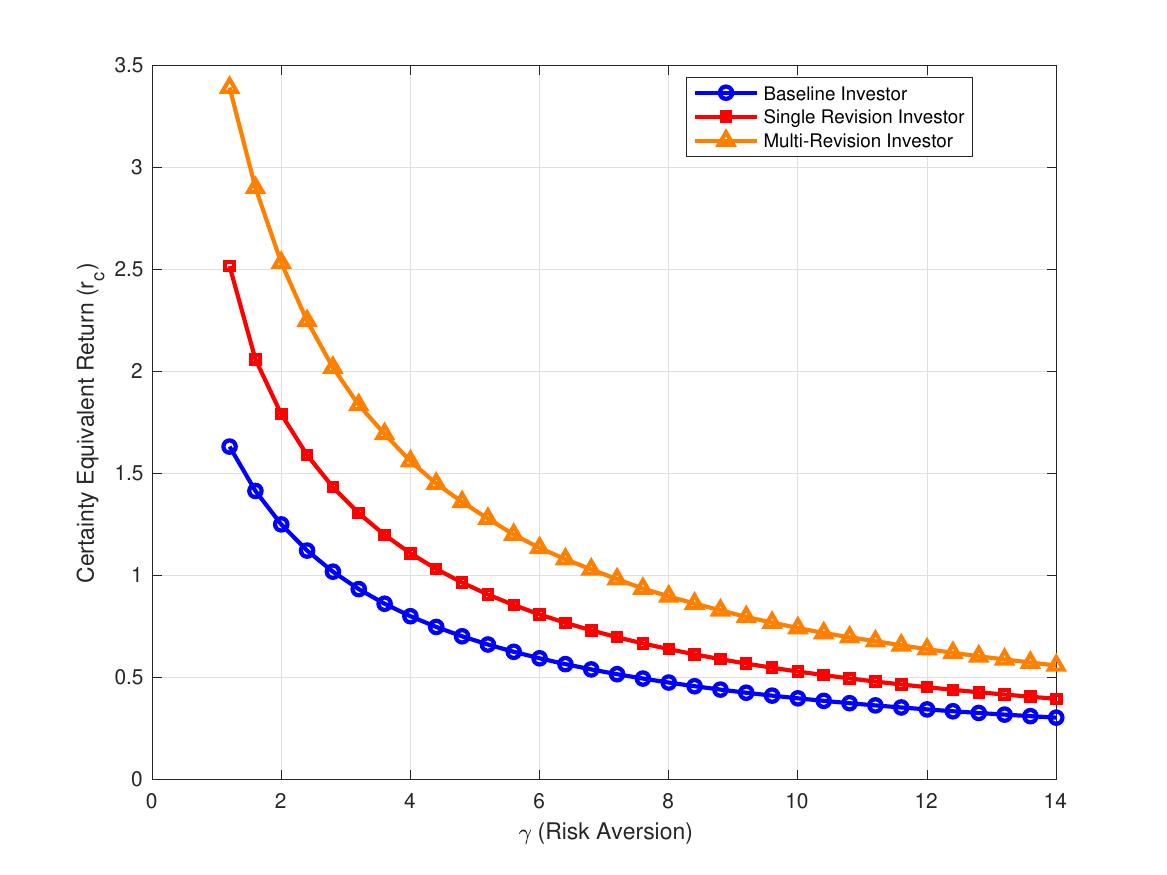}
         \label{fig:revisions_b}
     \end{subfigure}
        \caption{Certainty Equivalent Return comparison for three types of investors. \textmd{The figure illustrates how certainty equivalent returns vary for investors who receive no updates, a single mid-horizon update, or quarterly updates for different levels of risk-aversion and noise in the views.}}
        \label{fig:revisions}
\end{figure}

Figure \ref{fig:revisions} compares the certainty equivalent return for the three investors. Figure \ref{fig:revisions_a} compares the three investors for a fixed level of risk-aversion (\(\gamma = 5\)) and varying levels of noise \(\alpha \in (0,2]\). Observe that the certainty equivalent return increases with the number of anticipate updates. However, as the noise in the views and updates grows (\(\alpha \to 2\)), the difference between the investors diminish. This is intuitive: when views become uninformative (\(\alpha \to \infty\)), the certainty equivalent returns converge. Similar observations can be drawn from Figure \ref{fig:revisions_b}, which compares the certainty equivalent returns for a fixed noise level (\(\alpha = 0.6\)) and varying levels of risk aversion \(\gamma \in \{1, \dots, 14\}\).

\section{Conclusion and Further Research}
\label{sec8}
In this paper, we formulate a dynamic version of the Black-Litterman model with forward-looking expert views. We derive the dynamics of the conditional price process when asset prices are log-normal using techniques from Kalman smoothing, and show additionally that the conditional log-returns process can be written in terms of a multi-dimensional Brownian bridge with drift. In the process, we define a generalized notion of Brownian bridge where noisy estimates of the terminal value (in place of exact values) of a multi-dimensional Brownian motion are given, characterize its distributional properties and derive the stochastic differential equation for its evolution. We also show that the components of the generalized bridge are correlated one-dimensional Brownian bridges with  hitting times that are endogenously determined by the correlation structure between the elements of the original Brownian motion and the noisy observations. The conditional price process is now an affine model with the conditional log-returns playing the role of a predictor. We formulate a dynamic portfolio choice problem in terms of the conditional price process for an investor with the expert views. Although it is a factor model, we are able to derive (quite surprisingly) very explicit expressions for the optimal dynamic portfolio, which consists of a mean-variance holding and a hedging demand. 

The approach we present for incorporating forward-looking views into asset prices can be extended to views on market signals that drive returns and, more generally, random processes that a decision maker is modeling. 
Extensions in this direction will be reported elsewhere.

\section*{Acknowledgments}

Anas Abdelhakmi is partially supported by `Fondation Ibn Rochd Pour Les Sciences et L'Innovation (FIRSI)', under its 2021 SEEDS Initiative. Andrew Lim is supported by the  Ministry of Education, Singapore, under its 2021 Academic Research Fund Tier 2 grant call (Award ref: MOE-T2EP20121-0014).




\bibliographystyle{plainnat} 
\bibliography{reference}

\newpage

\ECSwitch

\setcounter{section}{0}








\ECHead{Prerequisites}
We recall some properties of conditioning over Gaussian vectors  $(X,Y) \in \mathbb{R}^{N  + K}$, where $X \in \mathbb{R}^N$ and $Y \in \mathbb{R}^K$ with respective means $\mu_X$ and $\mu_Y$ and covariance matrices $\Sigma_{XX}$ and $\Sigma_{YY}$. Denote by $\Sigma_{XY}$ the cross covariance between $X$ and $Y$. The conditional distribution of $X$ given $Y = y$ remains Gaussian, with conditional mean given by
\begin{equation}
    \mathbb{E}[X\,|\, Y =  y] = \mu_X + \Sigma_{XY}\Sigma^{-1}_{YY} (y - \mu_Y)
    \label{A}
\end{equation}
and conditional variance
\begin{equation}
    \mathbb{V}[X\,|\,Y = y] = \Sigma_{XX} - \Sigma_{XY}\Sigma^{-1}_{YY}\Sigma_{XY}^\top.
    \label{B}
\end{equation}
The Woodbury matrix identity (\cite{Woodburry}) will be used repeatedly:  Let $\Sigma \in \mathbb{R}^{N \times N}$ and $\Omega \in \mathbb{R}^{K \times K}$ be  invertible square matrices and  $U \in \mathbb{R}^{N \times K}$, $V \in \mathbb{R}^{K \times N}$. Then
\begin{equation*}
    (\Sigma + U\Omega V)^{-1} = \Sigma^{-1} - \Sigma^{-1}U(\Omega^{-1} + V \Sigma^{-1} U)V \Sigma^{-1}.
\end{equation*}

\section{Section \ref{sec3} -- Proof of Proposition \ref{proposition1}}
The proof of the proposition can be split into two parts: We first derive the drift and volatility of the conditional log-returns process, then use these expressions to show that the process can be written as a solution of a Stochastic Differential Equation (SDE). Finally, we derive the SDE of the conditional price process using Itô's Lemma.

We first start by deriving the conditional mean and covariance of the log-returns process.

\subsection{Conditional Mean and Covariance}
\label{ec11}

The price process $S(t)$, log-returns $X(t)$ and views $Y(0, T)$ are given by equations \eqref{eq9}--\eqref{eq10}, respectively. The investment horizon if $T$ and Assumption \ref{ass-covariance} is assumed to hold.

The vector $(X(t), Y(0,T))$ is then Gaussian with 
   \begin{equation*}
       \begin{pmatrix}
           X(t) \\
           Y(0,T)
       \end{pmatrix} \sim \mathcal{N}\left(\begin{pmatrix}
t \mu^x \\
T P \mu^x
\end{pmatrix},M\right),
   \end{equation*}
where the covariance matrix $M \in \mathbb{R}^{(N + K) \times (N + K)}$ is positive definite with block representation
\begin{equation*}
    M = \begin{pmatrix}
         t \Sigma & &t \Sigma P ^\top \\
         t  P \Sigma & &T (P \Sigma P^\top + \Omega)
    \end{pmatrix}.
\end{equation*}
Since the log-returns vector and views are jointly Gaussian, the distribution of any subset of the log-return conditional on any subset of the views will also be Gaussian. It follows that the conditional distribution of the log-returns given expert views is fully specified by its mean and covariance. 

By \eqref{A}, the conditional mean
\begin{equation}
\begin{split}
    \mathbb{E}[X(t) \,|\,Y(0,T) = y] &= \mathbb{E}[X(t)] + \Cov(X(t), Y(0,T)) \mathbb{V}[Y(0,T)]^{-1} (y - \mathbb{E}[Y(0,T)])  \\
    &= t \mu^x + \frac{t}{T} \Sigma P^\top ( P \Sigma P ^\top + \Omega)^{-1} (y - T P \mu^x)\\ 
    &= t \mu^x + t \beta_1 (y - TP \mu^x)
\end{split}
\textbf{}\label{eq:meanX}
\end{equation}
where 
\begin{equation*}
    \beta_1 = \frac{1}{T}\Sigma P^\top (P \Sigma P^\top + \Omega)^{-1}.
\end{equation*}
By \eqref{B} the conditional covariance 
\begin{equation*}
\begin{split}
    \mathbb{V}[X(t) \,|\,Y(0,T)= y] &= \mathbb{V}[X(t)] - \Cov(X(t), Y(0,T)) \mathbb{V}[Y(0,T)]^{-1} \Cov(Y(0,T), X(t))  \\
    &= t  \Sigma - \frac{t^2}{T} \Sigma P^\top ( P \Sigma P ^\top + \Omega)^{-1} P \Sigma. \\ 
\end{split}
\end{equation*}
A similar argument shows that the covariance between the log-return at times $t \leq T$ and $\tau \leq T$ conditioned on the views $Y(0,T) = y$ is 
\begin{equation*}
\begin{split}
        \Cov(X(t), X(\tau) \,|\, Y(0,T) = y) &= \mathbb{E}\big[ (X(t) - \mathbb{E}[X(t)\,|\, Y(0,T) = y])(X(\tau) - \mathbb{E}[X(s)\,|\,y]) \,|\,  Y(0,T) = yy  \big]\\
        &= \mathbb{E}[X(t)X(\tau)\,|\, Y(0,T) = y] - \mathbb{E}[X(t)\,|\,y]\mathbb{E}[X(\tau)\,|\, Y(0,T) = y]\\
        &= \min\{t,\tau\}  \Sigma - \frac{\tau t}{T} \Sigma P^\top ( P \Sigma P ^\top + \Omega)^{-1} P\Sigma.
\end{split}
\end{equation*}
We now use these results to derive the distribution of the conditional dynamics $dX(t) \,|\, Y(0,T) = y$.
\subsection{Conditional Dynamics}
\label{ec12}
Consider an increment of the process $X(t)$ over $[t, t +dt]$ given the history $\{X(\tau), \tau \leq t\}$ and views $Y(0,T) = y$:
\begin{equation*}
    dX(t) \,|\, (Y(0,T) = y,  \{X(\tau)_{\tau \leq t}\}) = \lim_{dt \to 0 }\big(X(t + dt) - X(t)\big) \,|\, \big(Y(0,T) = y, \{X(\tau)_{\tau \leq t}\}\big).
\end{equation*}
 Since the log-returns vector is driven by a Brownian motion $W(t)$, it is Markovian and the information contained in the historical data $\{X(\tau), \tau \leq t\}$ is all stored in the last state $X(t)$, thus
\begin{equation*}
   \big( X(t + dt) - X(t) \big) \,|\, \big(Y(0,T) = y, \{X(\tau)_{\tau \leq t}\}\big) = \big( X(t + dt) - X(t) \big) \,|\, \big(Y(0,T) = y, X(t) \big).
\end{equation*}
Since $X(t)$ and $Y(0,T)$ are Gaussian,  $\big(X(t+dt) \,|\, Y(0,T) = y, X(t)\big)$ is also Gaussian and fully identified by its mean and covariance matrix which we derive next.

\paragraph{\textbf{Mean of the conditional dynamics.}}
Consider the random variable $Z(t) = (X(t), Y(0,T))^\top$, and its realization $z = (x,y)^\top$. The vector $(X(t+dt), Z(t))^\top$ is jointly Gaussian. By \eqref{A} the conditional expectation of the log-returns at time $t + dt$ given $Z(t) = z$ is
\begin{equation*}
\begin{split}
        \mathbb{E}\big[ X(t+dt) \,|\, X(t) = x, Y(0,T)=y  \big] &= \mathbb{E}\big[ X(t+dt) \,|\, Z(t)=z  \big]\\
        &=\mathbb{E}\big[X(t+dt) \big]  + \Cov(X(t+dt), Z(t)) \mathbb{V}[Z(t)]^{-1} (z - \mathbb{E}[Z(t)]) \\
\end{split}
\end{equation*}
where the covariance matrix of $Z(t)$ is
\begin{equation}
\label{eqA1}
    \begin{split}
        \mathbb{V}[Z(t)] &= \begin{pmatrix}
            t  \Sigma & & t  \Sigma P^\top\\
            t  P \Sigma & & T  (P \Sigma P^\top + \Omega),
        \end{pmatrix}
    \end{split}
\end{equation}
 the covariance between $X(t+dt)$ and $Z(t)$ is 
\begin{equation*}
    \begin{split}
        \Cov(X(t+dt), Z(t)) &= \begin{pmatrix}
            t  \Sigma \\  (t + dt)  \Sigma P^\top
        \end{pmatrix}^\top
    \end{split}
\end{equation*}
and
\begin{eqnarray*}
    z - {\mathbb E}[Z(t)] = 
    \left[\begin{array}{cc}x-t \mu^x \\ y - T P \mu^x\end{array}\right].
    \end{eqnarray*}
We now derive the explicit expression of the inverse of the covariance matrix \eqref{eqA1}. By \cite{LU2002119}, the inverse of a $2 \times 2$ block matrix $R$ is
\begin{equation*}
   R^{-1} = \begin{pmatrix}
           A &  B\\
            C &D
        \end{pmatrix}^{-1} = \begin{pmatrix}
            (A - BD^{-1}C)^{-1} & & - (A - BD^{-1}C)^{-1} BD^{-1}\\
            -D^{-1}C(A - BD^{-1}C)^{-1} & &  D^{-1} + D^{-1} C (A - BD^{-1}C)^{-1} BD^{-1}
        \end{pmatrix}
\end{equation*}
where $R$ has size $N + K \times 2N$, $A$ and $B$ are $N \times N$, $C$ and $D$ are $K \times N$, and $D$ is a non-singular matrix. The Woodbury matrix identity gives
\begin{equation*}
    (A - BD^{-1}C)^{-1} = A^{-1} - A^{-1}B(D + C A^{-1} B)C A^{-1}.
\end{equation*}
Letting $A = B = t \Sigma$, $C = t P \Sigma$, and $D = T(P\Sigma P^\top + \Omega)$,
we have
\begin{eqnarray*}
    {\mathbb V}[Z(t)] =  \mathbb{V}[Z(t)] &= \begin{pmatrix}
            t  \Sigma & & t  \Sigma P^\top\\
            t  P \Sigma & & T  (P \Sigma P^\top + \Omega)
        \end{pmatrix}
\end{eqnarray*}
and hence
\begin{eqnarray*}
    \lefteqn{\Cov(X(t+dt), Z(t)) \mathbb{V}[Z(t)]^{-1} }\\ &  =  &\begin{bmatrix}
         I_N- dt \, \Sigma P^\top  ((T - t) P\Sigma P^\top + T \Omega)^{-1} P , \, \,  dt \, \Sigma P^\top   ((T - t) P\Sigma P^\top + T \Omega)^{-1}
    \end{bmatrix}.
\end{eqnarray*}
It follows that the conditional mean is
\begin{eqnarray}
       \lefteqn{\mathbb{E}\big[ X(t+dt) - X(t) \,|\, X(t) = x, Y(0,T)=y  \big]} \nonumber\\ &  = & dt \, \Big( \mu^x  -  \Sigma P^\top ((T - t) P\Sigma P^\top + T \Omega)^{-1} P (X(t) - t \mu^x) \nonumber \\  & & + \Sigma P^\top  ((T - t) P\Sigma P^\top + T \Omega)^{-1} (y - TP \mu^x) \big).
       \label{eqA2}
\end{eqnarray}
Woodbury's matrix identity gives
\begin{equation*}
    \begin{split}
        \Sigma P^\top  ((T - t) P\Sigma P^\top + T \Omega)^{-1} = \big( I_N + t \beta_2(t) \big) \beta_1
    \end{split}
\end{equation*}
where 
\begin{equation*}
    \beta_1 = \frac{1}{T}\Sigma P^\top (P \Sigma P^\top + \Omega)^{-1}
\end{equation*}
and
\begin{equation}
\label{eq:beta2}
    \beta_2(t) = \Sigma P^\top \left((T-t) P \Sigma P^\top + T \Omega\right)^{-1}P.
\end{equation}
Recalling \eqref{eq:meanX}, it can be shown after a bit of algebra that \eqref{eqA2} and hence the drift of the conditional process $dX(t) \,|\, Y(0,T) = y$ is
\begin{equation*}
\begin{split}
        \mathbb{E}\big[ X(t+dt)  - X(t) \,|\, X(t) = x, Y(0,T)=y \big] = dt \big(\mu^x +  \beta_1 (y - TP \mu^x) -  \beta_2(t) (X(t) - \mathbb{E}[X(t) \,|\, y])\big),
         \end{split}
\end{equation*}
where
\begin{equation*}
\begin{split}
    \mathbb{E}[X(t) \,|\,Y(0,T) = y] = t \mu^x + \beta_1 (y - TP \mu^x).
\end{split}
\end{equation*}
Since \(X(t)\) is continuous and has finite increments, by taking the limit as \(dt \to 0\), we obtain
\begin{equation}
\label{eqA3}
\begin{split}
        \mathbb{E}\big[ dX(t) \,|\, X(t) = x, Y(0,T)=y \big] = dt  \big(\mu^x +  \beta_1 (y - TP \mu^x) -  \beta_2(t) (X(t) - \mathbb{E}[X(t) \,|\, y])\big).
\end{split}
\end{equation}

\paragraph{\textbf{Covariance of the conditional dynamics.}}
A similar argument shows that the covariance of the conditional dynamics is
\begin{equation*}
\begin{split}
        \mathbb{V}\big[ X(t+dt) \,|\, X(t) = x, Y(0,T)=y  \big] &= \mathbb{V}\big[ X(t+dt) \,|\, Z=z  \big]\\
        &=\mathbb{V}\big[X(t+dt) \big]  - \Cov(X(t+dt), Z) \mathbb{V}[Z]^{-1} \Cov(X(t+dt), Z)^\top\\
        &= (dt) \Sigma - (dt)^2 \cdot \Sigma P^\top ((T - t) P \Sigma P^\top + T \Sigma)^{-1} P \Sigma\\
        &= (dt)  \Sigma + o(dt).
\end{split}
\end{equation*}
By taking the limit as  \(dt \to 0\), we get
\begin{equation}
\label{eqA4}
\begin{split}
        \mathbb{V}\big[ dX(t) \,|\, X(t) = x, Y(0,T)=y  \big] = (dt)\,    \Sigma + o(dt).
      \end{split}
\end{equation}

Since \(X(t)\) has finite increments, equations \eqref{eqA3} and \eqref{eqA4} are well defined and they characterize the drift and v olatility of the conditional process \mbox{$dX(t)\, | \,(X(t) , Y(0,T) = y)$}. We now derive the SDE that is satisfied by  the process $X(t)\,|\,Y(0,T) = y$.

\subsection{SDE of the conditional log-returns process}
We now show that the conditional process $X(t) \,|\, Y(0,T) = y$ is the solution to an SDE with drift \eqref{eqA3} and volatility  \eqref{eqA4}. We first define
\begin{equation*}
    \beta_3(t) = t \Big(\mu^x + \beta_1 ( Y(0,T) - TP \mu^x)\Big) - \int_{0}^t  \beta_2(s) (X(s) - \mathbb{E}[X(s) \,|\, Y(0,T)] \, ds, \, \, \, \text{for} \, \, t \in [0,T].
\end{equation*}
By showing that 
\begin{equation*}
    W^y(t) = X(t) -  \beta_3(t)
\end{equation*}
is a Brownian motion in the enlarged filtration $\mathcal{F}^Y_t := \sigma (\mathcal{F}_t \vee \sigma(Y(0,T)))$, we obtain
\begin{equation*}
    dW^y(t) = dX(t)\,|\,y - d\beta_3(t), \, \, \, \text{for} \, \, t \in [0,T]  
\end{equation*}
which gives the SDE representation of $X(t) \,|\, Y(0,T) = y$.

To prove that $W^y(t)$ is a Brownian motion in the filtration $\mathcal{F}_t^y$, we refer to Levy's Characterization of a Brownian motion (see for example \cite{Durret}).
\begin{theorem}[Levy's characterization of a Brownian motion]\label{levys}
    Let the stochastic process $W^y = (W_1^y, \dots, W_N^y)$ be a $N$--dimensional local continuous martingale with $W^y(0) = 0$. Then, the following is equivalent:
    \begin{enumerate}
        \item $W^y$ is a Brownian motion on the underlying filtered probability space with $W^y(t) \sim \mathcal{N}\big(0, t\Sigma\big)$.
        \item $W^y$ has quadratic covariations $[W_i^y(t), W_j^y(t)] = \Sigma_{ij}t$ for $1 \leq i,j \leq N$.
    \end{enumerate}
\end{theorem}

Since $X(t)$ and $\beta_3(t)$ are both continuous, $W^y$ is also continuous. Now we show that is it a local martingale in the filtration $\mathcal{F}_t^Y$, that is for $s \leq t$
\begin{equation*}
    \mathbb{E}[W^y(t) - W^y(s) \,|\, \mathcal{F}_s^Y] = 0.
\end{equation*}
We first have
\begin{equation*}
    \begin{split}
        \mathbb{E}[X(t) - X(s) \,|\,  \mathcal{F}_s^Y] &\stackrel{(a)}{=} \mathbb{E}[X(t) - X(s) \,|\, X(s), Y(0,T)]\\
        & \stackrel{(b)}{=}   (t-s)  \big(\mu^x + \beta_1 (y - TP \mu^x) - \beta_2(s) (X(s) - \mathbb{E}[X(s) \,|\, Y(0,T)])\big)
    \end{split}
\end{equation*}
where $(a)$ follows from the Markov property of $X(t)$ and $(b)$ is derived from \eqref{eqA2}. It follows that
\begin{equation*}
    \begin{split}
         \mathbb{E}[W^y(t) - W^y(s) \,|\, \mathcal{F}_s^Y] &=  \mathbb{E}[W^y(t) - W^y(s) \,|\,X(s), Y(0,T)]\\
         &=  \mathbb{E}[X(t) - X(s) \,|\,X(s), Y(0,T)] -  \mathbb{E}[\beta_3(t) - \beta_3(s) \,|\,X(s), Y(0,T)]\\
         &= \underbrace{- (t-s) \beta_2(s) (X(s) - \mathbb{E}[X(s) \,|\, Y(0,T)])}_{RHS_1}\\
         &+ \underbrace{\mathbb{E}\big[\int_s^t  \beta_2(u) (X(u) - \mathbb{E}[X(u) \,|\, Y(0,T)] du \,|\,X(s), Y(0,T)\big]}_{RHS_2}.
    \end{split}
\end{equation*}
By Fubini's theorem, the second term of the right-hand side
\begin{equation*}
    \begin{split}
        RHS_2 &= \int_s^t  \beta_2(u)  \mathbb{E}\Big[\Big(X(u) - \mathbb{E}\big[X(u) \,|\, Y(0,T)\big]\Big)\,\Big|\,X(s), Y(0,T)\Big] du\\
        &= \int_s^t \beta_2(u) \Big( \mathbb{E}\big[X(u) \,|\,X(s), Y(0,T)\big] - \mathbb{E}\big[X(u) \,|\, Y(0,T)\big]\Big) du\\
        & \stackrel{(c)}{=} \int_s^t \beta_2(u) \Big(I_N + (u-s) \beta_2(s)\Big)\Big(X(s) - \mathbb{E}\big[X(s) \,|\, Y(0,T)\big]\Big) du\\
        &= \biggl\{ \int_s^t \beta_2(u) \Big(I_N + (u-s) \beta_2(s)\Big) du \biggl\}\Big(X(s) - \mathbb{E}[X(s) \,|\, Y(0,T)]\Big)
    \end{split}
\end{equation*}
where $(c)$ can be obtained (after a bit of algebra) from 
\begin{equation*}
\begin{split}
        \mathbb{E}\big[ X(u) \,|\, X(s), Y(0,T)\big] = X(s) +  (u-s) \cdot \big(\mu^x +  \beta_1 (y - TP \mu^x) -  \beta_2(s) (X(s) - \mathbb{E}[X(s) \,|\, Y(0,T)])\big)
\end{split}
\end{equation*}
and
\begin{equation*}
\begin{split}
    \mathbb{E}[X(u) \,|\,Y(0,T) = y] = u \, \mu^x + \beta_1 (y - TP \mu^x).
\end{split}
\end{equation*}
By showing
\begin{equation*}
     \beta_2(u) (I_N + (u-s) \beta_2(s))  = \beta_2(s), \, \, \, \forall u \geq s
\end{equation*}
we will  have $RHS_1 + RHS_2 = 0$ and hence
\begin{equation*}
     \mathbb{E}[W^y(t) - W^y(s) \,|\, \mathcal{F}_s^Y] = 0
\end{equation*}
and $W^y(t)$ will be a local martingale in $\mathcal{F}_t^Y$. 

Recall from \ref{eq:beta2} that
\begin{equation*}
             \beta_2(u) = \Sigma P^\top ((T - u) P\Sigma P^\top + T \Omega)^{-1} P, \, \, \, \forall u \geq 0.
\end{equation*}
It follows that for $0 \leq s \leq u \leq T$ 
\begin{equation*}
    \begin{split}
         \beta_2(u) = \Sigma P^\top ((T -s) P\Sigma P^\top + T \Omega - (u - s) P \Sigma P^\top)^{-1} P.
    \end{split}
\end{equation*}
Woodbury's identity implies
\begin{equation*}
    \begin{split}
        \beta_2(u) &=  \beta_2(s) + (u-s) \beta_2(s) (I_N - (u-s)\beta_2(s))^{-1} \beta_2(s)\\
        &= \beta_2(s) (I_N - (u-s) \beta_2(s))^{-1}
    \end{split}
\end{equation*}
which gives
\begin{equation*}
      \beta_2(u)(I_N + (u-s) \beta_2(s)) = \beta_2(s), \quad \forall \; u \geq s
\end{equation*}
and hence
\begin{equation*}
      \mathbb{E}[W^y(t) - W^y(s) \,|\, \mathcal{F}_s^Y] = 0, \quad \forall \; 0 \leq s \leq t \leq T,
\end{equation*}
so $W^y$ is a local martingale in the filtration $\mathcal{F}^Y_t$. 



It  follows from \eqref{eqA4} that the quadratic variation of $W^y$ satisfies \[
[W_i^y(t), W_j^y(t)] = \Sigma_{ij}\,t, \quad \text{for } 1\le i,j \le N.
\] Thus, by Theorem \ref{levys}, $W^y$ is a Brownian motion with respect to the filtration $\mathcal{F}_t^Y$. 

It follows from \eqref{eqA3} -- \eqref{eqA4} that
\begin{equation*}
    X(t)\,|\,y = W^y(t) + \beta_3(t) \, | \, y,
\end{equation*}
and
\begin{equation*}
\begin{split}
        dX(t)\,|\,y &= dW^y(t) + d \beta_3(t) \,|\, y \\
        &=dt \big(\mu^x + \beta_1 ( y - TP \mu^x) - \beta_2(t) (X(t)\,|\,y - \mathbb{E}[X(t) \,|\, y] \big) + dW^y(t).
\end{split}
\end{equation*}

\subsection{Conditional Price Process}
The stock price process can be obtained directly from the log-returns by noting that
\begin{equation}
\label{eqA5}
    S(t)\,|\,(Y(0,T) = y) = S(0) \exp\big(X(t) \,|\, (Y(0,T) = y)\big).
\end{equation}
We apply Itô's lemma to \eqref{eqA5}. For $i \in [N]$, we have
\begin{equation*}
\begin{split}
        dS_i(t) \,|\, y &= S_i(0) e^{X_i(t) \,|\, y} dX_i(t)\,|\,y + \frac{1}{2}  S_i(0) e^{X_i(t) \,|\, y} (dX_i(t)\,|\,y)^2\\
        &= S_i(t) \,|\, y \cdot (dX_i(t)\,|\,y + (dX_i(t)\,|\,y)^2),
\end{split}
\end{equation*}
with
\begin{equation*}
    (dX_i(t)\,|\,y)^2 = (dW_i^y(t))^2 = \sigma_i^2 dt.
\end{equation*}
Therefore, the conditional dynamics of asset prices are 
\begin{equation*}
\begin{split}
        dS(t)\,|\,y &= D(S(t)\,|\,y) \big( dX(t)\,|\,y + \frac{1}{2}\diag(\Sigma) dt   \big)\\
        &= D(S(t)\,|\,y) \big( \tilde{\mu}(t,X(t)\,|\,y)dt + dW^y(t) \big),
\end{split}
\end{equation*}
with drift
\begin{equation*}
\begin{split}
        \tilde{\mu}(t,x) &= \mu^x+ \beta_1 \big(y - TP \mu^x \big) - \beta_2(t) \big(x - \mathbb{E}[X(t)\,|\,y]\big) + \frac{1}{2}\diag(\Sigma)\\
        &=\mu + \beta_1 \big(y - TP \mu^x \big) - \beta_2(t) \big(x - \mathbb{E}[X(t)\,|\,y]\big).
\end{split}
\end{equation*}
This completes the proof. \hfill $\square$

\subsection{Novikov's condition} 
\label{app:VF}


For \eqref{eq:CoM} to define a change of measure we require
${\mathbb E}_{\mathbb P}\left[\frac{d\mathbb Q}{d {\mathbb P}}\right]=1$. A sufficient condition is Novikov's condition 
(see, e.g., Lemma 8.6.2 in \cite{OkSendal2003})
\begin{equation*}
\mathbb{E}_\mathbb{P}\Biggl[\exp\Bigl(\frac{1}{2}\int_0^T \,|\,k(t)\,|\,^2\, dt\Bigr)\Biggr] < \infty.
\end{equation*}

Define
\begin{equation*}
\tilde{X}(t) := X^y(t) - \mathbb{E}[X^y(t)].
\end{equation*}
From \eqref{eq11}, \(\tilde{X}(t)\) satisfies the SDE
\begin{equation}\label{eq:U_SDE}
d\tilde{X}(t) = -\beta_2(t)\tilde{X}(t)\,dt + dW(t),\quad \tilde{X}(0)=0.
\end{equation}
Moreover, since
\begin{equation*}
\beta_2(t)= \Sigma P^\top \Bigl((T-t)P\Sigma P^\top + T\Omega\Bigr)^{-1}
\end{equation*}
is bounded on \([0,T]\), there exists a constant \(M>0\) such that
\begin{equation*}
|\beta_2(t)| \le M,\quad \forall\, t\in [0,T].
\end{equation*}

Thus, we get
\begin{equation*}
|k(t)|^2 \le 2\left|\beta_1\bigl(y-TP\mu^x\bigr)\right|^2 + 2\,|\beta_2(t)|^2\,|\tilde{X}(t)|^2 
\leq 2 M^2\,|\tilde{X}(t)|^2.
\end{equation*}
The Novikov condition reduces to showing that
\begin{equation*}
\mathbb{E}_\mathbb{P}\Biggl[\exp\Bigl(M^2\int_0^T |\tilde{X}(t)|^2\,dt\Bigr)\Biggr] < \infty.
\end{equation*}

The solution to \eqref{eq:U_SDE} is given by
\begin{equation*}
\tilde{X}(t)= \exp\Bigl(-\int_0^t \beta_2(s)\,ds\Bigr)\int_0^t \exp\Bigl(\int_0^s \beta_2(r)\,dr\Bigr)dW(s),
\end{equation*}
which shows that \(\tilde{X}(t)\) is a zero–mean Gaussian process with uniformly bounded variance (see, e.g., \cite{OkSendal2003, RevuzYor1999}). Thus, the quadratic function
\begin{equation*}
\int_0^T |\tilde{X}(t)|^2\,dt
\end{equation*}
admits finite exponential moments.
Thus, we conclude that
\begin{equation*}
\mathbb{E}_\mathbb{P}\Biggl[\exp\Bigl(\frac{1}{2}\int_0^T |k(t)|^2\,dt\Bigr)\Biggr] < \infty,
\end{equation*}
which verifies the Novikov condition. 

\newpage

\section{Section \ref{sec4}}

\subsection{Proof of Proposition \ref{proposition2}} 

\medskip

Let $W(t) \in \mathbb{R}$ be a standard Brownian motion with variance $\mathbb{V}[W(t)] = t$ and initial value $W(0) = a$. At time $t =0$, we observe a sample $y$ of the random variable $Y(0,T) = W(T) + \epsilon$ where $\epsilon \sim \mathcal{N}\big(0, T\omega^2\big)$ is independent of $W$. The following result shows that the conditional process $\{B(t) = (W(t) \,|\, Y(0,T) = y), t\in [0,T]\}$ is a restriction of a Brownian bridge from $a$ to $y$ with hitting time $\tilde{T} = T(1 + \omega^2)$ to the interval $[0,T]$. 

From Proposition \ref{SDE:BB}, a process $B(t)$ is defined as a Brownian bridge from $a$ to $y$ with hitting time $\tilde{T}=T(1+\omega^2)$ if it satisfies
 \begin{enumerate}
        \item $B(0) = a$, and $B(\tilde{T}) = y$ (with probability $1$),
        \item $\{B(t), t \in [0,\tilde{T}]\}$ is a Gaussian process,
        \item $\mathbb{E}[B(t)] = a + \frac{t}{\tilde{T}}(y-a)$ for $t \in [0,\tilde{T}]$,
        \item $\Cov\big(B(t), B(s) \big) = \min\{s,t \} -\dfrac{st}{\tilde{T}}$, for $s,t \in [0,\tilde{T}]$,
        \item With probability $1$, $t \to B(t)$ is continuous in $[0,\tilde{T}]$.
    \end{enumerate}

We now prove that $B(t) = (W(t) \,|\, Y(0,T) = y)$ satisfies the above properties for $t \in [0,T]$.

We first have $B(0) = W(0) = a$, and since the vector $(W(t) , Y(0,T))$ is jointly Gaussian, the conditional process $B(t) = (W(t)\,|\,Y(0,T))$ is normally distributed and satisfies $2$. It is therefore fully identified by its mean and variance, and we have
\begin{equation*}
\begin{split}
\mathbb{E}[B(t)] &= \mathbb{E}[W(t) \,|\, Y(0,T) = y] \\
&= \mathbb{E}[W(t)] + \Cov(W(t), Y(0,t)) \mathbb{V}^{-1}[Y(0,T)] (y - \mathbb{E}[Y(0,T)])\\
    &= a + \frac{t}{T(1 + \omega^2)} (y - a)\\
    &=  a + \frac{t}{\tilde{T}} (y - a), \, \, \, \text{for} \, \, t \in [0,T].
\end{split}
\end{equation*}
Thus, $B(t)$ satisfies $3$. Now let $s,t \in \mathbb{R}$ with $s \leq t$, we have
\begin{equation}
\label{B1}
    \begin{split}
        \Cov(B(t), B(s)) &= \Cov(W(t), W(s) \,|\, Y(0,T) = y)\\
                &= \mathbb{E}\big[\big(W(t) - \mathbb{E}[W(t) \,|\, Y(0,T) = y]\big)\big(W(s) - \mathbb{E}[W(s) \,|\, Y(0,T) = y]\big) \,|\, Y(0,T) = y\big]\\
                &= \mathbb{E}\big[W(t)W(s) \,|\, Y(0,T) = y \big] - \mathbb{E}\big[W(s) \,|\, Y(0,T) = y \big]\mathbb{E}\big[W(t) \,|\, Y(0,T) = y \big],
    \end{split}
\end{equation}
by the law of total expectation we get
\begin{equation*}
    \begin{split}
        \mathbb{E}\big[W(t)W(s) \,|\, Y(0,T) = y \big] &= \mathbb{E}\big[W(s) \mathbb{E}[W(t) \,|\, W(s), Y(0,T) = y] \,|\, Y(0,T) = y \big].
    \end{split}
\end{equation*}
Since \(W(t)\) and \(Y(0,T)\) are jointly Gaussian, we write
\begin{align*}
\mathbb{E}[W(t) \mid W(s), Y(0,T) = y]
&= \mathbb{E}[W(t) \mid W(s)]\\
&\quad +\, \Cov\bigl(W(t),Y(0,T)\,\big|\;W(s)\bigr)\,\mathbb{V}\bigl[Y(0,T)\,\big|\;W(s)\bigr]^{-1}\\
&\quad \cdot \Bigl[y \;-\;\mathbb{E}\bigl(Y(0,T)\,\big|\;W(s)\bigr)\Bigr]\\
&= W(s) \;+\; \frac{t-s}{T(1+\omega^2)\;-\;s}\,\bigl(y \;-\;W(s)\bigr)\\
&= \frac{\tilde{T} - t}{\tilde{T} - s}\;W(s) \;+\; \frac{t - s}{\tilde{T} - s}\;y.
\end{align*} 
Therefore, the conditional expectation of the product $W(t)W(s)$ is
\begin{equation}
\label{B2}
    \begin{split}
        \mathbb{E}\big[W(t)W(s) \,|\, Y(0,T) = y \big] &= \frac{\tilde{T} - t}{\tilde{T} - s} \mathbb{E}[W^2(s)\,|\,Y(0,T) = y] + \frac{t - s}{\tilde{T} - s}y\cdot\mathbb{E}[W(s)\,|\,Y(0,T) = y].
    \end{split}
\end{equation}
It follows \eqref{B1} and \eqref{B2} that
\begin{equation*}
\begin{split}
         \Cov(B(t), B(s)) &= \min\{s,t\} - \frac{st}{\tilde{T}},  \, \, \, \text{for} \, \, \, t \in [0,T],
\end{split}
\end{equation*}
and $B(t)$ satisfies $4$. This implies that ${\mathbb V}[B(\tilde{T})]=0$ so by 3, $B(\tilde{T})=y$. Since $W(t)$ has continuous sample paths, so too does $B(t)$ and hence $5$ holds. Therefore, $\{B(t) = (W(t) \,|\, Y(0,T) = y), t\in [0,T]\}$ is a restriction of a Brownian bridge from $a$ to $y$ with hitting time $\tilde{T} = T(1 + \omega^2)$ to the interval $[0,T]$. This concludes the proof. \hfill $\square$

\subsection{Example \ref{example2}: Detailed calculations}
\label{App:example2}
Let $W_1(t)$ and $W_2(t)$ be two standard Brownian motions with correlation $\rho \in (0,1]$. It can easily be shown that there exist a standard Brownian motion $W_3(t)$ such that:
\begin{enumerate}
    \item $W_3(t)$ is independent of $W_1(t)$,
    \item $W_2(t) = \rho W_1(t) + \sqrt{1 - \rho^2} W_3(t) $.
\end{enumerate}

Given a view   $y_2$ sampled from $Y_2(0,T) = W_2(T) + \epsilon$ where $\epsilon$ is zero-mean Gaussian and with variance $\omega^2$ and independent of $W_2(t)$, we show that the two processes $B_1(t) = \big(W_1(t) \,|\, Y_2(0,T) = y_2 \big)$ and $B_2(t) = \big(W_2(t) \,|\, Y_2(0,T) = y_2\big)$ are Brownian bridges restricted to the interval $[0,T]$. The latter is directly deduced from Proposition \ref{proposition2}, where $B_2(t) = (W_2(t) \,|\,Y_2(0,T) = y_2)$ is a restriction of a Brownian bridge from $0$ to $y_2$ with hitting time $\tilde{T}_2 = T + \omega^2$ to $[0,T]$. Here, we prove the same for $\{B_1(t), t \in [0,T]\}$. 

We start by showing how the view $y_2$ of the Brownian motion $W_2(T)$ can be transformed to a view $y_1$ about the Brownian motion $W_1(T)$. Consider the random variable $Y_1(0, T)$ derived from $Y_2(0,T)$ such that
\begin{equation*}
    Y_1(0,T) = \frac{1}{\rho} Y_2(0,T).
\end{equation*}
We have
\begin{equation*}
    \begin{split}
    Y_1(0,T) &= \frac{1}{\rho} Y_2(0,T)\\
        &=\frac{1}{\rho}( W_2(T) +  \epsilon) \\
        &\stackrel{(a)}{=}  W_1(T) +  \frac{\sqrt{1 - \rho^2}}{\rho} W_3(T)  + \frac{1}{\rho}\epsilon\\
        &= W_1(T) + \bar{\epsilon},
    \end{split}
\end{equation*}
where $(a)$ follows from the decomposition of the Brownian motion $W_2(t)$ into $W_1(t)$ and $W_3(t)$, and 
\begin{equation*}
    \bar{\epsilon}   \sim \mathcal{N}\big(0 ,\dfrac{\omega^2 + (1 - \rho^2) T}{\rho^2} \big)
\end{equation*} 
is the noise term in the view $Y_1(0,T)$.
Therefore, we have
\begin{equation*}
\begin{split}
        B_1(t) &= W_1(t) \,|\, (Y_2(0,T) = y_2) \\
    &= W_1(t) \,|\, (Y_1(0,T) = y_1).
\end{split}
\end{equation*}
From Proposition \ref{proposition2}, the conditional process $B_1(t) = (W_1(t) \,|\, Y_1(0,T) = y_1)$ is a Brownian bridge from $0$ to $y_1$ with hitting time $\tilde{T}_1$ restricted to the interval $[0,T]$, with
\begin{equation*}
    \begin{split}
        \tilde{T}_1 &= T + \mathbb{V}[\bar{\epsilon}]\\
        &= \frac{\omega^2 + T}{\rho^2}.
    \end{split}
\end{equation*}
When $\rho = 0$, notice that $W_1(t)$ and $Y_2(0,T)$ are independent,and  therefore
\begin{equation*}
    \begin{split}
        B_1(t) &= W_1(t) \,|\, (Y_2(0,T) = y_2)\\
        &= W_1(t),
    \end{split}
\end{equation*}
this is also equivalent to having $\tilde{T}_1 = \infty$ (notice that a Brownian bridge with infinite hitting time is a Brownian motion).  \hfill $\square$

\subsection{Example \ref{example2}: Plots}
\label{sec:BBexample}


 \begin{figure}[ht]
 \begin{center}
       \includegraphics[width= 0.5\textwidth]{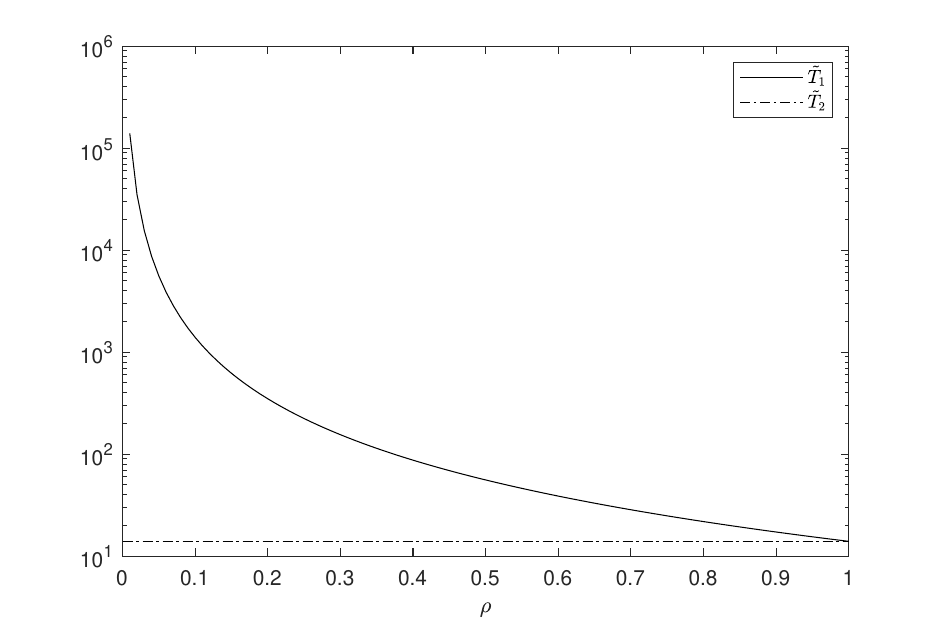}
     \caption{Impact of the correlation on the hitting time of the Bb. \textmd{Comparison of the hitting times of the two Brownian bridges for a fixed level of uncertainty in the view ($\omega^2 = 4$) as a function of the correlation between the Brownian motions.}}
     \label{fig:figure10}
 \end{center}
\end{figure}

Figure \ref{fig:figure10} compares the hitting times of both Brownian bridges for a fixed level of view uncertainty ($\omega^2 = 4$). The hitting time of the Brownian bridge for asset $2$ is $\tilde{T}_2 = T + \omega^2 = 14$, and that of asset $1$ depends on the correlation between the two assets. When they are independent ($\rho = 0$), the hitting time $\tilde{T}_1$ is infinite because the view on asset $2$ provides no information about asset $1$. In this case, the Bb for asset 1 remains a Brownian motion. As the correlation between the two processes increases, the difference between the hitting times $\tilde{T}_1$ and $\tilde{T}_2$ diminishes because the view of asset $2$ provides information about asset $1$. When the two assets are perfectly correlated ($\rho = 1$) the  hitting times are equal.
  \begin{figure}[ht]
 \begin{center}
       \includegraphics[width= 0.5\textwidth]{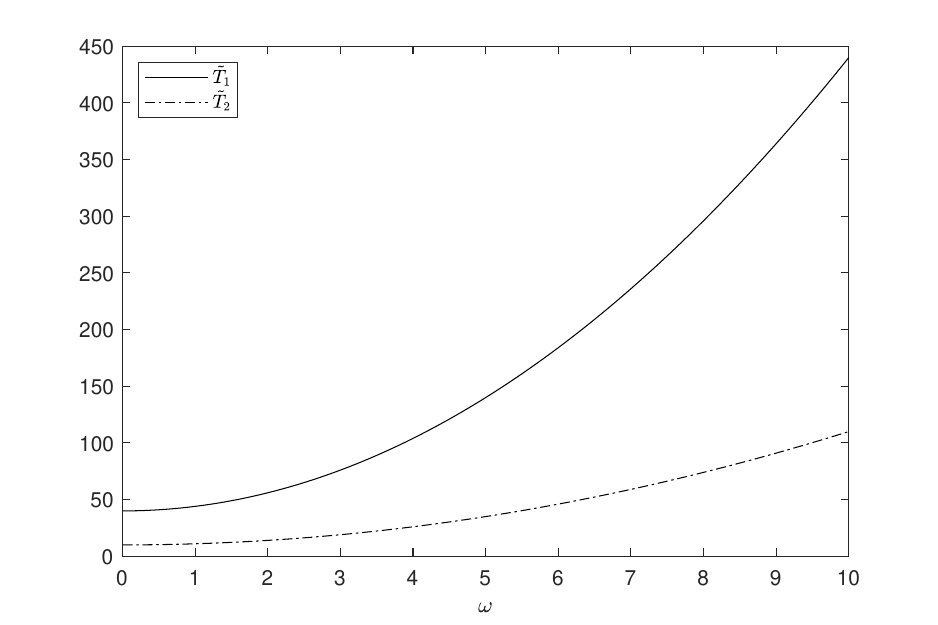}
     \caption{Impact of the noise on the hitting time of the Bb. \textmd{Comparison of the hitting times of the two Brownian bridges for a fixed level of correlation between the Brownian motions ($\rho = 0.5$) as a function of the noise in the view.
     }}
       \label{fig:figure11}
 \end{center}
 \end{figure}
 
 In figure \ref{fig:figure11}, we fix the correlation coefficient $\rho$ to be $0.5$, and compare the hitting times of the Brownian bridges for varying degrees of uncertainty in the view ($\omega^2 \in [0,10^2]$). When the view is certain ($\omega^2 = 0$),  $\tilde{T}_2 = T$ as we know for sure the terminal value of asset $2$. $\tilde{T}_1$ is always greater than $T$  due to the imperfect correlation between the two assets. 
 
 As we increase the view uncertainty, $\tilde{T}_1$ increases at a faster rate than $\tilde{T}_2$; The information about asset $1$ disappears faster than the information about asset $2$ due to the correlation between the two assets being less than $1$.

\subsection{Proof of Proposition \ref{proposition3}}
Let $W(t)$ be an $N-$dimensional Brownian motion starting at $a \in \mathbb{R}^N$ with
\begin{equation*}
    W(t) \sim \mathcal{N}\big(a, t \Sigma).
\end{equation*}
At $t = 0$, we have $Y(0, T) =y$ where
\begin{equation*}
    Y(0,T) = P W(T) + \epsilon,
\end{equation*}
$P \in \mathbb{R}^{K \times N}$ is a linear mapping such that $P L_j \neq 0$ for $j \in [N]$, and $\epsilon$ is normally distributed with $\epsilon \sim \mathcal{N}\big(0, T \Omega \big)$. We define the conditional process $B(t) \in \mathbb{R}^N$ as the Brownian motion $W(t)$ conditioned on the forward-looking views $Y(0,T) = y$
\begin{equation*}
    B(t) = W(t) \,|\, (Y(0,T) = y), \, \, \, \text{for}\, \, \, t \in [0,T].
\end{equation*}
Since the vector $(W(t), Y(0,T))$ is Gaussian, the conditional process $B(t)$ is also Gaussian, and with probability $1$, we have
\begin{equation*}
    B(0) = W(0) \,|\, (Y(0,T) = y) = a.
\end{equation*}
Additionally, the conditional expectation is
\begin{equation*}
\begin{split}
        \mathbb{E}[B(t)] &= \mathbb{E}[W(t) \,|\, Y(0,T) = y]\\
        &= \mathbb{E}[W(t)] + \Cov(W(t), Y(0,T)) \mathbb{V}^{-1}[Y(0,T)] \big(y - \mathbb{E}[Y(0,T)]\big)\\
        &= a + t \Sigma P^\top (T P\Sigma P^\top + T\Omega)^{-1}(y - Pa)\\
        &= a + \frac{t}{T} \Sigma P^\top (P \Sigma P^\top + \Omega)^{-1}(y - Pa).
\end{split}
\end{equation*}
For $s,t \in \mathbb{R}$ with $s \leq t$, the covariance between $B(t)$ and $B(s)$ is
\begin{equation*}
    \begin{split}
        \Cov(B(t), B(s)) &= \Cov(W(t), W(s) \,|\, Y(0,T) = y)\\
                &= \Cov(W(t), W(s)) - \Cov(W(t), Y(0,T)) \mathbb{V}\bigl[Y(0,T) \bigr]^{-1} \Cov(Y(0,T), W(s))\\
                &= s \, \Sigma - (t  \Sigma P^\top ) \, \left( T\, P \Sigma P^\top + T \, \Omega\right)^{-1}  (s  P \Sigma)\\
                &= s  \Sigma - \frac{st}{T} \Sigma P^\top(P\Sigma P^\top + \Omega)^{-1} P\Sigma.
    \end{split}
\end{equation*}
We define
\begin{equation*}
    H := \frac{1}{T}(PL)^\top (P\Sigma P^\top + \Omega)^{-1} PL \in \mathbb{R}^{N \times N},
\end{equation*}
and recall that \(\Sigma = L L^\top\), thus
\begin{equation*}
\begin{split}
        \Cov(B(t), B(s)) &= s  \Sigma - \frac{st}{T} \Sigma P^\top(P\Sigma P^\top + \Omega)^{-1} P\Sigma\\
        &=L \big(s  I_N - \frac{st}{T}(PL)^\top(P\Sigma P^\top + \Omega)^{-1} PL)L^\top\\
        &= L \big( s I_N - st H\big)L^\top.
 \end{split}
\end{equation*}
As the covariance matrices $\Sigma$ and $\Omega$ are positive definite, and the linear mapping matrix $P$ satisfies the condition $PL_i \neq 0$, for $i \in [N]$, it is easy to see that for a non-zero vector $z \in \mathbb{R}^N$, we have
\begin{equation*}
        z^\top H z = \frac{1}{T} (PLz)^\top (P\Sigma P^\top + \Omega)^{-1} PLz \geq 0,
\end{equation*}
therefore, $H$ is positive semi-definite\footnote{It is not positive definite as there can exist a vector $z \in \mathbb{R}^N$ such that $z \neq 0$ and $PLz = 0$.}. Finally, as the Brownian motion $t \to W_i(t)$ is continuous for every $i \in [N]$, and the view $y$ is given at time $0$ and expires at time $T$ (the view causes no jumps in the interval $(0,T)$), the process $t \to B_i(t) = (W_i(t) \,|\, Y(0,T) = y)$ is also continuous for $t \in [0,T]$ and  $i \in [N]$. This concludes the proof. \hfill $\square$

\subsection{Proof of Theorem \ref{theorem1}}
\label{App:thm1}
We define the stochastic process $\{\bar{B}(t), t \in [0,T]\}$ as
\begin{equation*}
    \bar{B}(t) = L^{-1}(B(t) - \mathbb{E}[B(t)]),
\end{equation*}
where $\{B(t), t \in [0,T]\}$ satisfies the properties in Theorem \ref{proposition5}. It is easy to see that
\begin{enumerate}
     \item $\bar{B}(0) = 0$ (with probability $1$), 
    \item $\bar{B}$ is a Gaussian process,
    \item $\mathbb{E}[\bar{B}(t)] = 0$  , for $t \in [0,T]$,
    \item $cov\big(\bar{B}_i(t), \bar{B}_j(s) \big) =  \begin{cases}
        \min\{s,t \} - \dfrac{st}{\tilde{T}_i}, \, \, \, \text{if} \, \, \, i = j,\\
        - \dfrac{st}{H_{i,j}}, \, \, \, \text{if} \, \, \, i \neq j,\\
    \end{cases}$
    \item With probability $1$, $t \to \bar{B}_i(t)$ is continuous in $[0,T]$ for $i \in [N]$.
\end{enumerate}
It follows  that each $\bar{B}_i(t)$, where $i \in [N]$, satisfies Definition \ref{definition1} and is therefore a Brownian bridge from $0$ to $0$ with hitting time $\tilde{T}_{i}$ restricted to the interval $[0,T]$. Thus
\begin{equation*}
    d\bar{B}_i(t) = \frac{y - \bar{B}_i(t)}{\tilde{T}_i - t} dt + dW_i^y(t), \, \, \, \text{for} \, \, i \in [N].
\end{equation*}
However, these Brownian bridges are correlated with 
\begin{equation*}
    \Cov(\bar{B}_i(t), \bar{B}_j(t)) = -\frac{t^2}{\tilde{T}_{ij}}\, \, \, \, \text{for} \, \, i \neq j, \, \, \, t \in [0,T]
\end{equation*}
so the SDE  of the multidimensional $\bar{B}(t)$ is not obtained  from ``stacking" the marginals.

We now derive the SDE representation of $\bar{B}(t)$. As in the proof of Proposition \ref{proposition1}, we first derive the drift and volatility of $d\bar{B}(t)$, then show that it admits an SDE representation. Let $\{V_j(t), j \in [N]\}$ be a vector of $N$-independent Brownian motions with 
\begin{equation*}
    V(t) \sim \mathcal{N}\big(0, t I_N\big),
\end{equation*}
and consider the forward-looking views
\begin{equation*}
    \bar{Y}(0,T) = (PL)V(T) + \epsilon,
\end{equation*}
where $\epsilon \sim \mathcal{N}\big(0,T \Omega\big)$. It can be easily proven that the process $\bar{B}(t)$ has the same distribution as the Brownian motion $V(t)$ conditioned on the views $\bar{Y}(0,T) = 0$
\begin{equation*}
    \bar{B}(t) \stackrel{d}{=} V(t) \,|\, (\bar{Y}(0,T) = 0).
\end{equation*}
Conditional on  $\bar{B}(t)$ at time $t$
\begin{equation*}
\begin{split}
        \mathbb{E}[\bar{B}(t+dt) \,|\, \bar{B}(t)] &= \mathbb{E}[V(t+dt) \,|\, V(t), Y(0,T) = 0]\\
        &= \begin{pmatrix}
            (t + dt)  I_N & t  (PL)^\top
        \end{pmatrix} 
        \begin{pmatrix}
            t  I_N & & t  (PL)^\top\\
            (t + dt)  PL & & T (  (PL)(PL)^\top + \Omega)
        \end{pmatrix}^{-1} \begin{pmatrix}
            V(t) \\ 0
        \end{pmatrix}.
\end{split}
\end{equation*}
Further calculations show
\begin{equation*}
     \mathbb{E}[\bar{B}(t+dt) \,|\, \bar{B}(t)] = \bar{B}(t) - \frac{dt}{T} (PL)^\top \big( (1 - \frac{t}{T}) P \Sigma P^\top + \Omega)^{-1}PL \bar{B}(t).
\end{equation*}
Defining
\begin{equation*}
\begin{split}
        \bar{\beta}_2(t) = \frac{1}{T} (PL)^\top \big( (1 - \frac{t}{T}) P \Sigma P^\top + \Omega)^{-1}PL
\end{split}
\end{equation*}
it follows that
\begin{equation*}
    \mathbb{E}[\bar{B}(t + dt) - \bar{B}(t) \,|\, \bar{B}(t)] = - (dt) \bar{\beta}_2(t) \bar{B}(t),
\end{equation*}
by continuity of \(\bar{B}(t)\) we get by letting \(dt \to 0\)
\begin{equation}
\label{B3}
    \mathbb{E}[d\bar{B}(t)\,|\, \bar{B}(t)] = - (dt) \bar{\beta}_2(t) \bar{B}(t).
\end{equation}
A similar argument shows that the covariance of $\bar{B}(t)$ is
\begin{equation*}
\begin{split}
        \mathbb{V}[\bar{B}(t+dt) - \bar{B}(t) \,|\, \bar{B}(t)] &= \mathbb{V}[V(t+dt) \,|\, V(t), Y(0,T) = 0]\\
        &= \begin{pmatrix}
            t  I_N & t  (PL)^\top
        \end{pmatrix} 
        \begin{pmatrix}
            t  I_N & t  (PL)^\top\\
            (t + dt)  PL & T (  (PL)(PL)^\top + \Omega)
        \end{pmatrix}^{-1} \begin{pmatrix}
            t  I_N \\ t  PL
        \end{pmatrix}\\
        &= (dt) \cdot I_N + o(dt),
\end{split}
\end{equation*}
thus, by letting \(dt \to 0\), we get
\begin{equation}
    \label{B4}
\mathbb{V}[d\bar{B}(t) \,|\, \bar{B}(t)] = (dt) I_N + o(dt).
\end{equation}
We show that $\bar{B}(t)$ is a solution to a Stochastic Differential Equation where the drift \eqref{B3} and volatility \eqref{B4} using Levy's characterization of a Brownian motion as in Proposition \ref{proposition1}. Recall that
\begin{equation*}
   \bar{B}(t) = L^{-1}(B(t) - \mathbb{E}[B(t)]).
\end{equation*}
Additionally, we can write 
\begin{equation*}
\begin{split}
    X^y(t) &= \mu^x t + W(t) \,|\, (Y(0,T) = y)\\
    &= \mu^x t + B(t)
\end{split} 
\end{equation*}
so
\begin{equation*}
    \bar{B}(t) = L^{-1}(X^y(t) - \mathbb{E}[X^y(t)]).
\end{equation*}
 From \eqref{eq11}, \eqref{B3} and \eqref{B4}, $\bar{B}(t)$ satisfies
\begin{equation*}
    d\bar{B}(t) = - dt \cdot \bar{\beta}_2(t) \bar{B}(t) dt + dV^y(t)
\end{equation*}
where $V^y(t) = L^{-1}W^y(t)$ is a vector of $N$ independent Brownian motions. This concludes the proof. \hfill 

\hfill $\square$

\subsection{Proof of Proposition \ref{proposition4}}
\label{App:PropHittingTimes}
    Consider the conditional process $\{\bar{B}(t), t \in [0,T]\}$ satisfying \eqref{eq17}, we showed in the proof of Theorem \ref{theorem1} that each element $\bar{B}_i(t)$, $i \in [N]$, is the restriction to $[0, T]$ of a a Brownian bridge from $0$ to $0$ with hitting time 
    \begin{equation*}
        \tilde{T}_i = T \big((PL_i)^\top (P \Sigma P^\top + \Omega)^{-1} PL_i \big)^{-1} > 0,
    \end{equation*}
     Since $\Sigma$ and $\Omega$ are positive definite, so too is $(P \Sigma P^\top + \Omega)^{-1}$. Furthermore, as we assume that $PL_i \neq 0$, for $i \in [N]$, it follows that $\tilde{T}_i < \infty$. Now we show that the hitting times $\tilde{T}_i$ are strictly larger than the views horizon $T$. 
     
     For $i\in [N]$, we have
     \begin{equation*}
         \tilde{T}_i = \dfrac{1}{H_{ii}}
     \end{equation*}
     where
    \begin{equation*}
        \begin{split}
             H &= \frac{1}{T} (PL)^\top (P \Sigma P^\top + \Omega)^{-1} PL\\
             &\stackrel{(a)}{=} \frac{1}{T} L^\top P^\top\big( \Omega^{-1} - \Omega^{-1} P (\Sigma^{-1} + P^\top \Omega^{-1} P)^{-1}   P^\top \Omega^{-1}  \big) P L\\
             &= \frac{1}{T} L^\top \big( P^\top \Omega^{-1}P (I_N -  (\Sigma^{-1} + P^\top \Omega^{-1} P)^{-1}P^\top \Omega^{-1}P) \big) L\\
             &= \frac{1}{T} L^\top \big( I_N - \Sigma^{-1} (\Sigma^{-1} + P^\top \Omega^{-1} P)^{-1}  \big) L\\
             &= \frac{1}{T} \big(I_N - L^{-1} (\Sigma^{-1} + P^\top \Omega^{-1} P)^{-1} (L^{-1})^\top \big)
        \end{split}
    \end{equation*}
    with $(a)$ coming from the Woodbury matrix identity. Thus, the hitting time $\tilde{T}_i$ can be written as
\begin{equation*}
    \frac{1}{\tilde{T}_{i}} = \frac{1}{T} \big( 1 - \ell^{-1}_i (\Sigma^{-1} + P^\top \Omega^{-1} P)^{-1} (\ell^{-1}_i)^\top\big),
\end{equation*}
where $\ell^{-1}_i$ is the $i^{th}$ row of $L^{-1}$, the inverse of the Cholesky decomposition matrix. Since $(\Sigma^{-1} + P^\top \Omega^{-1} P)^{-1}$ is positive definite and $\ell_i^{-1} \neq 0$ (because $L^{-1}$ is invertible)
\begin{equation*}
    \ell^{-1}_i (\Sigma^{-1} + P^\top \Omega^{-1} P)^{-1} (\ell^{-1}_i)^\top > 0.
\end{equation*}
Therefore
\begin{equation}
\label{eqB6}
    \frac{1}{\tilde{T}_{i}} = \frac{1}{T} \Big( 1 - \ell^{-1}_i (\Sigma^{-1} + P^\top \Omega^{-1} P)^{-1} (\ell^{-1}_i)^\top\Big) < \frac{1}{T}.
\end{equation}
Since $\tilde{T_i} > 0$ it follows that 
\begin{equation*}
    \tilde{T}_{i} > T, \, \, \, \, \text{for} \, \, i \in [N].
\end{equation*}
We now show that the hitting times are increasing in the covariance matrix $\Omega$. Consider two positive definite matrices $\Omega^1$ and $\Omega^2$ such that $\Omega^1 \succeq \Omega^2$ (the matrix $\Omega^1 - \Omega^2$ is positive semi-definite). Let $\tilde{T}_i^1$ and $\tilde{T}_i^2$ be their respective hitting times. We first have
\begin{equation*}
   (\Omega^{2})^{-1}  \succeq (\Omega^{1})^{-1}.
\end{equation*}
Since $\Sigma$ is positive definite
\begin{equation*}
    (\Sigma^{-1} + P^\top (\Omega^1)^{-1} P)^{-1} \succeq  (\Sigma^{-1} + P^\top (\Omega^2)^{-1} P)^{-1}
\end{equation*}
and hence
\begin{equation*}
    \ell^{-1}_i (\Sigma^{-1} + P^\top (\Omega^1)^{-1} P)^{-1} (\ell^{-1}_i)^\top \geq \ell^{-1}_i (\Sigma^{-1} + P^\top (\Omega^2)^{-1} P)^{-1} (\ell^{-1}_i)^\top, \, \, \, \text{for} \, \, i \in [N].
\end{equation*}
It follows from \eqref{eqB6} that 
\begin{equation*}
    \tilde{T}_i^1 \geq \tilde{T}_i^2, \, \, \, \text{for} \, \, i \in [N].
\end{equation*}
If $\Omega^1 \succ \Omega^2$ (the matrix $\Omega^1 - \Omega^2$ is strictly positive definite)
\begin{equation*}
    \tilde{T}_i^1 > \tilde{T}_i^2, \, \, \, \text{for} \, \, i \in [N].
\end{equation*}

This concludes the proof. \hfill $\square$

\subsection{Proof of the Results in Section \ref{Application_BL}} Consider the log-returns process satisfying \eqref{eq8}
\begin{equation*}
    X(t) = t \mu^x + W(t),
\end{equation*}
where $W(t) \sim \mathcal{N}\big(0,t\Sigma\big)$ a $N-$dimensional Brownian motion. Let $y$ be the expert views vector sampled from \eqref{eq10}. Conditional on $Y(0,T) = y$, we have
\begin{equation*}
    \begin{split}
        X^y(t) &= X(t) \,|\, (Y(0,T) = y)\\
        &= t\mu^x + W(t)\,|\, (Y(0,T) = y)\\
        &= t \mu^x + B(t),
   \end{split}
\end{equation*}
where $B(t)$ satisfies \eqref{eq16}. From Theorem \ref{theorem1}, we can write
\begin{equation*}
    B(t) = \mathbb{E}[B(t)] + L \bar{B}(t),
\end{equation*}
where $\bar{B}(t)$ is a zero mean stochastic process satisfying the SDE \eqref{eq17}. Furthermore, we have
\begin{equation*}
    \begin{split}
        dX^y(t) &= (dt)\mu^x + dB(t)\\
        &\stackrel{(a)}{=} (dt) \mu^x + (dt) \beta_1 (y - T P \mu^x) + L d\bar{B}(t)\\
        &= (dt) \big(\mu^x + \beta_1 (y - T P \mu^x) - L\bar{\beta}_2(t) \bar{B}(t)\big) + LdV^y(t)\\
        &\stackrel{(b)}{=} (dt) \big(\mu^x +  \beta_1 (y - T P \mu^x) - L\bar{\beta}_2(t) L^{-1} (B(t) - \mathbb{E}[B(t)])\big) + LdV^y(t),
    \end{split}
\end{equation*}
where $(a)$ follows from \eqref{eq17} and $(b)$ from \eqref{eq16}. Since
\begin{equation*}
    X^y(t) - \mathbb{E}[X^y(t)] = B(t) - \mathbb{E}[B(t)],
\end{equation*}
and 
\begin{equation*}
    L \bar{\beta}_2(t) L^{-1} = \beta_2(t),
\end{equation*}
the conditional log-returns is a solution to the following SDE
\begin{equation*}
    dX^y(t) = (dt) \big(\mu^x +  \beta_1 (y - T P \mu^x) - \beta_2(t)(X^y(t) - \mathbb{E}[X^y(t)]\big) + LdV^y(t).
\end{equation*}
Since \(V^y(t)\) is defined by $L V^y(t) =W^y(t)$ (see Equation \eqref{eq9}), it follows that
\begin{equation*}
    dX^y(t) = (dt) \big(\mu^x +  \beta_1 (y - T P \mu^x) - \beta_2(t)(X^y(t) - \mathbb{E}[X^y(t)]\big) + dW^y(t).
\end{equation*}
\hfill $\square$

\newpage

\section{Section \ref{sec5}}

\medskip

\subsection{Proof of Value function and Proposition \ref{prop:VF1}} \label{app:policy}

We show that the value function  \eqref{eq:V-base} with
 \(A(t) \in \mathbb{R}^{N \times N}\), \(b(t) \in \mathbb{R}^N\), and \(c(t)\) being solutions of \eqref{eq19}--\eqref{eq20b}
 is the solution of the HJB equation \eqref{eq:HJB-base}.

The optimal policy is the maximizer in the HJB \eqref{eq:HJB-base}:
\begin{equation*}
        \pi^*(t) = - \frac{\nabla_z V}{ z \nabla_z^2 V} \Sigma^{-1}(\tilde{\mu}(t,x) - r_f \mathbf{1}_N) - \frac{ \nabla^2_{x,z}V}{ z \nabla_z^2 V}
\end{equation*}
where
\begin{equation*}
\begin{split}
            \tilde{\mu}(t, x) &=  \mu  + \beta_1 \big(y - T P \mu^x\big) - \beta_2(t) \big(x - \mathbb{E}[X(t) | y]\big))\\
        &= \alpha_t + \Sigma \eta_t x
\end{split}
\end{equation*}
with \(\eta_t : [0,T] \to \mathbb{R}^{N\times N}\) is
\begin{equation}
 \label{eq:ECeta}
 \begin{split}
                  \eta_t &= -\Sigma^{-1}\beta_2(t)\\
                  &= -P^\top ((T - t) P \Sigma P^\top + T \Omega)^{-1}P
 \end{split}
    \end{equation}
and \(\alpha_t : [0,T] \to \mathbb{R}^{N}\) is
    \begin{equation}
    \label{eq:ECalpha}
            \alpha_t = \mu + \beta_1 (y - TP \mu^x) - \Sigma \eta_t \mathbb{E}[X^y(t)].
    \end{equation}
 Observe that $\eta_t$   is a symmetric matrix. From \eqref{eq:V-base}, we have
\begin{equation*}
    \begin{cases}
        \dfrac{\partial V}{\partial t} &=  (\dfrac{1}{2}x^\top A'(t) x + x^\top b'(t) + c'(t)) V,\\
        \nabla_z V &= \dfrac{1 - \gamma}{z}V,\\
        \nabla_x V &= (A(t) x + b(t)) V, \\
        \nabla_z^2V &= \dfrac{-\gamma(1 - \gamma)}{z^2}V,\\
        \nabla_x^2 V &= (A(t) + (A(t)x + b(t))(A(t)x+b(t))^\top)V,\\
        \nabla_{x,z}^2 V &=  \dfrac{1-\gamma}{z}(A(t)x + b(t))V.\\
    \end{cases}
\end{equation*}
The HJB equation \eqref{eq:HJB-base} becomes
\begin{eqnarray*}
 0 & = &   \frac{1}{2} x^\top\left\{A'(t) + \frac{1 - \gamma}{\gamma} \eta_t \Sigma \eta_t + \frac{1}{\gamma} (A(t) \Sigma \eta_t + \eta_t \Sigma A(t))  + \frac{1}{\gamma} A(t) \Sigma A(t)\right\}x \\
 & & +    x^\top\left\{b'(t) + \frac{1}{\gamma}\big(\eta_t + A(t)\big) \Sigma b(t) + \frac{1 - \gamma}{\gamma}\big( \eta_t +A(t) \big)(\alpha_t - r_f \mathbf{1}_N) + A(t) \big(\alpha_t - \frac{\diag(\Sigma)}{2} \big)\right\} \\ 
    & & + c'(t) + (1 - \gamma) r_f + \frac{1}{2} \Tr\big(A(t) \Sigma \big) + \frac{1 - \gamma}{2 \gamma}\big(\alpha_t - r_f \mathbf{1}_N \big)^\top \Sigma^{-1} \big(\alpha_t - r_f \mathbf{1}_N\big) + \big(\alpha_t - \frac{\diag(\Sigma)}{2}\big)^\top b(t) \\
         & & + \frac{1 - \gamma}{\gamma} \big(\alpha_t - r_f \mathbf{1}_N\big)^\top b(t) + \frac{1}{2\gamma} b^\top(t) \Sigma b(t)
\end{eqnarray*}
with terminal conditions $A(T)=0$, $B(T)=0$ and $c(t)=0$. It follows that if $A(t)$, $B(t)$ and $c(t)$ satisfy \eqref{eq19}--\eqref{eq20b}, that \eqref{eq:V-base} is the solution of \eqref{eq:HJB-base}.

To see that $A(t)$ is strictly negative definite when $t<T$, consider the linear quadratic problem
\begin{eqnarray*}
    \tilde{V}(t, x) = \min_u \int_t^T \left\{ x_t^\top\left(-\frac{1-\gamma}{\gamma}\eta_t^\top\Sigma\eta_t\right)x_t + \frac{1}{\gamma} u_t^\top \Sigma u_t \right\}dt
\end{eqnarray*}
where
\begin{eqnarray*}
    dx_t & = & \frac{1}{\gamma}\Sigma \eta_t x_t + \frac{1}{\gamma}{\Sigma}u_t \\
    x_t& = & x.
\end{eqnarray*}
Since $-\frac{1-\gamma}{\gamma} \eta_t^\top\Sigma\eta_t \geq0$ and $\Sigma>0$ when $t<T$, $\tilde{V}(t, x)>0$ unless $x=0$. It can be shown that $\tilde{V}(t, x) = - x^\top A(t) x$ which implies that $A(t)$ is strictly negative definite when $t<T$.

Given \eqref{eq:V-base}, the optimal investment policy is 
\begin{equation*}
        \pi^*(t) = \frac{1}{\gamma} \big(\Sigma^{-1} (\tilde{\mu}(t,x) - r_f \mathbf{1}_N) + A(t)x + b(t) \big).
\end{equation*}  
which completes the proof. \hfill $\square$

\subsection{Proof of Theorem \ref{proposition5}}
\label{App:thm2}


We derive explicit expressions for the ODEs \eqref{eq19}, \eqref{eq20} and the hedging demand \eqref{eq22a}.

\subsubsection*{Explicit solution of Riccati equation \eqref{eq19}.}

The following results will be useful.
\begin{lemma}
\label{lemmaC1}
    Suppose Assumption \ref{ass-covariance} holds and $\eta_t$ is given by \eqref{eq:ECeta}. Then $\eta_t$  is symmetric and negative semi-definite for $t \in [0,T]$ and 
\begin{eqnarray*}
    \frac{d \eta_t}{d t} & = & - \eta_t \Sigma \eta_t \\ 
     \eta_T & = & -P^\top (T\Omega)^{-1}P.
\end{eqnarray*}
If 
\begin{equation}
\label{eq:zeta}
    \zeta_t = - P^\top \Big(\frac{1}{\gamma}(T-t)P\Sigma P^\top + T \Omega\Big)^{-1}P
\end{equation}
then
\begin{equation}
\label{eq:RicattiQ}
\begin{cases}
        \zeta'_t + \frac{1}{\gamma} \zeta_t \Sigma \zeta_t = 0\\ \vspace{-0.25cm}\\
\zeta_T = \eta_T.
\end{cases}
\end{equation}
\end{lemma}

\proof{Proof}

It is clear that $\eta_t$ is symmetric and negative semi-definite. Additionally, for an invertible matrix $R(t) \in \mathbb{R}^{K \times K}$
\begin{equation}
\label{C5}
    \frac{d}{d t} ( R^{-1}(t)) = - R^{-1}(t)\frac{d}{d t} (R(t)) R^{-1}(t).
\end{equation}
If
\begin{equation*}
    R(t) = (T-t) P \Sigma P^\top + T \Omega \in \mathbb{R}^{K \times K},
\end{equation*}
\begin{equation*}
    \begin{split}
        \frac{d}{d t} \eta_t &= - P^\top \frac{d}{d t} ( R^{-1}(t)) P\\
        &= P^\top R^{-1}(t)\frac{d}{d t} (R(t)) R^{-1}(t) P\\
       &=  - P^\top \big((T-t) P \Sigma P^\top + T \Omega \big)^{-1} P \Sigma P^\top \big((T-t) P \Sigma P^\top + T \Omega \big)^{-1}\\
       &= - \eta_t \Sigma \eta_t.
    \end{split}
\end{equation*}
The ODE for $\zeta_t$ can be derived similarly.
\hfill $\square$
\endproof 

Let 
\begin{equation}
\label{eq:solutions_Ricatti}
A(t) = \zeta_t - \eta_t.
\end{equation}
Since $A(T) = 0$ and
\begin{eqnarray*}
    \frac{d}{dt}A(t) & = & - \frac{1}{\gamma}\zeta_t \Sigma \zeta_t + \eta_t\Sigma\eta_t \\
    & = & -\frac{1}{\gamma}\big(A(t) + \eta_t\big)\Sigma\big(A(t) + \eta_t) + \eta_t\Sigma\eta_t \\
    & = & -\frac{1-\gamma}{\gamma}\eta_t\Sigma\eta_t - \frac{1}{\gamma}\Big(A(T) \Sigma\eta_t + \eta_t\Sigma A(t)\Big) - \frac{1}{\eta}A(t) \Sigma A(t)
\end{eqnarray*}
it follows that the solution of \eqref{eq19} is \eqref{eq:solutions_Ricatti}.

Now we show that 
\begin{equation*}
    A(t) = \zeta_t - \eta_t = M(t) \eta_t
\end{equation*}
where 
\begin{equation}
\label{eq:ECM}
    M(t) = (\gamma - 1) (1 - \frac{t}{T}) P^\top \Omega^{-1}P \Big(\gamma\Sigma^{-1} + (1 - \frac{t}{T}) P^\top \Omega^{-1} P\Big)^{-1} \in \mathbb{R}^{N \times N}.
\end{equation}
Observe from \eqref{eq:solutions_Ricatti} that
\begin{equation*}
    A(t) = - P^\top (C(t)^{-1} - F(t)^{-1})P,
\end{equation*}
where 
\begin{equation*}
    C(t) = \frac{T-t}{\gamma} P \Sigma P^\top + T \Omega,
\end{equation*}
and
\begin{equation*}
    F(t) = (T-t) P \Sigma P^\top + T \Omega.
\end{equation*}
Since
\begin{equation*}
\begin{split}
        C(t)^{-1} &= \Big(\frac{1 - \gamma}{\gamma} (T-t) P\Sigma P^\top + F(t)\Big)^{-1}\\
        & \stackrel{(a)}{=} F(t)^{-1} - F(t)^{-1}P \Big(\frac{\gamma}{(1-\gamma)(T-t)}\Sigma^{-1} + P^\top F(t)^{-1}P\Big)^{-1}P^\top F(t)^{-1},
\end{split}
\end{equation*}
where $(a)$ follows from the Woodbury  identity, it follows that
\begin{equation*}
    \begin{split}
        A(t) &= - P^\top (C(t)^{-1} - F(t)^{-1})P\\
        &=  P^\top F(t)^{-1}P\Big(\frac{\gamma}{(1-\gamma)(T-t)}\Sigma^{-1} + P^\top F(t)^{-1}P\Big)^{-1} P^\top F(t)^{-1}P\\
        &= - \eta_t \Big(\frac{\gamma}{(\gamma - 1)(T-t)}  \Sigma^{-1} + \eta_t \Big)^{-1}\eta_t.
    \end{split}
\end{equation*}
The Woodbury identity also implies that
\begin{equation}
\begin{split}
        - \eta_t \Big(\frac{\gamma}{(\gamma - 1)(T-t)}\Sigma^{-1} + \eta_t \Big)^{-1} &= (\gamma - 1) (1 - \frac{t}{T}) P^\top \Omega^{-1}P \Big(\gamma\Sigma^{-1} + (1 - \frac{t}{T}) P^\top \Omega^{-1} P\Big)^{-1}\\
        &= M(t)
\end{split}
\label{eq:M1}
\end{equation}
so
\begin{equation*}
    A(t) = M(t) \eta_t,
\end{equation*}
where $M(t)$ satisfies \eqref{eq:ECM}.

\subsubsection*{Explicit  solution of ODE \eqref{eq20}.}

The following result will be useful.

\begin{lemma}
\label{lemmaC3}
Let
\begin{equation*}
    s(t) = \Big(\frac{\gamma}{(\gamma - 1)(T-t)}\Sigma^{-1} + \eta_t \Big)^{-1} \in \mathbb{R}^{N \times N}.
\end{equation*}
    Then the matrices $\eta_t \Sigma \in \mathbb{R}^{N\times N}$ and $\eta_t s(t) \in \mathbb{R}^{N\times N}$ commute for all $t \in [0, T)$, i.e.,
    \begin{equation*}
        (\eta_t \Sigma)(\eta_t s(t)) = (\eta_t s(t))(\eta_t \Sigma), \, \, \, \text{for} \, \, \, t \in [0,T).
    \end{equation*}
\end{lemma}

\proof{Proof}
We have
\begin{equation*}
    s(t) = \big(\lambda_t \Sigma^{-1} + \eta_t \big)^{-1} \in \mathbb{R}^{N \times N},
\end{equation*}
with 
\begin{equation*}
    \lambda_t = \frac{\gamma}{(\gamma - 1)(T-t)}.
\end{equation*}
We can then write
\begin{equation*}
    \begin{split}
        \eta_t s(t) \eta_t \Sigma &= \eta_t s(t) \big(\eta_t + \lambda_t \Sigma^{-1} - \lambda_t \Sigma^{-1} \big) \Sigma\\
        &= \eta_t (I_N - \lambda_t s(t) \Sigma^{-1})\Sigma\\
        &= \eta_t \Sigma - \lambda_t \eta_t s(t)\\
        &= \eta_t \Sigma (I_N - \lambda_t \Sigma^{-1} s(t))\\
        &= \eta_t \Sigma \big(\eta_t + \lambda_t \Sigma^{-1} - \lambda_t \Sigma^{-1} \big)s(t)\\
        &= \eta_t \Sigma \eta_t s(t)
    \end{split}
\end{equation*}
which completes the proof. \hfill $\square$
\endproof

We now show that 
    \begin{equation*}
    b(t) = M(t) \Sigma^{-1}(\alpha_t - r_f \mathbf{1}_N)
\end{equation*}
is the unique solution of \eqref{eq20}. (Recall that $A(t) = M(t) \eta_t$). 

Since $M(T) = 0$,  $b(T) = 0$,  the terminal condition is satisfied. Differentiating with respect to $t$
\begin{equation*}
    \begin{split}
        b'(t) &= M'(t) \Sigma^{-1} (\alpha_t - r_f \mathbf{1}_N) + M(t) \Sigma^{-1} \frac{\partial \alpha_t}{\partial t}.\\
    \end{split}
\end{equation*}
From \eqref{eq:ECalpha} and Lemma \ref{lemmaC1}
\begin{equation*}
    \frac{d \alpha_t}{d t} = - \eta_t (\alpha_t - \frac{1}{2}\diag(\Sigma)),
\end{equation*}
so
\begin{equation*}
    b'(t) = M'(t) \Sigma^{-1} (\alpha_t - r_f \mathbf{1}_N) - A(t) (\alpha_t - \frac{1}{2}\diag(\Sigma)).
\end{equation*}
Therefore, we show $b(t)$ satisfies \eqref{eq:b} by showing
\begin{equation}
\label{eq:equalitytoprove}
    M'(t) \Sigma^{-1} (\alpha_t - r_f \mathbf{1}_N) = - \frac{1}{\gamma}\big(\eta_t + A(t)\big) \Sigma q(t) - \frac{1 - \gamma}{\gamma}\big( \eta_t +A(t) \big)(\alpha_t - r_f \mathbf{1}_N).
\end{equation}

Define
\begin{equation*}
    s(t) = \big(\frac{\gamma}{(\gamma - 1)(T-t)}\Sigma^{-1} + \eta_t \big)^{-1} \in \mathbb{R}^{N \times N}.
\end{equation*}

Observe $s(t)$ is invertible and $M(t) = - \eta_t s(t)$ \eqref{eq:M1}.
It follows that
\begin{equation}
\label{eq:ECM'}
    \begin{split}
        M'(t) &= -(\eta_t s(t))'\\
        &\stackrel{(a)}{=} \eta_t \Sigma \eta_t - \eta_t s'(t)\\
        &\stackrel{(b)}{=} \eta_t \Sigma \eta_t + \eta_t s(t) (s^{-1}(t))' s(t)\\
        &\stackrel{}{=} \eta_t \Sigma \eta_t - \frac{1-\gamma}{\gamma} \eta_t \big(I_N - s(t) \eta_t \big) \Sigma \big(I_N - \eta_t s(t) \big),
    \end{split} 
\end{equation}
where $(a)$ follow from Lemma \ref{lemmaC1} and $(b)$ from the expression of the derivative of the inverse. Because of Lemma \ref{lemmaC3}
\begin{equation*}
    \begin{split}
        M'(t) 
 &\stackrel{(a)}{=} \frac{1}{\gamma} (\eta_t - \eta_t s(t) \eta_t) \Sigma \eta_t s(t) - \frac{1 - \gamma}{\gamma} (\eta_t - \eta_t s(t) \eta_t) \Sigma\\
 &\stackrel{(b)}{=} -\frac{1}{\gamma}(\eta_t + M(t) \eta_t) \Sigma M(t) - \frac{1 - \gamma}{\gamma}(\eta_t + M(t) \eta_t) \Sigma\\
 &\stackrel{(c)}{=} -\frac{1}{\gamma}(\eta_t + A(t)) \Sigma M(t) - \frac{1 - \gamma}{\gamma}(\eta_t + A(t)) \Sigma,\\
    \end{split}
\end{equation*}
where $(b)$ from the definition of $s(t)$ and $(c)$ from the expression of $A(t)$. It follows that 
\begin{equation*}
\begin{split}
        M'(t) \Sigma^{-1}(\alpha_t - r_f \mathbf{1}_N) &= -\frac{1}{\gamma}(\eta_t + A(t)) \Sigma M(t) \Sigma^{-1}(\alpha_t - r_f \mathbf{1}_N)- \frac{1 - \gamma}{\gamma}(\eta_t + A(t)) (\alpha_t - r_f \mathbf{1}_N)\\
        &= -\frac{1}{\gamma}(\eta_t + A(t)) \Sigma q(t)- \frac{1 - \gamma}{\gamma}(\eta_t + A(t)) (\alpha_t - r_f \mathbf{1}_N).
\end{split}
\end{equation*}
which shows that \eqref{eq:equalitytoprove} holds and hence that $q(t)$ is a solution of \eqref{eq20}. Since \eqref{eq20} is a linear ODE with coefficients that are bounded, this solution is unique.

\subsubsection*{Hedging demand \eqref{eq22a}.}

From \eqref{eq:b} the hedging demand can be written
\begin{equation*}
    \begin{split}
        \frac{1}{\gamma}\frac{\partial g}{\partial x}(t,x) &= \frac{1}{\gamma}( A(t)x + b(t))\\
        &= \frac{1}{\gamma}(M(t) \eta_t x + M(t) \Sigma^{-1} (\alpha_t - r_f))\\
        &= \frac{1}{\gamma}(M(t) (\eta_t x + \Sigma^{-1} (\alpha_t - r_f)))\\
        &= \frac{1}{\gamma}(M(t) \Sigma^{-1} (\alpha_t + \Sigma \eta_t x - r_f))\\
        &\stackrel{(b)}{=} \frac{1}{\gamma}(M(t) \Sigma^{-1} (\tilde{\mu}(t,x) - r_f))\\
        &\stackrel{(c)}{=}M(t) \pi^*_{MV}(t), 
    \end{split}
\end{equation*}
where $(b)$ follows from the definition of $\alpha_t$ and $\eta_t$, and $(c)$ from the definition of the mean-variance term
\begin{equation*}
    \pi^*_{MV}(t) = \frac{1}{\gamma}\Sigma^{-1} (\tilde{\mu}(t,x) - r_f).
\end{equation*}

\subsection{Proof of Corollary \ref{theorem2}}
From \eqref{eq22} in Theorem \ref{proposition5}
\begin{equation*}
    \frac{\partial g}{\partial x} = M(t) \Sigma^{-1} (\tilde{\mu}(t,x) - r_f).
\end{equation*}
where $M(t)$ is given by \eqref{eq:ECM}. 
It follows from \eqref{eq21} that
\begin{equation*}
     \pi^*(t) =  \frac{1}{\gamma}\big(I_N + M(t) \big) \Sigma^{-1}  (\tilde{\mu}(t,x) - r_f).
\end{equation*}
Now we show that 
\begin{equation*}
    \begin{split}
        \Big(\big(I_N + M(t) \big)\Sigma^{-1}\Big)^{-1} 
        & \equiv \big( \Sigma^{-1} + (1 - \frac{t}{T}) P^\top \Omega^{-1} P\big)^{-1} + \frac{1}{\gamma} \big(\Sigma - \big( \Sigma^{-1} + (1 - \frac{t}{T}) P^\top \Omega^{-1} P\big)^{-1}\big) \\ &= \Sigma_{\text{DBL}}.
    \end{split}
\end{equation*}
From \eqref{eq:M1}
\begin{equation*}
\begin{split}
        M(t) &= - \eta_t \big(\frac{\gamma}{(\gamma - 1)(T-t)}\Sigma^{-1} + \eta_t \big)^{-1}\\
        &= - I_N + \frac{\gamma}{(\gamma - 1)(T-t)}\Sigma^{-1}\big(\frac{\gamma}{(\gamma - 1)(T-t)}\Sigma^{-1} + \eta_t \big)^{-1}.
\end{split}
\end{equation*}
Since $M(t)$ is positive semi-definite when $\gamma > 1$ (see \eqref{eq:ECM}),  follows that $I_N + M(t)$ is positive definite invertible. We can then write
\begin{equation*}
    \begin{split}
        \Big(\big(I_N + M(t) \big)\Sigma^{-1}\Big)^{-1} &= \Sigma \big(I_N + M(t) \big)^{-1}\\
        &= \frac{(\gamma-1)(T-t)}{\gamma}\Sigma \left(\frac{\gamma}{(\gamma - 1)(T-t)}\Sigma^{-1} + \eta_t \right) \Sigma\\
        &= \frac{(\gamma-1)(T-t)}{\gamma}\Sigma \left(\frac{\gamma}{(\gamma - 1)(T-t)}\Sigma^{-1} - P^\top ((T-t)P \Sigma P^\top + T \Omega)^{-1} P \right) \Sigma.
    \end{split}
\end{equation*}
Woodbury's identity implies
\begin{equation*}
    P^\top ((T-t)P \Sigma P^\top + T \Omega)^{-1} P = \frac{1}{T - t} \Sigma^{-1} - \frac{1}{T-t}\Sigma^{-1} \big(\Sigma^{-1} + (1-\frac{t}{T}) P^\top \Omega^{-1}P\big)^{-1}\Sigma^{-1}
\end{equation*}
from which it follows that
\begin{equation*}
    \begin{split}
        \Big(\big(I_N + M(t) \big)\Sigma^{-1}\Big)^{-1} &= \frac{(\gamma-1)(T-t)}{\gamma}\Sigma \left(\frac{\gamma}{(\gamma - 1)(T-t)}\Sigma^{-1} - P^\top ((T-t)P \Sigma P^\top + T \Omega)^{-1} P \right) \Sigma\\
        &= \Sigma - \frac{\gamma - 1}{\gamma} \left(\Sigma - \Big(\Sigma^{-1} + (1-\frac{t}{T}) P^\top \Omega^{-1}P\Big)^{-1} \right)\\
        &= \frac{1}{\gamma} \Sigma + (1 - \frac{1}{\gamma}) \left(\Sigma^{-1} + (1-\frac{t}{T}) P^\top \Omega^{-1}P\right)^{-1}\\
        &= \Sigma_{\text{DBL}}.
    \end{split}
\end{equation*}
Therefore, we have
\begin{equation*}
     \pi^*(t) =  \frac{1}{\gamma} \Sigma_{\text{DBL}}^{-1}  (\tilde{\mu}(t,x) - r_f),
\end{equation*}
which completes the proof. \hfill $\square$

\subsection{Single-Period Black-Litterman}
\label{App:SinglePeriod_BL}

From \eqref{eq8}, the log-returns follow
\[
X(t) = t\mu^x + W(t),
\]
where \(W(t)\) is a Brownian motion with covariance matrix \(\Sigma\). Then for \(T\ge t\) we have
\begin{equation}
\label{eq:EC_XT}
X(T) = X(t) + (T-t)\mu^x + \bigl(W(T)-W(t)\bigr).
\end{equation}

The investor’s views are modeled by
\[
Y(0,T) = P\,X(T) + \epsilon,
\]
with \(\epsilon\) being independent Gaussian noise wit \(0\) mean and covariance \(T\Omega\). Inserting \eqref{eq:EC_XT} into the view, we write
\begin{equation}
\label{eq:EC_Y}
Y(0,T) = P\Bigl( X(t) + (T-t)\mu^x + \bigl(W(T)-W(t)\bigr) \Bigr) + \epsilon.
\end{equation}

Since \(X(T)\) and \(Y(0,T)\) are jointly Gaussian, the conditional expectation is given by
\begin{eqnarray}
\label{eq:EC_condExpGeneral}
\lefteqn{
\mathbb{E}\Bigl[ X(T) \,\Big|\, X(t),\, Y(0,T)=y\Bigr] 
= \mathbb{E}\Bigl[ X(T) \,\Big|\, X(t)\Bigr]} \nonumber \\
& &\quad + \Cov\Bigl(X(T),Y(0,T) \,\big|\, X(t)\Bigr)
\mathbb{V}\Bigl[Y(0,T) \,\big|\, X(t)\Bigr]^{-1}\Bigl( y - \mathbb{E}\bigl[Y(0,T) \,\big|\, X(t)\bigr] \Bigr).
\end{eqnarray}

\subsubsection*{Derivation of the Conditional Mean}

From \eqref{eq:EC_XT} we have
\[
\mathbb{E}\Bigl[ X(T) \,\Big|\, X(t)\Bigr] = X(t) + (T-t)\mu^x.
\]
It follows from \eqref{eq:EC_Y}  that
\begin{equation*}
\begin{split}
\mathbb{E}\Bigl[ Y(0,T) \,\Big|\, X(t)\Bigr] &= P\Bigl(X(t)+(T-t)\mu^x\Bigr),\\
\mathbb{V}\Bigl[ Y(0,T) \,\Big|\, X(t)\Bigr] &= (T-t)P\Sigma P^\top + T\Omega,
\end{split}
\end{equation*}
and
\[
\Cov\Bigl(X(T),Y(0,T) \,\Big|\, X(t)\Bigr) 
= \operatorname{Cov}\Bigl( W(T)-W(t), \, P\bigl(W(T)-W(t)\bigr)\Bigr)
= (T-t)\Sigma P^\top.
\]

Substituting in \eqref{eq:EC_condExpGeneral} it can be shown that 
\begin{eqnarray}
\label{eq:EC_condMeanIncrement}
\mathbb{E}\Bigl[ X(T)-X(t) \,\Big|\, X(t),\, Y(0,T)=y\Bigr] = (T-t)\,\tilde{\mu}^x(t,X(t))
\end{eqnarray}
where
\begin{eqnarray*}
    \tilde{\mu}^x(t,x)&= &\mu^x + \beta_1 (y - TP \mu^x) + \beta_2(t) (\mathbb{E}[X^y(t)] - x) \\ [5pt]
    \beta_1 & = & \frac{1}{T}\Sigma P^\top (P \Sigma P^\top + \Omega)^{-1},\\ [5pt]
            \beta_2(t) & = & \Sigma P^\top \left((T-t) P \Sigma P^\top + T\Omega\right)^{-1} P
\end{eqnarray*}


\subsubsection*{Derivation of the Conditional Covariance}

For jointly Gaussian variables the conditional covariance is given by
\begin{eqnarray*}
\lefteqn{\mathbb{V}\Bigl[X(T) \,\Big|\, X(t),\, Y(0,T)=y\Bigr] 
= \mathbb{V}\Bigl[X(T) \,\Big|\, X(t)\Bigr]} \\ \quad
& &- \Cov\Bigl(X(T),Y(0,T)\,\Big|\,X(t)\Bigr)
\mathbb{V}\Bigl[Y(0,T)\,\Big|\,X(t)\Bigr]^{-1}
\Cov\Bigl(Y(0,T),X(T)\,\Big|\,X(t)\Bigr).
\end{eqnarray*}
Since
\[
\operatorname{Var}\Bigl[X(T) \,\Big|\, X(t)\Bigr] = (T-t)\Sigma,
\]
and using the expressions
\[
\operatorname{Cov}\Bigl(X(T),Y(0,T)\,\Big|\,X(t)\Bigr) = (T-t)\Sigma P^\top,\quad
\mathbb{V}\Bigl[Y(0,T)\,\Big|\,X(t)\Bigr] = (T-t)P\Sigma P^\top + T\Omega,
\]
we deduce that
\begin{equation}
\label{eq:EC_condVarDetailed}
\mathbb{V}\Bigl[X(T)-X(t) \,\Big|\, X(t),\, Y(0,T)=y\Bigr] 
= (T-t)\Sigma - (T-t)^2\Sigma P^\top \Bigl((T-t)P\Sigma P^\top + T\Omega\Bigr)^{-1}P\Sigma.
\end{equation}

By applying Woodbury’s matrix identity to the right-hand side of \eqref{eq:EC_condVarDetailed}, we write
\[
\mathbb{V}\Bigl[X(T)-X(t) \,\Big|\, X(t),\, Y(0,T)=y\Bigr]
= (T-t)\left(\Sigma^{-1} + \left(1-\frac{t}{T}\right)P^\top\Omega^{-1}P\right)^{-1}.
\]
Thus, we define
\[
\Sigma_{BL|t} = \left(\Sigma^{-1} + \left(1-\frac{t}{T}\right)P^\top\Omega^{-1}P\right)^{-1}.
\]

\subsubsection*{Optimal Portfolio Policy}

In the single-period setting, an investor with risk-aversion parameter \(\gamma>0\) maximizes the objective
\begin{equation*}
\begin{split}
\max_{\pi} \quad & \pi^\top\,\mathbb{E}\Bigl[X(T)-X(t)\,\Big|\, X(t),\, Y(0,T)=y\Bigr] + \Bigl(1-\pi^\top \mathbf{1}_N\Bigr)r_f \\
& \quad - \frac{\gamma}{2}\,\pi^\top\,\mathbb{V}\Bigl[X(T)-X(t)\,\Big|\, X(t),\, Y(0,T)=y\Bigr]\pi,
\end{split}
\end{equation*}
where \(r_f\) is the risk-free rate. Using \eqref{eq:EC_condMeanIncrement} and \eqref{eq:EC_condVarDetailed}, the optimal portfolio is
\[
\pi^*_{BL|t} = \frac{1}{\gamma}\,(\Sigma_{BL|t})^{-1}\Bigl(\tilde{\mu}^x(t,x)-r_f\mathbf{1}_N\Bigr).
\] 
\hfill $\square$

\newpage

\section{Section \ref{sec:Extensions}}
\label{Appendix:DetailedAnalysis}

\medskip



\subsection{Section \ref{sec:revisions}}
\label{Sec:EC_ViewsRevision}





\subsubsection{Proof of Proposition \ref{cor:UpdatedViews}}
Recall that the state variable \(I(t) = Y^j(t_j,T)\) is the most recent expert view and 
\begin{equation*}
\bar{X}(t) = X(t) - X(t_j)
\end{equation*}
is the log-returns over the horizon \([t_j, t_{j+1})\). The following result shows that \(\bar{X}(t)\) given the most recent view \(I(t)\)  is independent of earlier views. 
\begin{lemma}
\label{Lemma:ObsoleteViews}
For \( t \in [t_j, t_{j+1})\), given the most recent view \(Y^j(t_j,T)\), the process $\bar{X}(t)$ is independent of all previous views \(Y^k(t_k,T)\), \(k \in \{0 ,\dots, j-1\}\).
\end{lemma}

The proof is given in Appendix \ref{proof_lemma:ObsoleteViews}.

From Lemma \ref{Lemma:ObsoleteViews}, it follows that
\begin{equation*}
     \bar{X}^y(t) \triangleq \bar{X}(t) \,|\, \mathcal{F}_t^Y = \bar{X}(t) \,|\, I(t).
\end{equation*}
At time \(t = t_j\), the investor receives the view \(Y^j(t_j,T) = y^j\) with covariance matrix \(\Omega^j\). It follows from Proposition \ref{proposition1} that the dynamics of \( \bar{X}^y(t) \) on $[t_j, t_{j+1})$ are given by
\begin{equation}
\label{eq:ECBarX_updated}
    d\bar{X}^y(t) =   \bigg(\mu^x + \beta_1^j \big(y^j - (T-t_j) P \mu^x \big) + \beta_2^j(t) \big(\mathbb{E}[\bar{X}^y(t)] - \bar{X}^y(t)\big)\bigg) dt + dW^y(t)
\end{equation}
where
\begin{eqnarray*}
\beta_1^j & = & \Sigma P^\top \big((T-t_j) P \Sigma P^\top+\Omega^j\big)^{-1} \in  \mathbb{R}^{N \times K},\\ [8pt]
\beta_2^j(t) & = & \Sigma P^\top \big((T-t) P \Sigma P^\top + \Omega^j\big)^{-1} P  \in \mathbb{R}^{N \times N},
\end{eqnarray*}
and
\begin{equation*}
\mathbb{E}[\bar{X}^y(t)]  = (t - t_j) \big(\mu^x +  \beta_1^j \big(y^j -(T-t_j) P \mu^x \big)\big).
\end{equation*}
This completes the proof. \hfill $\square$

\subsubsection{Proof of Theorem \ref{thm:updatedViews}}
\label{EC_Views_Revision_Control}


We show the value function is of the form 
\begin{equation*}
    V(t,z,\bar{x},y) = U(z) \exp\left(g^j(t,\bar{x},y)\right), \quad t \in [t_j, t_{j+1})
\end{equation*}
where  \(g^j : \mathbb{R} \times \mathbb{R}^N \times \mathbb{R}^K \to \mathbb{R}\) is quadratic in the log-returns vector \(x\) and  views  \(y\)
\begin{equation}
\label{eq:EC_g_function_Update}
    g^j(t,\bar{x},y) = \frac{1}{2} \bar{x}^\top A^j(t) \bar{x} + \bar{x}^\top \left(B^j(t) y + \bar{b}^j(t)\right) + \left( - \frac{1}{2} y^\top \bar{C}^j(t) y + y^\top \hat{c}^j(t) + \bar{c}^j(t) \right).
\end{equation}
We derive ODEs for the coefficients of this function using dynamic programming; specifically, the value function solves the HJB equation in each interval $(t_j, t_{j+1})$ with the terminal condition $V(t_{j+1}^-, \bar{x}, y)$ determined from the value function for the next interval $[t_{j+1}, t_{j+2}]$ at time $t_{j+1}$  \eqref{eq:continuation}. The terminal condition for the last interval $[t_M, T]$ is $V(T, \bar{x}, y)=0$.

We begin with the last interval $[t_{M}, T]$ ($j=M$). Since $V(T, \bar{x}, y)=0$, \[g(T,\bar{x},y) = 0\]  so \begin{eqnarray*}A^M(T) = \mathbf{0}_{N \times N}, \,
         B^M(T) = \mathbf{0}_{N \times K},\,
         C^M(T) = \mathbf{0}_{K \times K}, \,
         \bar{b}^M(T) = \mathbf{0}_{N},\,
         \hat{c}^M(T) = \mathbf{0}_{K}, \,
         \bar{c}^M(T) = 0.\end{eqnarray*} 
For the last interval $[t_{M}, T]$, we can use Proposition \ref{prop:VF1} directly with the views $Y(t_{M}, T)=y$ and views covariance \(\Omega^j\).  It follows that \(A^j(t)\) ($j=M$) is the solution of the Riccati equation \eqref{eq19}, $B^j(t) y + \bar{b}^j(t)$ satisfies the ODE \eqref{eq20}, and \[- \frac{1}{2} y^\top \bar{C}^j(t) y + y^\top \hat{c}^j(t) + \bar{c}^j(t)\] solves \eqref{eq20b}. 
That is, \(A^j(t) \in \mathbb{R}^{N\times N}\) is negative semi-definite with
  \begin{equation}
    \label{eq:EC_Ricatti_update}
        A^{j\prime}(t) + \dfrac{1 - \gamma}{\gamma} \eta^{j}_t \Sigma \eta^{j}_t + \dfrac{1}{\gamma} \left(A^j(t) \Sigma \eta^{j}_t + \eta^{j}_t \Sigma A^j(t)\right) + \dfrac{1}{\gamma} A^j(t) \Sigma A^j(t) = 0 
    \end{equation}
    where
    \begin{equation*}
             \eta^j_t = -P^\top ((T - t) P \Sigma P^\top + \Omega^j)^{-1} P;
    \end{equation*}
    \(B^j(t) \in \mathbb{R}^{N \times K}\) and
    $\bar{b}^j(t) \in \mathbb{R}^N$ are solutions of
    \begin{equation}
\label{eq:EC_Bj_update}
    B^{j'}(t) + \frac{1}{\gamma} (\eta_t^j + A^j(t)) \Sigma B^j(t) + \frac{1-\gamma}{\gamma} \eta_t^j \alpha_1^j(t) + \frac{1}{\gamma}A^j(t) \alpha_1^j(t) = 0,
\end{equation}
and 
\begin{equation}
\label{eq:EC_barbj_update}
    \begin{split}
        \bar{b}^{j'}(t) &+ \frac{1}{\gamma} (\eta_t^j + A^j(t)) \left(\Sigma \bar{b}^j(t) + \alpha_0^j(t) - r_f \mathbf{1}_N \right) - \eta_t^j \left( \alpha_0^j(t) - r_f \mathbf{1}_N \right)
        + A^j(t) \left(r_f \mathbf{1}_N - \frac{1}{2} \diag(\Sigma) \right) = 0,
     \end{split}
\end{equation}
where 
\begin{equation*}
    \begin{cases}
        \alpha_1^j(t) = - \Sigma P^\top \left( (T-t) P \Sigma P^\top + \Omega^j\right)^{-1} \in \mathbb{R}^{N \times K},\\
        
        \alpha_0^j(t) = \mu - (T-t_j) \beta_1^j P \mu^x - (t-t_j) \Sigma \eta_t^j \left(\mu^x - (T-t_j) \beta_1^j P \mu^x\right) \in \mathbb{R}^N;
    \end{cases}
\end{equation*}
and $C^j(t) \in \mathbb{R}^{K \times K}$, $\hat{c}^j(t) \in \mathbb{R}^K$ and $\bar{c}^j(t) \in \mathbb{R}$ are solutions of
\begin{equation}
\label{eq:EC_Cj_update}
    C^{j'}(t) - \frac{1}{\gamma} \left(B^j(t) + \Sigma^{-1} \alpha_1^j(t)\right)^\top \Sigma \left(B^j(t) + \Sigma^{-1} \alpha_1^j(t)\right) + \alpha_1^j(t)^\top \Sigma^{-1} \alpha_1^j(t) = 0,
\end{equation}
\begin{equation}
\label{eq:EC_hatcj_update}
    \begin{split}
        \hat{c}^{j'}(t) &+ \frac{1-\gamma}{\gamma}\alpha_1^j(t) \Sigma^{-1}\left(\alpha_0^j(t) - r_f \mathbf{1}_N \right) + \frac{1}{\gamma} \alpha_1^j(t)^\top \bar{b}^j(t) + B^j(t)^\top (r_f \mathbf{1}_N - \frac{1}{2} \diag(\Sigma))\\
        &+ \frac{1}{\gamma}B^{j}(t)^\top \left(\Sigma \bar{b}^j(t) + \alpha_0^j(t) - r_f \mathbf{1}_N \right) = 0,
    \end{split}
\end{equation}
and 
\begin{equation}
\label{eq:EC_barcj_update}
\begin{split}
        \bar{c}^{j'}(t) &+ (1-\gamma)r_f + \frac{1}{2} \Tr(A^j(t) \Sigma) + \frac{1-\gamma}{ 2 \gamma} \left(\alpha_0^j(t) - r_f \mathbf{1}_N\right)^\top \Sigma^{-1}\left(\alpha_0^j(t) - r_f \mathbf{1}_N\right)\\
        & + \left(\alpha_0^j(t) - \frac{1}{2}\diag(\Sigma)\right)^\top \bar{b}^j(t) + \frac{1-\gamma}{ \gamma}\left(\alpha_0^j(t) - r_f \mathbf{1}_N\right)^\top \bar{b}^j(t) + \frac{1}{2\gamma} \bar{b}^{j}(t)^\top \Sigma \bar{b}^j(t) = 0.
\end{split}
\end{equation}

For earlier time intervals $[t_j, t_{j+1})$ ($j=0, \cdots, M-1$), $g^j(t,\bar{x},y)$ can be derived in a similar manner by solving the dynamic programming equation over the interval $[t_j, t_{j+1}]$. This gives us the same ODEs for $A^j(t)$, $B^j(t)$, $\bar{b}^j(t)$, $C^j(t)$, $\hat{c}^j(t)$ and $\bar{c}^j(t)$ when $j=M$ though with different terminal conditions  because they depend on the view $Y^{j+1}(t_{j+1}, T)$ and the non-zero value function $V(t,z,\bar{x},y)$ at $t_{j+1}$  through the Principle of Optimality \eqref{eq:continuation}. The following result computes the expectation in \eqref{eq:continuation}. The proof can be found in Appendix \ref{proof_lemma:EC_ContinuationValueFunction_Revision}.

 

 \begin{lemma}
    \label{lemma:EC_ContinuationValueFunction_Revision}
    \begin{eqnarray}
    V(t_{j+1}^-,z,\bar{x},y)&  = & \mathbb{E}[V(t_{j+1}, z, 0, Y^{j+1}) \, | \, \bar{X}^y(t_{j+1}^-) = \bar{x}, I(t) = y], \, \, \, \text{for} \, \, j \in \{0, \dots, M-1\} \nonumber\\
    & = & U(z) \exp\left(g^j(t_{j+1},\bar{x},y)\right), \quad t \in [t_j, t_{j+1})
 \label{eq:EC_continuationVF_update}
\end{eqnarray}
where
\begin{equation*}
    g^j(t_{j+1},\bar{x},y) = \frac{1}{2} \bar{x}^\top A^{j}(t_{j+1}) \bar{x} + \bar{x}^\top \left(B^{j}(t_{j+1}) y + \bar{b}^{j}(t_{j+1})\right) + \left( - \frac{1}{2} y^\top \bar{C}^{j}(t_{j+1}) y + y^\top \hat{c}^{j}(t_{j+1}) + \bar{c}^{j}(t_{j+1}) \right)
\end{equation*}
with
    \begin{equation}
        \label{terminal:C}
    \begin{split}
    C^j(t_{j+1}) &= \bar{\beta}_0^{j\top} \left(C^{j+1}(t_{j+1})^{-1} +  \bar{\Omega}^{j+1 \, | \, j} \ \right)^{-1} \bar{\beta}_0^j,\\
      \hat{c}^j(t_{j+1}) &= \bar{\beta}_0^{j\top} \left(C^{j+1}(t_{j+1})^{-1} +  \bar{\Omega}^{j+1 \, | \, j}  \right)^{-1}\left(C^{j+1}(t_{j+1})^{-1} \hat{c}^{j+1}(t_{j+1}) - \bar{\alpha}_0^j\right),\\
       \bar{c}^j(t_{j+1}) & = \bar{c}^{j+1}(t_{j+1}) + (\bar{\beta}_1^{j})^\top \hat{c}^{j+1}(t_{j+1}) - \frac{1}{2} \bar{\beta}_1^{j\top} C^{j+1}(t_{j+1}) \bar{\beta}_1^{j} + \ln\left(\det(I_K + \bar{\Omega}^{j+1} C^{j+1}(t_{j+1}))^{-\frac{1}{2}}\right)\\ 
            & \quad + \frac{1}{2} (\hat{c}^{j+1}(t_{j+1}) -  C^{j+1}(t_{j+1})\bar{\beta}_1^{j})^\top \big((\bar{\Omega}^{j+1})^{-1}  + C^{j+1}(t_{j+1})\big)^{-1}(\hat{c}^{j+1}(t_{j+1}) -  C^{j+1}(t_{j+1})\bar{\beta}_1^{j}),
    \end{split}
    \end{equation}
    \begin{equation}
    \label{terminal:A}
        \begin{split}
             A^j(t_{j+1}) &= -P^\top C^j(t_{j+1}) P
        \end{split}
    \end{equation}
 and
    \begin{equation}
    \label{terminal:B}
    \begin{split}
         \bar{b}^j(t_{j+1}) & = - P^\top \hat{c}^j(t_{j+1}),\\
             B^j(t_{j+1}) &= P^\top C^j(t_{j+1}),
         \end{split}
    \end{equation}
    where 
    \begin{equation*}
        \begin{split}
           \bar{\alpha}_0^j &= (T-t_{j+1}) (\Omega^j - \Omega^{j+1})\left((T-t_{j+1}) P \Sigma P^\top + \Omega^j \right)^{-1} P \mu^x,\\
           \bar{\beta}_0^j &= I_K - (\Omega^j - \Omega^{j+1}) \left((T-t_{j+1}) P \Sigma P^\top + \Omega^j \right)^{-1},\\
            \bar{\Omega}^{j+1 \, | \, j} &= (\Omega^j - \Omega^{j+1}) \left((T-t_{j+1}) P\Sigma P^\top + \Omega^j\right)^{-1}\left((T-t_{j+1})P\Sigma P^\top + \Omega^{j+1}\right),
        \end{split}
    \end{equation*}
    are constants. 
\end{lemma}

For the interval $[t_j, t_{j+1}]$ ($j=0, \cdots, M-1$) \eqref{terminal:C} gives the  terminal conditions for the ODEs \eqref{eq:EC_Cj_update}--\eqref{eq:EC_barcj_update}, \eqref{terminal:A} for the Riccati equation \eqref{eq:EC_Ricatti_update}, and \eqref{terminal:B} for the ODEs \eqref{eq:EC_Bj_update}--\eqref{eq:EC_barbj_update}. This provides us with enough information to solve the 6 ODEs \eqref{eq:EC_Ricatti_update}--\eqref{eq:EC_barcj_update}.

The terminal condition for $A^j(t)$, $B^j(t)$ and $\bar{b}^j(t)$ suggest that it might be possible to express the solutions of their respective ODEs in terms of  $C^j(t)$ and $\hat{c}^j(t)$. The following result shows that  this is indeed the case, so only 3 of these ODEs instead of 6 need to be solved numerically to compute $g^j(t, \bar{x}, y)$ on $[t_j, t_{j+1}]$. The proof can be found in Appendix \ref{proof_lemma:EC_ContinuationValueFunction_Revision}.

\begin{lemma}
    \label{lemma:EC_update_similarities}
      For \(t \in [t_j, t_{j+1})\) and \(j \in \{0, \dots, M\}\), let \(A^j(t) \in \mathbb{R}^{N \times N}\), \(B^j(t) \in \mathbb{R}^{N \times K}\), and \(C^j(t) \in \mathbb{R}^{K \times K}\) be the solutions of \eqref{eq:EC_Ricatti_update}, \eqref{eq:EC_Bj_update}, and \eqref{eq:EC_Cj_update}, respectively, with boundary conditions specified in Lemma \ref{lemma:EC_ContinuationValueFunction_Revision}. Then,
    \begin{eqnarray*}
    \begin{split}
                &A^j(t) =  - P^\top C^j(t) P, \quad \text{for} \quad t \in [t_j, t_{j+1}), \; j \in \{0, \dots, M\},\\
                &B^j(t) = P^\top C^j(t), \quad \text{for} \quad t \in [t_j, t_{j+1}), \; j \in \{0, \dots, M\}.
    \end{split}
    \end{eqnarray*}
    Furthermore, if \(\bar{b}^j(t)\) and \(\hat{c}^j(t)\) are the solutions of \eqref{eq:EC_barbj_update} and \eqref{eq:EC_hatcj_update}, respectively,
    \begin{equation*}
        \bar{b}^j(t) = -P^\top \hat{c}^j(t), \quad \text{for} \quad t \in [t_j, t_{j+1}), \; j \in \{0, \dots, M\}.
    \end{equation*}
\end{lemma}


It follows that
\begin{equation*}
\begin{split}
        g^{j}(t,\bar{x},y) &= \frac{1}{2} \bar{x}^\top A^j(t) \bar{x} + \bar{x}^\top \left(B^j(t) y + \bar{b}^j(t)\right) + \left( - \frac{1}{2} y^\top \bar{C}^j(t) y + y^\top \hat{c}^j(t) + \bar{c}^j(t) \right)\\ 
    &=-\frac{1}{2} (P\bar{x} - y)^\top C^{j}(t) (P \bar{x} - y) - (P\bar{x} - y)^\top \hat{c}^{j}(t) + \bar{c}^{j}(t)
    \end{split}
\end{equation*}
and the hedging demand is
 \begin{equation*}
 \begin{split}
               \frac{1}{\gamma} \frac{\partial g^j}{\partial \bar{x}}(t,\bar{x},y) &= \frac{1}{\gamma} \left(A^j(t) \bar{x} + B^j(t) y + \bar{b}^j(t)\right) \\ 
               \medskip
               &= \frac{1}{\gamma} P^\top\big( C^j(t) (y - P\bar{x}\big)  - \hat{c}^j(t)\big).
\end{split}
\end{equation*}
Hence, the hedging demand depends only on two coefficients, \(C^j(t)\) and \(\hat{c}^j(t)\) obtained by solving \eqref{eq:EC_Cj_update}--\eqref{eq:EC_hatcj_update} with boundary conditions \eqref{terminal:C}.

The following result shows that \(C^j(t)\) and \(\hat{c}^j(t)\) admit explicit solutions, which we derive. The proof can be found in Appendix \ref{proof_lemma:EC_ContinuationValueFunction_Revision}.

\begin{lemma}
    \label{Lemma:EC_Solutions_revision}
    Let $C^j(t) \in \mathbb{R}^{K \times K}$ and $\hat{c}^j(t) \in \mathbb{R}^K$ be the solutions to \eqref{eq:EC_Cj_update} and \eqref{eq:EC_hatcj_update}, respectively, with boundary conditions \eqref{terminal:C}. Then
    \begin{equation*}
    \begin{split}
              &C^j(t) = \bar{M}^j(t) P^\top \bar{\eta}_t^j,\\
        &\hat{c}^j(t) = \bar{M}^j(t) \left( \Sigma^{-1} (\mu - r_f \mathbf{1}_N)  + (T-t) P^\top \bar{\eta}_t^j P \mu^x\right),
    \end{split}
    \end{equation*}
    where 
    \begin{equation*}
        \bar{M}^j(t) =- (\gamma - 1) (T-t) (\Omega^j)^{-1}P \Big(\gamma\Sigma^{-1} + (T-t) P^\top (\Omega^j)^{-1} P\Big)^{-1} \in \mathbb{R}^{K \times N}.
    \end{equation*}
\end{lemma}

It follows from Lemma \ref{Lemma:EC_Solutions_revision} and \eqref{eq:EC_g_function_Update} that
   \begin{equation*}
    \begin{split}
               \frac{1}{\gamma} \frac{\partial g^j}{\partial \bar{x}}(t,\bar{x},y) &= \frac{1}{\gamma} P^\top\big( C^j(t) (y - P\bar{x}\big)  - \hat{c}^j(t)\big)\\
               &= - \frac{1}{\gamma} P^\top \bar{M}^j(t) \Sigma^{-1} \big(\tilde{\mu}^j(t,\bar{x},y) - r_f\big)\\
               &= \frac{1}{\gamma} P^\top M^j(t) \Sigma^{-1} \big(\tilde{\mu}^j(t,\bar{x},y) - r_f\big),
    \end{split}
    \end{equation*}
    where 
\begin{eqnarray*}
         M^j(t) = (\gamma - 1) (T-t) P^\top (\Omega^j)^{-1}P \Big(\gamma\Sigma^{-1} + (T-t) P^\top (\Omega^j)^{-1} P\Big)^{-1} \in \mathbb{R}^{N \times N}.
    \end{eqnarray*}

    And the optimal policy is 
       \begin{equation*}
        \pi^{j *}(t,\bar{x},y) = \frac{1}{\gamma} \Sigma^{-1} \big(\tilde{\mu}^j(t,\bar{x},y) - r_f \mathbf{1}_N\big) + \frac{1}{\gamma} \frac{\partial g^j}{\partial \bar{x}}(t,\bar{x},y), \, \,\, \text{for} \, \, \, t \in [t_j, t_{j+1}),
    \end{equation*}
    where
    \begin{equation*}
    \tilde{\mu}^j(t,\bar{x},y) = \mu + \beta_1^j (y - (T-t_j)P \mu^x) + \beta_2^j(t) (\mathbb{E}[\bar{X}^y(t)] - \bar{x}), \, \,\, \text{for} \, \, \, t \in [t_j, t_{j+1})
\end{equation*}
    is the drift of the conditional asset price. This concludes the proof. \hfill $\square$

\subsection{Section  \ref{sec:short-term_views}}
\label{Sec:EC_QuarterlyViews}


\subsubsection{Proofs of Conditional Market Dynamics} 
\label{ECsec:QuarterlyViewsDynamics}
\paragraph{\textbf{Proof of Proposition \ref{prop:independence_quarterly_views}.}}

Assume views $\{Y^j(T_j, T_{j+1}), \, j \in \{0, \dots, M\}\}$ satisfy \eqref{eq:Quarterlyviews}--\eqref{eq:quarterlyNoises}. Let 
\begin{equation*}
    \bar{X}^y(t) \triangleq X(t) \, | \, \left(Y^0(0,T_1), \dots, Y^j(T_j, T_{j+1})\right), \; t \in [T_j, T_{j+1})
\end{equation*}
be the conditional log-returns and refined views  $\{\bar{Y}^j(T_j, T_{j+1}), \, j \in \{0, \dots, M\}\}$ be defined by \eqref{eq:refined_view}.
It follows that
\begin{eqnarray*}
    \bar{Y}^j(T_j, T_{j+1}) \, | \, (X(T_j), X(T_{j+1})) = P \big(X(T_{j+1}) - X(T_j)\big) + \epsilon^{j,0} \sim \mathcal{N}\big( P \big(X(T_{j+1}) - X(T_j)\big), \Omega^{j,0} \big).
\end{eqnarray*}
Since \(\{\epsilon^{j,0} , \, j \in \{1, \dots, M\}\}\) are independent and  
\[
X(t) = t \mu^x + W(t)
\]  
has independent increments, 
\[
\{\bar{Y}^j(T_j, T_{j+1}) \mid j \in \{0, \dots, M\}\}
\]  
are mutually independent.  
Let  
\begin{eqnarray*}
    \bar{X}(t) = X(t) - X(T_j), \quad t \in [T_j, T_{j+1})
\end{eqnarray*}  
denote the log-return between the time of the last view \(T_j\) and the current time \(t\). Since the  increments of \(X(t)\) are independent,  \(\bar{X}(t)\) is independent of  
\[
\{\bar{Y}^k(T_k, T_{k+1}) \mid k \in \{0, \dots, j-1\}\}.
\]  
It follows that
\begin{eqnarray*}
\begin{split}
            \bar{X}(t) \mid \left(Y^0(0,T_1), \dots, Y^j(T_j, T_{j+1})\right)  
            &= \bar{X}(t) \mid \left(\bar{Y}^0(0,T_1), \dots, \bar{Y}^j(T_j, T_{j+1})\right)\\
            &= \bar{X}(t) \mid \bar{Y}^j(T_j, T_{j+1}).
\end{split}
\end{eqnarray*}
This concludes the proof. \hfill $\square$

\paragraph{\textbf{Proof of Proposition \ref{cor:QuarterlyViews}.}}
From Proposition \ref{prop:independence_quarterly_views}, we have
\begin{equation*}
    \begin{split}
        \bar{X}^y(t) & \triangleq \bar{X}(t) \, | \, \left(Y^0(0,T_1), \dots, Y^j(T_j, T_{j+1})\right)\\ 
        &= \bar{X}(t) \, | \, \bar{Y}^j(T_j, T_{j+1}).
    \end{split}
\end{equation*}
This setting is analogous to that in Section \ref{sec3} except 
\begin{itemize}
    \item The horizon is now \([T_j, T_{j+1}]\) instead of \([0,T]\).
    \item The log-returns are given by \(\bar{X}(t)\) instead of \(X(t)\).
    \item The views vector is \(\bar{Y}^j(T_j,T_{j+1}) = \bar{y}^j\) instead  \(Y(0,T) = y\).
\end{itemize}
It follows from Proposition \ref{proposition1} that
 \begin{equation*}
        d\bar{X}^y(t) =   \bigg(\mu^x + \beta_1^j \big(\bar{y}^j - (T_{j+1}-T_j) P \mu^x \big) + \beta_2^j(t) \big(\mathbb{E}[\bar{X}^y(t)] - \bar{X}^y(t)\big)\bigg) dt + dW^y(t), \, \,\, \text{for} \, \, \, t \in [T_j, T_{j+1})
    \end{equation*}
    where
    \begin{eqnarray*}
        \bar{y}^j = y^j - \sum_{i=1}^{\bar{p}} \Phi^i (y^{j-i} - P \bar{x}^{j-i})
    \end{eqnarray*}
    with
\begin{eqnarray*}
    \beta_1^j & = & \Sigma P^\top \big((T_{j+1}-T_j) P \Sigma P^\top+\Omega^{j,0}\big)^{-1} \in  \mathbb{R}^{N \times K},\\ [8pt]
    \beta_2^j(t) & = & \Sigma P^\top \big((T_{j+1}-t) P \Sigma P^\top + \Omega^{j,0}\big)^{-1} P  \in \mathbb{R}^{N \times N},
\end{eqnarray*}
and
\begin{equation*}
   \mathbb{E}[\bar{X}^y(t)]  = (t - T_j) \big(\mu^x +  \beta_1^j \big(\bar{y}^j -(T_{j+1}-T_j) P \mu^x \big)\big).
\end{equation*}
This completes the proof. \hfill \(\square\)

\subsubsection{Proof of Theorem \ref{thm:QuarterlyViews}}
\label{EC_Views_Quarterly_Control}

The proof of this theorem closely follows that of Theorem \ref{thm:updatedViews} in that we solve for the value function recursively over intervals $[T_{M}, T_{M+1}], [T_{M-1}, T_M], \cdots, [T_0, T_1]$ with the value functions of adjacent intervals being tied together by the Principal of Optimality. 

Once again, we assume that \begin{equation*}
    V(t, z, \bar{x}, \bar{y}) = U(z) \exp\left(g^j(t, \bar{x}, \bar{y})\right), \quad t \in [T_j, T_{j+1}),
\end{equation*}
where the function \( g^j : \mathbb{R} \times \mathbb{R}^N \times \mathbb{R}^K \to \mathbb{R} \) is quadratic in both the log-returns vector \( \bar{x} \) and the transformed views \( \bar{y} \).  For \( t \in [T_j, T_{j+1}) \), we define  
\begin{equation}
\label{eq:EC_g_function_quarterly}
     g^j(t, \bar{x}, \bar{y}) = \frac{1}{2} \bar{x}^\top A^j(t) \bar{x} + \bar{x}^\top \left(B^j(t) \bar{y} + \bar{b}^j(t)\right) 
    + \left(- \frac{1}{2} \bar{y}^\top \bar{C}^j(t) \bar{y} + \bar{y}^\top \hat{c}^j(t) + \bar{c}^j(t) \right).
\end{equation}

For $j=M$ (final interval $[T_M, T]$), the terminal condition $g(T,\bar{x},\bar{y}) = 0$ so 
\begin{eqnarray*}
A^M(T) = \mathbf{0}_{N \times N}, \,
         B^M(T) = \mathbf{0}_{N \times K},\,
         C^M(T) = \mathbf{0}_{K \times K}, \,
         \bar{b}^M(T) = \mathbf{0}_{N},\,
         \hat{c}^M(T) = \mathbf{0}_{K}, \,
         \bar{c}^M(T) = 0.
\end{eqnarray*}
Substituting \eqref{eq:EC_g_function_quarterly} into the HJB equation over the investment interval $[T_M, T]$, it can be shown that
\(A^j(t) \in \mathbb{R}^{N\times N}\)  satisfies
  \begin{equation}
    \label{eq:EC_Ricatti_quarterly}
        A^{j\prime}(t) + \dfrac{1 - \gamma}{\gamma} \eta^{j}_t \Sigma \eta^{j}_t + \dfrac{1}{\gamma} \left(A^j(t) \Sigma \eta^{j}_t + \eta^{j}_t \Sigma A^j(t)\right) + \dfrac{1}{\gamma} A^j(t) \Sigma A^j(t) = 0 , 
    \end{equation}
    where
    \begin{equation*}
             \eta^j_t = -P^\top ((T_{j+1} - t) P \Sigma P^\top + \Omega^{j,0})^{-1} P.
    \end{equation*}
    Note that $A^j(t)$ is negative semi-definite. Likewise, 
     \(B^j(t) \in \mathbb{R}^{N \times K}\) and $\bar{b}^j(t) \in \mathbb{R}^N$ satisfy
    \begin{equation}
\label{eq:EC_Bj_quarterly}
    B^{j'}(t) + \frac{1}{\gamma} (\eta_t^j + A^j(t)) \Sigma B^j(t) + \frac{1-\gamma}{\gamma} \eta_t^j \alpha_1^j(t) + \frac{1}{\gamma}A^j(t) \alpha_1^j(t) = 0
\end{equation}
and 
\begin{equation}
\label{eq:EC_barbj_quarterly}
    \begin{split}
        \bar{b}^{j'}(t) &+ \frac{1}{\gamma} (\eta_t^j + A^j(t)) \left(\Sigma \bar{b}^j(t) + \alpha_0^j(t) - r_f \mathbf{1}_N \right) - \eta_t^j \left( \alpha_0^j(t) - r_f \mathbf{1}_N \right)
        + A^j(t) \left(r_f \mathbf{1}_N - \frac{1}{2} \diag(\Sigma) \right) = 0,
     \end{split}
\end{equation}
where 
\begin{equation*}
    \begin{cases}
        \alpha_1^j(t) = - \Sigma P^\top \left( (T_{j+1}-t) P \Sigma P^\top + \Omega^{j,0}\right)^{-1} \in \mathbb{R}^{N \times K},\\
        
        \alpha_0^j(t) = \mu - (T_{j+1}-T_j) \beta_1^j P \mu^x - (t-T_j) \Sigma \eta_t^j \left(\mu^x - (T_{j+1}-T_j) \beta_1^j P \mu^x\right) \in \mathbb{R}^N,
    \end{cases}
\end{equation*}
and $C^j(t) \in \mathbb{R}^{K \times K}$, $\hat{c}^j(t) \in \mathbb{R}^K$  and $\bar{c}^j(t) \in \mathbb{R}$ solve 
\begin{equation}
\label{eq:EC_Cj_quarterly}
    C^{j'}(t) - \frac{1}{\gamma} \left(B^j(t) + \Sigma^{-1} \alpha_1^j(t)\right)^\top \Sigma \left(B^j(t) + \Sigma^{-1} \alpha_1^j(t)\right) + \alpha_1^j(t)^\top \Sigma^{-1} \alpha_1^j(t) = 0,
\end{equation}
\begin{equation}
\label{eq:EC_hatcj_quarterly}
    \begin{split}
        \hat{c}^{j'}(t) &+ \frac{1-\gamma}{\gamma}\alpha_1^j(t) \Sigma^{-1}\left(\alpha_0^j(t) - r_f \mathbf{1}_N \right) + \frac{1}{\gamma} \alpha_1^j(t)^\top \bar{b}^j(t) + B^j(t)^\top (r_f \mathbf{1}_N - \frac{1}{2} \diag(\Sigma))\\
        &+ \frac{1}{\gamma}B^{j}(t)^\top \left(\Sigma \bar{b}^j(t) + \alpha_0^j(t) - r_f \mathbf{1}_N \right) = 0,
    \end{split}
\end{equation}
and
\begin{equation}
\label{eq:EC_barcj_quarterly}
\begin{split}
        \bar{c}^{j'}(t) &+ (1-\gamma)r_f + \frac{1}{2} \Tr(A^j(t) \Sigma) + \frac{1-\gamma}{ 2 \gamma} \left(\alpha_0^j(t) - r_f \mathbf{1}_N\right)^\top \Sigma^{-1}\left(\alpha_0^j(t) - r_f \mathbf{1}_N\right)\\
        & + \left(\alpha_0^j(t) - \frac{1}{2}\diag(\Sigma)\right)^\top \bar{b}^j(t) + \frac{1-\gamma}{ \gamma}\left(\alpha_0^j(t) - r_f \mathbf{1}_N\right)^\top \bar{b}^j(t) + \frac{1}{2\gamma} \bar{b}^{j}(t)^\top \Sigma \bar{b}^j(t) = 0.
\end{split}
\end{equation}
For $j=0, \cdots, M-1$, $g^j(t, \bar{x}, \bar{y})$ satisfies \eqref{eq:EC_g_function_quarterly} with coefficients satisfying the ODEs \eqref{eq:EC_Ricatti_quarterly}--\eqref{eq:EC_barcj_quarterly} except now with terminal conditions that depend on the coefficients of $g^{j+1}(t, \bar{x}, \bar{y})$ which we now specify. 

\begin{lemma}
    \label{lemma:EC_ContinuationValueFunction_quarterly}
    Let  
    \begin{eqnarray*}
        V(t,z,\bar{x}, \bar{y}) = \max_{\pi \in \mathcal{A}} \mathbb{E}\big[ U(Z(T)) | \bar{X}(t) = \bar{x}, Z(t) = z, I(t) = \bar{y}  \big]
    \end{eqnarray*}
    be the value function of the investor at time \(t \in [T_j, T_{j+1})\), with \(j \in \{0, \dots, M-1\}\). The continuation condition of the value function at \(t = T_{j+1}\) is given by
            \begin{equation}
    \label{eq:EC_continuationVF_quarterly}
    \begin{split}
    V(T_{j+1}^-,z,\bar{x},\bar{y}) &= \mathbb{E}[V(T_{j+1}, z, 0, \bar{Y}^{j+1}) \, | \, \bar{X}^y(T_{j+1}^-) = \bar{x}, I(T_{j+1}^-) = \bar{y}]\\
    &= \mathbb{E}[V(T_{j+1}, z, 0, \bar{Y}^{j+1})]\\
    &= U(z) \exp\left(g^j(T_{j+1},\bar{x},\bar{y}) \right)
    \end{split}
\end{equation}
    where
\begin{equation*}
     g^j(t, \bar{x}, \bar{y}) = \frac{1}{2} \bar{x}^\top A^j(t) \bar{x} + \bar{x}^\top \left(B^j(t) \bar{y} + \bar{b}^j(t)\right) 
    + \left(- \frac{1}{2} \bar{y}^\top \bar{C}^j(t) \bar{y} + \bar{y}^\top \hat{c}^j(t) + \bar{c}^j(t) \right).
\end{equation*}  
The boundary conditions of the coefficients are
\begin{eqnarray*}A^j(T_{j+1}) = \mathbf{0}_{N \times N}, \,
         B^j(T_{j+1}) = \mathbf{0}_{N \times K},\,
         C^j(T_{j+1}) = \mathbf{0}_{K \times K}, \,
         \bar{b}^j(T_{j+1}) = \mathbf{0}_{N},\,
         \hat{c}^j(T_{j+1}) = \mathbf{0}_{K}
         \end{eqnarray*} 
and 
\begin{equation}
\label{eq:bar_c_terminal_quarterly}
    \begin{split}
        \bar{c}^j(T_{j+1}) & =  \ln\left(\det(I_K + \bar{\Omega}^{j+1} C^{j+1}(T_{j+1}))^{-\frac{1}{2}}\right)\\ 
            & \quad + \frac{1}{2} \left(\hat{c}^{j+1}(T_{j+1}) - C^{j+1}(T_{j+1})\bar{\alpha}_0^{j+1}\right)^\top \left(C^{j+1}(T_{j+1})^{-1} +  \bar{\Omega}^{j+1} \ \right)^{-1}  \left(\hat{c}^{j+1}(T_{j+1}) - C^{j+1}(T_{j+1})\bar{\alpha}_0^{j+1}\right)^\top\\
                &\quad + \frac{1}{2} \bar{\alpha}_0^{j+1\top} C^{j+1}(T_{j+1}) \bar{\alpha}_0^{j+1} + \bar{\alpha}_0^{j+1\top} \left(\hat{c}^{j+1}(T_{j+1}) - C^{j+1}(T_{j+1})\bar{\alpha}_0^{j+1}\right) + \bar{c}^{j+1}(T_{j+1})
    \end{split}
\end{equation}
where
\begin{equation*}
    \begin{split}
    \bar{\alpha}_0^{j+1} &= \mathbb{E}\bigl[\bar{Y}^{j+1}(T_{j+1}, T_{j+2}) \bigr]= (T_{j+2}-T_{j+1}) P \mu^x,\\
        \bar{\Omega}^{j+1}
        &= \mathbb{V}[\bar{Y}^{j+1}(T_{j+1}, T_{j+2})]
        = (T_{j+2}-T_{j+1}) P \Sigma P^\top + \Omega^{j+1,0},
    \end{split}
\end{equation*}
    are the mean and the covariance of the transformed view \(\bar{Y}^{j+1}(T_{j+1}, T_{j+2})\), respectively.
    
    \end{lemma}
The proof is given in Appendix \ref{proof_lemma:EC_ContinuationValueFunction_quarterly}.

Observe that many of the boundary conditions are zero because the pair $\{(\bar{X}(t), \bar{Y}^j(T_j,T_{j+1})), t\in[t_j, t_{j+1})\}$   across intervals  $j=0, \cdots M$ are independent.   The following result gives $A^j(t)$, $B^j(t)$ and $\bar{b}^j(t)$ in terms of $C^j(t)$ and $\hat{c}^j(t)$. The proof  follows that of Lemma \ref{lemma:EC_update_similarities} and is omitted.
\begin{lemma}
    \label{lemma:EC_quarterly_similarities}
      For \(t \in [T_j, T_{j+1})\) and \(j \in \{0, \dots, M\}\), let \(A^j(t) \in \mathbb{R}^{N \times N}\), \(B^j(t) \in \mathbb{R}^{N \times K}\), and \(C^j(t) \in \mathbb{R}^{K \times K}\) be the solutions of \eqref{eq:EC_Ricatti_quarterly}, \eqref{eq:EC_Bj_quarterly}, and \eqref{eq:EC_Cj_quarterly}, respectively, with boundary conditioned specified in Lemma \ref{lemma:EC_ContinuationValueFunction_quarterly}. Then,
    \begin{eqnarray*}
    \begin{split}
                &A^j(t) =  - P^\top C^j(t) P, \quad \text{for} \quad t \in [T_j, T_{j+1})\\
                &B^j(t) = P^\top C^j(t), \quad \text{for} \quad t \in [T_j, T_{j+1}).
    \end{split}
    \end{eqnarray*}
    Furthermore, let \(\bar{b}^j(t)\) and \(\hat{c}^j(t)\) be the solutions of \eqref{eq:EC_barbj_quarterly} and \eqref{eq:EC_hatcj_quarterly}, respectively. Then,
    \begin{equation*}
        \bar{b}^j(t) = -P^\top \hat{c}^j(t), \quad \text{for} \quad t \in [T_j, T_{j+1}).
    \end{equation*}
 \end{lemma}

From Lemma \ref{lemma:EC_quarterly_similarities}, it follows that we can express \eqref{eq:EC_g_function_quarterly} as
\begin{equation*}
    g^{j}(t,\bar{x},\bar{y}) =  -\frac{1}{2} (P\bar{x} - \bar{y})^\top C^{j}(t) (P \bar{x} - \bar{y}) - (P\bar{x} - \bar{y})^\top \hat{c}^{j}(t) + \bar{c}^{j}(t).
\end{equation*}
The following result gives closed form solutions of \(C^j(t)\) and \(\hat{c}^j(t)\).

\begin{lemma}
    \label{Lemma:EC_Solutions_quarterly}
    Let $C^j(t) \in \mathbb{R}^{K \times K}$ and $\hat{c}^j(t) \in \mathbb{R}^K$ be the solutions to \eqref{eq:EC_Cj_quarterly} and \eqref{eq:EC_hatcj_quarterly}, respectively, with boundary conditions given by Lemma \ref{lemma:EC_ContinuationValueFunction_quarterly}. Then
    \begin{equation*}
    \begin{split}
              &C^j(t) = \bar{M}^j(t) P^\top \bar{\eta}_t^j,\\
        &\hat{c}^j(t) = \bar{M}^j(t) \left( \Sigma^{-1} (\mu - r_f \mathbf{1}_N)  + (T_{j+1}-t) P^\top \bar{\eta}_t^j P \mu^x\right),
    \end{split}
    \end{equation*}
    where 
    \begin{equation*}
        \bar{M}^j(t) =- (\gamma - 1) (T_{j+1}-t) (\Omega^{j,0})^{-1}P \Big(\gamma\Sigma^{-1} + (T_{j+1}-t) P^\top (\Omega^{j,0})^{-1} P\Big)^{-1} \in \mathbb{R}^{K \times N}.
    \end{equation*}
\end{lemma}

The proof is given in Appendix \ref{proof_lemma:EC_Solutions_quarterly}.

It follows from Lemma \ref{Lemma:EC_Solutions_quarterly} and \eqref{eq:EC_g_function_quarterly} that
   \begin{equation*}
    \begin{split}
               \frac{1}{\gamma} \frac{\partial g^j}{\partial \bar{x}}(t,\bar{x},\bar{y}) &= \frac{1}{\gamma} P^\top\big( C^j(t) (\bar{y} - P\bar{x}\big)  - \hat{c}^j(t)\big)\\
               &= - \frac{1}{\gamma} P^\top \bar{M}^j(t) \Sigma^{-1} \big(\tilde{\mu}^j(t,\bar{x}, \bar{y}) - r_f\big)\\
               &= \frac{1}{\gamma} P^\top M^j(t) \Sigma^{-1} \big(\tilde{\mu}^j(t,\bar{x}, \bar{y}) - r_f\big),
    \end{split}
    \end{equation*}
    where 
\begin{eqnarray*}
         M^j(t) = (\gamma - 1) (T_{j+1}-t) P^\top (\Omega^{j,0})^{-1}P \Big(\gamma\Sigma^{-1} + (T_{j+1}-t) P^\top (\Omega^{j,0})^{-1} P\Big)^{-1} \in \mathbb{R}^{N \times N}.
    \end{eqnarray*}

    And the optimal policy is 
       \begin{equation*}
        \pi^{j *}(t,\bar{x},\bar{y}) = \frac{1}{\gamma} \Sigma^{-1} \big(\tilde{\mu}^j(t,\bar{x},\bar{y}) - r_f \mathbf{1}_N\big) + \frac{1}{\gamma} \frac{\partial g^j}{\partial \bar{x}}(t,\bar{x},\bar{y}), \, \,\, \text{for} \, \, \, t \in [T_j, T_{j+1}),
    \end{equation*}
    where
    \begin{equation*}
    \tilde{\mu}^j(t,\bar{x},\bar{y}) = \mu + \beta_1^j (\bar{y} - (T_{j+1}-T_j)P \mu^x) + \beta_2^j(t) (\mathbb{E}[\bar{X}^y(t)] - \bar{x}), \, \,\, \text{for} \, \, \, t \in [T_j, T_{j+1})
\end{equation*}
    is the drift of the conditional stock price, which concludes the proof. \hfill $\square$

\newpage

\section{Proofs: Results in Section \ref{Appendix:DetailedAnalysis}}
\label{sec:EC_proofs_detailedanalysis}

\medskip

\subsection{Proofs: Results in \ref{Sec:EC_ViewsRevision}}

\paragraph{\textbf{Proof of Lemma \ref{Lemma:ObsoleteViews}.}}\mbox{}
\label{proof_lemma:ObsoleteViews}
Revised views \eqref{eq:UpdatedViews}  satisfy
\begin{equation*}
    Y^j(t_j,T) \, | \, (X(t_j), X(T)) = P \left( X(T) - X(t_j) \right) + \epsilon^j \sim \mathcal{N}\big(P \left(X(T) - X(t_j)\right), \Omega^j\big)
\end{equation*}
where log-returns $X(t) = t \mu^x + W(t)$ from \eqref{eq9} 
has independent increments.
For each \(j \in \{1, \dots, M\}\) and \(t \in [t_j, t_{j+1})\) (with \(t_{M+1} = T\)) define
$\bar{X}(t) = X(t) - X(t_j)$.
Since \(X(t)\) has independent increments, \(\bar{X}(t)\) is independent of \(\{Y^0(0,T), \dots, Y^{j-1}(t_{j-1},T)\}\), which concludes the proof. \hfill $\square$

\medskip

\paragraph{\textbf{Proof of Lemma \ref{lemma:EC_ContinuationValueFunction_Revision}}}\mbox{}
\label{proof_lemma:EC_ContinuationValueFunction_Revision}
From the Principle of Optimality 
\[
   V\bigl(t_{j+1}^-, z, \bar{x}, y\bigr)
   \;=\;
   \mathbb{E}_{\,Y^{j+1}}\Bigl[
      V\bigl(t_{j+1}, z, 0, Y^{j+1}\bigr)
      \;\bigg|\;
     Z(t_{j+1}^-) = z, \bar{X}^y\bigl(t_{j+1}^-\bigr)=\bar{x},\;
      I(t_{j+1}^-)=y
   \Bigr].
\]
We now evaluate this expectation. Recall that
\begin{equation}
\label{eq:EC_V(t,z,0)_update}
\begin{split}
        V(t_{j+1},z,0, y) &= U(z) \exp\left(g^j(t_{j+1}, \bar{x},y)\right)\\
        &= U(z) \exp\left( -\frac{1}{2}y^{\top}C^j(t_{j+1}) y + y^{\top} \hat{c}^j(t_{j+1}) + \bar{c}^j(t_{j+1}) \right).
\end{split}
\end{equation}
From \eqref{eq:UpdatedViews} -- \eqref{eq:noise-redundant-views},  the random variable $Y^{j+1}(t_{j+1}, T) \, | \, \left(\bar{X}(t_{j+1}^-) = \bar{x}, Y^j(t_j, T) = y\right)$ is Gaussian with conditional mean and covariance 
\begin{equation*}
    \begin{split}
        &\bar{\mu}^{j+1 \, | \, j} = \bar{\alpha}_0^j - \bar{\beta}_0^j (P \bar{x} - y),\\
        &\bar{\Omega}^{j+1 \, | \, j} = (\Omega^j - \Omega^{j+1}) \left((T-t_{j+1}) P\Sigma P^\top + \Omega^j\right)^{-1}\left((T-t_{j+1})P\Sigma P^\top + \Omega^{j+1}\right)
    \end{split}
\end{equation*}
where 
\begin{equation*}
        \begin{split}
           &\bar{\alpha}_0^j = (T-t_{j+1}) (\Omega^j - \Omega^{j+1})\left((T-t_{j+1}) P \Sigma P^\top + \Omega^j \right)^{-1} P \mu^x,\\
           &\bar{\beta}_0^j = I_K - (\Omega^j - \Omega^{j+1}) \left((T-t_{j+1}) P \Sigma P^\top + \Omega^j \right)^{-1}.
        \end{split}
    \end{equation*}
It follows that 
\begin{eqnarray*}
\lefteqn{V(t_{j+1}^-, z, \bar{x},y) =\mathbb{E}_{\,Y^{j+1}}\Bigl[
      V\bigl(t_{j+1}, z, 0, Y^{j+1}\bigr)
      \;\bigg|\;
      Z(t_{j+1}^-) = z, \bar{X}^y\bigl(t_{j+1}^-\bigr)=\bar{x},\;
      Y^j\bigl(t_j,T\bigr)=y
   \Bigr]}\\
   &= & \int (2\pi)^{-K/2} \det(\bar{\Omega}^{j+1 \, | \, j})^{-1/2} \exp\left(-\frac{1}{2}(y^{j+1} -\bar{\mu}^{j+1 \, | \, j})^\top (\bar{\Omega}^{j+1 \, | \, j})^{-1}(y^{j+1} -\bar{\mu}^{j+1 \, | \, j}) \right) V(t_{j+1},z,0) dy^{j+1}\\
&  = & U(z) \int (2\pi)^{-K/2} \det(\bar{\Omega}^{j+1 \, | \, j})^{-1/2} \exp\bigl(
-\frac{1}{2}(y^{j+1} -\bar{\mu}^{j+1 \, | \, j})^\top (\bar{\Omega}^{j+1 \, | \, j})^{-1}(y^{j+1} -\bar{\mu}^{j+1 \, | \, j})\\
& &-\frac{1}{2}y^{j+1\top}C^j(t_{j+1}) y^{j+1} + y^{j+1\top} \hat{c}^j(t_{j+1}) + \bar{c}^j(t_{j+1})) \bigr) dy^{j+1}
\end{eqnarray*}
where the last equality follows from \eqref{eq:EC_V(t,z,0)_update}. By rearranging the terms in the exponential
\begin{eqnarray}
\label{eq:EC_Value_F_Revisions_Int}
  \lefteqn{V(t_{j+1}^-, z, \bar{x},y) = U(z) \det\left(I_N  + \bar{\Omega}^{j+1 \, | \, j} C^{j+1}(t_{j+1})\right)^{-1/2} \exp\left(\epsilon(y, \bar{x})\right)}\\ [5pt]
        & & \times \int (2\pi)^{-K/2} \det(\hat{\Omega}^{j+1 \, | \, j})^{-1/2} \exp\left(-\frac{1}{2}(y^{j+1} -\hat{\mu}^{j+1 \, | \, j})^\top (\hat{\Omega}^{j+1 \, | \, j})^{-1}(y^{j+1} -\hat{\mu}^{j+1 \, | \, j}) \right) dy^{j+1} \nonumber
\end{eqnarray}
where 
  \begin{equation}
  \label{eq:EC_epsilon_VF_update}
        \begin{split}
            \epsilon(y, \bar{x}) =& -\frac{1}{2} \left(P\bar{x} - y\right)^\top \bar{\beta}_0^{j\top} \left(C^{j+1}(t_{j+1})^{-1} +  \bar{\Omega}^{j+1 \, | \, j} \ \right)^{-1} \bar{\beta}_0^j \left(P\bar{x} - y\right)\\
            &- (P \bar{x} - y)^\top \bar{\beta}_0^{j\top} \left(C^{j+1}(t_{j+1})^{-1} +  \bar{\Omega}^{j+1\, | \, j}  \right)^{-1}\left(C^{j+1}(t_{j+1})^{-1} \hat{c}^{j+1}(t_{j+1}) - \bar{\alpha}_0^j\right)  + \epsilon_0,
        \end{split}
    \end{equation}
     \begin{equation*}
    \begin{split}
                \epsilon_0 = &\frac{1}{2} \left(\hat{c}^{j+1}(t_{j+1}) - C^{j+1}(t_{j+1})\bar{\alpha}_0^j\right)^\top \left(C^{j+1}(t_{j+1})^{-1} +  \bar{\Omega}^{j+1 \, | \, j} \ \right)^{-1}  \left(\hat{c}^{j+1}(t_{j+1}) - C^{j+1}(t_{j+1})\bar{\alpha}_0^j\right)^\top\\
                &+ \frac{1}{2} \bar{\alpha}_0^{j\top} C^{j+1}(t_{j+1}) \bar{\alpha}_0^j + \bar{\alpha}_0^{j\top} \left(\hat{c}^{j+1}(t_{j+1}) - C^{j+1}(t_{j+1})\bar{\alpha}_0^j\right) + \bar{c}^{j+1}(t_{j+1}),
    \end{split}
    \end{equation*}
    and
\begin{equation*}
    \begin{split}
        &\hat{\Omega}^{j+1 \, | \, j} = \left((\bar{\Omega}^{j+1 \, | \, j})^{-1} + C^{j+1}(t_{j+1})\right)^{-1} \in \mathbb{R}^{K \times K},\\
        &\hat{\mu}^{j+1 \, | \, j} = \hat{\Omega}^{j+1 \, | \, j} \left(\hat{c}^{j+1}(t_{j+1}) - C^{j+1}(t_{j+1}) \left(\alpha_0^j - \beta_0^j ( P \bar{x} - y)\right)\right) + \bar{\mu}^{j+1 \, | \, j} \in \mathbb{R}^K. \\
    \end{split}
\end{equation*}
Notice that the term inside the integral in \eqref{eq:EC_Value_F_Revisions_Int} is the density function of a multivariate Gaussian random variable with mean $\hat{\mu}^{j+1 \, | \, j}$ and covariance $\hat{\Omega}^{j+1 \, | \, j}$ so it integrates to \(1\). It follows that
\begin{equation}
\label{eq:EC_boundaries_VF_update}
    V(t_{j+1}^-, z, \bar{x},y) = U(z) \det\left(I_N  + \bar{\Omega}^{j+1 \, | \, j} C^{j+1}(t_{j+1})\right)^{-1/2} \exp\left(\epsilon(y, \bar{x})\right)
\end{equation}
where \(\epsilon(y, \bar{x})\) given by \eqref{eq:EC_epsilon_VF_update}.

We now derive the boundary conditions of the ODEs \eqref{eq:EC_Ricatti_update}--\eqref{eq:EC_barcj_update}. From  \eqref{eq:EC_boundaries_VF_update}
\begin{eqnarray*}
        \lefteqn{V(t_{j+1}^-, z, \bar{x},y)}\\ & =  & U(z) \exp\left(\frac{1}{2} \bar{x}^\top A^j(t_{j+1}) x + \bar{x}^\top \left(B^j(t_{j+1}) y + \bar{b}^j(t_{j+1})\right) -\frac{1}{2}y^{\top}C^j(t_{j+1}) y + y^{\top} \hat{c}^j(t_{j+1}) + \bar{c}^j(t_{j+1})  \right) \\&= & U(z) \det\left(I_N  + \bar{\Omega}^{j+1 \, | \, j} C^{j+1}(t_{j+1})\right)^{-1/2} \exp\left(\epsilon(y, \bar{x})\right).
\end{eqnarray*}
It follows that
\begin{equation}
\label{eq:EC_conditions_boundaries_update_1}
    \frac{1}{2} \bar{x}^\top A^j(t_{j+1}) \bar{x} + \bar{x}^\top \left(B^j(t_{j+1}) y + \bar{b}^j(t_{j+1})\right) -\frac{1}{2}y^{ \top}C^j(t_{j+1}) y + y^{ \top} \hat{c}^j(t_{j+1}) = \epsilon(y, \bar{x}) - \epsilon_0,
\end{equation}
and 
\begin{equation}
\label{eq:EC_conditions_boundaries_update_2}
\exp\left(\bar{c}^j(t_{j+1}) \right)  = \det\left(I_N  + \bar{\Omega}^{j+1 \, | \, j} C^{j+1}(t_{j+1})\right)^{-1/2} \exp\left(\epsilon_0 \right).
\end{equation}
It follows from \eqref{eq:EC_conditions_boundaries_update_1}  that
\begin{equation*}
\begin{split}
     A^j(t_{j+1}) &= -P^\top\bar{\beta}_0^{j\top} \left(C^{j+1}(t_{j+1})^{-1} +  \bar{\Omega}^{j+1 \, | \, j} \ \right)^{-1} \bar{\beta}_0^j P,\\
      B^j(t_{j+1}) &= P^\top\bar{\beta}_0^{j\top} \left(C^{j+1}(t_{j+1})^{-1} +  \bar{\Omega}^{j+1 \, | \, j} \ \right)^{-1} \bar{\beta}_0^j,\\
    C^j(t_{j+1}) &= \bar{\beta}_0^{j\top} \left(C^{j+1}(t_{j+1})^{-1} +  \bar{\Omega}^{j+1 \, | \, j} \ \right)^{-1} \bar{\beta}_0^j,\\
      \bar{b}^j(t_{j+1}) &= -P^\top \bar{\beta}_0^{j\top} \left(C^{j+1}(t_{j+1})^{-1} +  \bar{\Omega}^{j+1 \, | \, j}  \right)^{-1}\left(C^{j+1}(t_{j+1})^{-1} \hat{c}^{j+1}(t_{j+1}) - \bar{\alpha}_0^j\right),\\
      \hat{c}^j(t_{j+1}) &= \bar{\beta}_0^{j\top} \left(C^{j+1}(t_{j+1})^{-1} +  \bar{\Omega}^{j+1 \, | \, j}  \right)^{-1}\left(C^{j+1}(t_{j+1})^{-1} \hat{c}^{j+1}(t_{j+1}) - \bar{\alpha}_0^j\right).
      \end{split}
\end{equation*}
Thus, we have $A^j(t_{j+1}) =- P^\top C^j(t_{j+1}) P$, $B^j(t_{j+1}) =  C^j(t_{j+1})P$ and $\bar{b}^j(t_{j+1}) = -P^\top \hat{c}^j(t_{j+1})$.
Additionally, from \eqref{eq:EC_conditions_boundaries_update_2} we directly get
\begin{eqnarray*}
    \lefteqn{\bar{c}^j(t_{j+1})  = \bar{c}^{j+1}(t_{j+1}) + (\bar{\beta}_1^{j})^\top \hat{c}^{j+1}(t_{j+1})} \\ [5pt] &  & - \frac{1}{2} \bar{\beta}_1^{j\top} C^{j+1}(t_{j+1}) \bar{\beta}_1^{j} + \ln\left(\det(I_K + \bar{\Omega}^{j+1 \, | \, j} C^{j+1}(t_{j+1}))^{-\frac{1}{2}}\right)\\ [5pt]
            & & \quad + \frac{1}{2} \big(\hat{c}^{j+1}(t_{j+1}) -  C^{j+1}(t_{j+1})\bar{\beta}_1^{j}\big)^\top \Big((\bar{\Omega}^{j+1 \, | \, j})^{-1}  + C^{j+1}(t_{j+1})\Big)^{-1}\big(\hat{c}^{j+1}(t_{j+1}) -  C^{j+1}(t_{j+1})\bar{\beta}_1^{j}\big).
\end{eqnarray*}
This concludes the proof. \hfill $\square$

\paragraph{\textbf{Proof of Lemma \ref{lemma:EC_update_similarities}}}\
\label{proofLemma_EC_update_similarities}

From Lemma~\ref{lemma:EC_ContinuationValueFunction_Revision} the matrix 
\(A^j(t) \in \mathbb{R}^{N\times N}\) (\(t \in [t_j,t_{j+1})\), \(j \in \{1,\dots,M\}\)) and  \(B^j(t) \in \mathbb{R}^{N \times K}\) are solutions of 
\begin{equation}
\label{eq:EC_A_corr}
\begin{cases}
A^{j'}(t) + \dfrac{1-\gamma}{\gamma}\eta_t^j \Sigma\, \eta_t^j 
+ \dfrac{1}{\gamma}\Bigl(A^j(t)\Sigma\, \eta_t^j + \eta_t^j \Sigma\, A^j(t)\Bigr)
+ \dfrac{1}{\gamma} A^j(t) \Sigma\, A^j(t) = 0,\\[1mm]
A^j(t_{j+1}) = -P^\top C^j(t_{j+1})P,\\[1mm]
A^M(T)=\mathbf{0}_{N\times N},
\end{cases}
\end{equation}
and
\begin{equation}
\label{eq:EC_B_corr}
\begin{cases}
B^{j'}(t) + \dfrac{1}{\gamma}\Bigl(\eta_t^j + A^j(t)\Bigr)\Sigma\,B^j(t)
+\dfrac{1-\gamma}{\gamma}\eta_t^j \alpha_1^j(t)
+\dfrac{1}{\gamma} A^j(t) \alpha_1^j(t) = 0,\\[1mm]
B^j(t_{j+1}) = P^\top C^j(t_{j+1}),\\[1mm]
B^M(T)=\mathbf{0}_{N \times K},
\end{cases}
\end{equation}
respectively, 
where
\[
\alpha_1^j(t) = -\Sigma P^\top\Bigl((T-t)P\Sigma P^\top+\Omega^j\Bigr),
\]
and
\[
\eta_t^j = -P^\top\Bigl((T-t)P\Sigma P^\top + \Omega^j\Bigr)P
= \Sigma^{-1}\,\alpha_1^j(t)P.
\]

By direct substitution, it is easy to show that 
\begin{eqnarray}
A^j(t) = -B^j(t)P,\quad \text{for } t \in [t_j,t_{j+1}],\; j\in\{0,\dots,M\}.
\label{eq:tempA}
\end{eqnarray}
is a solution of \eqref{eq:EC_A_corr}.
It follows that \eqref{eq:EC_B_corr} can be rewritten as
\begin{equation}
\label{eq:EC_B_corr_2}
\begin{cases}
B^{j'}(t) + \dfrac{1}{\gamma}\Bigl(\eta_t^j - B^j(t)P\Bigr)\Sigma\,B^j(t)
+\dfrac{1-\gamma}{\gamma}\eta_t^j \alpha_1^j(t)
-\dfrac{1}{\gamma}B^j(t)P\alpha_1^j(t)=0,\\[1mm]
B^j(t_{j+1}) = P^\top C^j(t_{j+1}),\\[1mm]
B^M(T)=\mathbf{0}_{N \times K}.
\end{cases}
\end{equation}

Next, from Lemma~\ref{lemma:EC_ContinuationValueFunction_Revision} we have that \(C^j(t)\in\mathbb{R}^{K\times K}\) is the solution of
\begin{equation}
\label{eq:EC_C_corr}
C^{j'}(t) - \frac{1}{\gamma}\Bigl(B^j(t)+\Sigma^{-1}\alpha_1^j(t)\Bigr)^\top\Sigma\Bigl(B^j(t)+\Sigma^{-1}\alpha_1^j(t)\Bigr)
+\alpha_1^j(t)^\top\Sigma^{-1}\alpha_1^j(t)=0
\end{equation}
with boundary conditions given by~\eqref{terminal:C} and 
\[
C^M(T) = \mathbf{0}_{K \times K}.
\]
It is easy to show that 
\[
B^j(t)=P^\top C^j(t),\quad t \in [t_j,t_{j+1}],\; j \in \{0,\dots,M\}.
\]
is a solution of \eqref{eq:EC_B_corr_2} and hence by \eqref{eq:tempA}, that $A^j(t)=-P^\top, C^j(t) P$.

Similarly, recall from Lemma~\ref{lemma:EC_ContinuationValueFunction_Revision} that 
\begin{equation}
\label{eq:EC_barbj_corr}
\begin{cases}
\bar{b}^{j'}(t)+\frac{1}{\gamma}\Bigl(\eta_t^j+A^j(t)\Bigr)\Bigl(\Sigma\, \bar{b}^j(t)+\alpha_0^j(t)-r_f\,\mathbf{1}_N\Bigr)
-\eta_t^j\Bigl(\alpha_0^j(t)-r_f\,\mathbf{1}_N\Bigr)\\[1mm]
\quad\quad\quad\quad+A^j(t)\Bigl(r_f\,\mathbf{1}_N-\frac{1}{2}\diag(\Sigma)\Bigr)=0,\\[1mm]
\bar{b}^j(t_{j+1})=-P^\top \hat{c}^j(t_{j+1}),\\[1mm]
\bar{b}^M(T)=0,
\end{cases}
\end{equation}
with 
\[
\alpha_0^j(t)= \mu-(T-t_j)\beta_1^jP\mu^x-(t-t_j)\Sigma\,\eta_t^j\Bigl(\mu^x-(T-t_j)\beta_1^jP\mu^x\Bigr)\in \mathbb{R}^N.
\]
If \(\hat{c}^j(t)\) is the solution to
\begin{equation}
\label{eq:EC_hatcj_corr}
\begin{split}
\hat{c}^{j'}(t) &+\dfrac{1-\gamma}{\gamma}\,\alpha_1^j(t)\Sigma^{-1}\Bigl(\alpha_0^j(t)-r_f\,\mathbf{1}_N\Bigr)
+\dfrac{1}{\gamma}\,\alpha_1^j(t)^\top \bar{b}^j(t)
+ B^j(t)^\top \Bigl(r_f\,\mathbf{1}_N-\frac{1}{2}\diag(\Sigma)\Bigr)\\[1mm]
&\quad+\dfrac{1}{\gamma}\,B^j(t)^\top\Bigl(\Sigma\,\bar{b}^j(t)+\alpha_0^j(t)-r_f\,\mathbf{1}_N\Bigr)= 0,
\end{split}
\end{equation}
with terminal condition \(\hat{c}^M(T)=0\), then 
\[
\bar{b}^j(t)=-P^\top \hat{c}^j(t),\quad t \in [t_j,t_{j+1}],\; j \in \{0,\dots,M\}.
\]
is a solution to \eqref{eq:EC_barbj_corr}.

This completes the proof. \hfill $\square$

\paragraph{\textbf{Proof of Lemma \ref{Lemma:EC_Solutions_revision}.}}
\label{proof_lemma:EC_Solutions_revision}

Recall from Lemma~\ref{lemma:EC_update_similarities} that \(C^j(t)\in\mathbb{R}^{K\times K}\) \(j\in\{0,\ldots,M\}\) satisfies the Riccati equation 
\begin{equation}
\label{eq:EC_Riccati_Cj_Withconditions}
\begin{cases}
C^{j\prime}(t) - \dfrac{1-\gamma}{\gamma}\,\bar{\eta}_t^j P\Sigma P^\top \bar{\eta}_t^j 
+ \dfrac{1}{\gamma}\Bigl(C^j(t)P\Sigma P^\top \bar{\eta}_t^j + \bar{\eta}_t^j P\Sigma P^\top C^j(t)\Bigr)
- \dfrac{1}{\gamma}C^j(t)P\Sigma P^\top C^j(t)=0,\\[1mm]
C^j(t_{j+1}) = \bar{\beta}_0^{j\top}\Bigl( C^{j+1}(t_{j+1})^{-1}+\bar{\Omega}^{j+1\,|\,j}\Bigr)^{-1}\bar{\beta}_0^j,
\end{cases}
\end{equation}
where
\[
\bar{\eta}_t^j = -\Bigl((T-t)P\Sigma P^\top+\Omega^j\Bigr)^{-1},
\]
\[
\bar{\beta}_0^j = I_K - (\Omega^j-\Omega^{j+1})\Bigl((T-t_{j+1})P\Sigma P^\top+\Omega^j\Bigr)^{-1},
\]
and
\[
\bar{\Omega}^{j+1\,|\,j} = (\Omega^j-\Omega^{j+1})\Bigl((T-t_{j+1})P\Sigma P^\top+\Omega^j\Bigr)^{-1}\Bigl((T-t_{j+1})P\Sigma P^\top+\Omega^{j+1}\Bigr).
\]
The terminal condition is given by \(C^M(T)=\mathbf{0}_{K\times K}\).

We first show that 
\begin{equation}
\label{eq:C_Q_eta}
C^j(t)
=\bar{\eta}_t^j+\left(\frac{T-t}{\gamma}P\Sigma P^\top+\Omega^j\right)^{-1}
\end{equation}
is a solution of \eqref{eq:EC_Riccati_Cj_Withconditions}.

To begin, observe that $\bar{\eta}_t^j$ satisfies the ODE
\begin{eqnarray*}
    \frac{d \bar{\eta}_t^j}{d t} = - \bar{\eta}_t^j P\Sigma P^\top \bar{\eta}_t^j.
\end{eqnarray*}
It follows that
\begin{equation*}
        C^j(t) = \bar{\eta}_t^j + Q^j(t)
    \end{equation*}
is the solution of \eqref{eq:EC_Riccati_Cj_Withconditions} if a symmetric matrix  $Q^j(t) \in \mathbb{R}^{K \times K}$ for $j=\{0, \cdots, M\}$  such that 
\begin{equation}
    \label{eq:EC_Q_Riccati_update}
        \begin{cases}
            Q^{j'}(t) - \dfrac{1}{\gamma} Q^j(t) P \Sigma P^\top Q^j(t) = 0,\\
            Q^j(t_{j+1}) = - \eta_{t_{j+1}}^j + \bar{\beta}_0^{j\top} \left(C^{j+1}(t_{j+1})^{-1} +  \bar{\Omega}^{j+1 \, | \, j} \ \right)^{-1} \bar{\beta}_0^j
        \end{cases}
    \end{equation}
     with terminal condition \(Q^M(T) = (\Omega^M)^{-1}\) can be found. To establish  \eqref{eq:C_Q_eta} we show that 
     \begin{equation}
        Q^j(t) = \left(\frac{T-t}{\gamma} P \Sigma P^\top + \Omega^j \right)^{-1}.
        \label{eq:Qj_temp}
    \end{equation}

Since
\begin{equation*}
    \begin{split}
        \frac{d}{d t}Q^j(t) &= \frac{d}{d t}\left(\frac{T-t}{\gamma} P \Sigma P^\top + \Omega^j \right)^{-1} \\&= -\left(\frac{T-t}{\gamma} P \Sigma P^\top + \Omega^j\right)^{-1} \left( -\frac{1}{\gamma} P\Sigma P^\top \right)\left(\frac{T-t}{\gamma} P \Sigma P^\top + \Omega^j \right)^{-1}\\
        &=\frac{1}{\gamma} Q^j(t) P\Sigma P^\top Q^j(t),
        \end{split}
    \end{equation*}
    \eqref{eq:Qj_temp} satisfies the differential equation in \eqref{eq:EC_Q_Riccati_update}. We now show that \eqref{eq:Qj_temp} satisfies the required terminal and boundary conditions.

We can see from \eqref{eq:Qj_temp} that $Q^M(T)=(\Omega^j)^{-1}$ so the terminal condition is satisfied so \eqref{eq:Qj_temp} is the solution of \eqref{eq:EC_Q_Riccati_update}  and $C^M(t)=\bar{\eta}^M_t + Q^M(t)$ on the interval $[t_M, T]$. 

Suppose there is a $j\in\{0, \cdots, M-1\}$ such that $Q^{j+1}(t)$ is the solution of \eqref{eq:EC_Q_Riccati_update} on  $[t_{j+1}, t_{j+2}]$ and hence 
\begin{eqnarray}
C^{j+1}(t)=\bar{\eta}^{j+1}_t + Q^{j+1}(t), \; t\in [t_{j+1}, t_{j+2}].
\label{eq:inductionj+1}
\end{eqnarray}
Using \eqref{eq:inductionj+1}, the definitions of $\bar{\eta}^j_t$, $\bar{\Omega}^{j+1|j}$ and $\bar{\beta}^j_0$, and the Woodbury identity
\begin{eqnarray*}
    - \bar{\eta}_{t_{j+1}}^j + \bar{\beta}_0^{j\top} \left(C^{j+1}(t_{j+1})^{-1} +  \bar{\Omega}^{j+1 \, | \, j} \ \right)^{-1} \bar{\beta}_0^j = \left(\frac{T-t_{j+1}}{\gamma} P \Sigma P^\top + \Omega^j \right)^{-1} = Q^j(t_{j+1})
\end{eqnarray*}
so \eqref{eq:Qj_temp} is the solution of \eqref{eq:EC_Q_Riccati_update} and $C^j(t)+\bar{\eta}^j_t+Q^{j}(t)$ on $[t_j, t_{j+1}]$. It follows from induction that \eqref{eq:Qj_temp} is the solution of \eqref{eq:EC_Q_Riccati_update}.

Finally, it follows from the Woodbury identity that
\[
C^j(t)=\bar{M}^j(t)P^\top\bar{\eta}_t^j,
\]
where
\begin{equation*}
\bar{M}^j(t)=- (\gamma-1)(T-t)(\Omega^j)^{-1}P\Bigl(\gamma\Sigma^{-1}+(T-t)P^\top(\Omega^j)^{-1}P\Bigr)^{-1}
\end{equation*}
which completes the first half of the proof of Lemma \ref{Lemma:EC_Solutions_revision}.

For the second half, recall from Lemma~\ref{lemma:EC_update_similarities} that for 
\(t\in[t_j,t_{j+1})\), \(\hat{c}^j(t)\) is the solution of
  \begin{equation}
\label{eq:EC_hatcj_update_withconditions}
\begin{cases}
        &\begin{split}
        \hat{c}^{j '}(t) & -\frac{1-\gamma}{\gamma} \bar{\eta}_t^j P(\alpha_0^j(t) - r_f \mathbf{1}_N) + \frac{1}{\gamma}\bar{\eta}_t^j P\Sigma P^\top \hat{c}^j(t) - \frac{1}{\gamma}C^j(t) P\Sigma P^\top \hat{c}^j(t)+ \frac{1-\gamma}{\gamma} C^j(t) P (\alpha_0^j(t) - r_f) \\ &+ C^j(t) P (r_f - \frac{1}{2}\diag(\Sigma)) = 0,
    \end{split}\\
    &\hat{c}^j(t_{j+1}) = \bar{\beta}_0^{j\top} \left(C^{j+1}(t_{j+1})^{-1} +  \bar{\Omega}^{j+1 \, | \, j}  \right)^{-1}\left(C^{j+1}(t_{j+1})^{-1} \hat{c}^{j+1}(t_{j+1}) - \bar{\alpha}_0^j\right),
\end{cases}
\end{equation}
where 
\[
C^j(t)=\bar{M}^j(t)\,P^\top \bar{\eta}_t^j,
\]
and
\begin{equation*}
\begin{split}
\alpha_0^j(t) &= \mu+(T-t)\,\Sigma\, P^\top \bar{\eta}_t^j\, P\mu^x,\\[1mm]
\bar{\alpha}_0^j &= - (T-t_{j+1})\,(\Omega^j - \Omega^{j+1})\,\bar{\eta}_{t_{j+1}}^j\, P\mu^x,\\
\bar{M}^j(t)&=- (\gamma-1)(T-t)(\Omega^j)^{-1}P\Bigl(\gamma\Sigma^{-1}+(T-t)P^\top (\Omega^j)^{-1}P\Bigr)^{-1}.
\end{split}
\end{equation*}
with the terminal condition \(\hat{c}^M(T)=0\). It can be shown that
\begin{equation*}
\hat{c}^j(t)=\bar{M}^j(t)\Sigma^{-1}(\mu-r_f\mathbf{1}_N)
+(T-t)\,C^j(t)\,P\mu^x
=\bar{M}^j(t)\Bigl(\Sigma^{-1}(\mu-r_f\mathbf{1}_N)
+(T-t)P^\top \bar{\eta}_t^j\,P\mu^x\Bigr),
\end{equation*}
is the solution of  \eqref{eq:EC_hatcj_update_withconditions} by a direct substitution of this expression into the differential equation and boundary conditions.

\hfill \(\square\)

\subsection{Proofs: Results in \ref{Sec:EC_QuarterlyViews}}

\paragraph{\textbf{Proof of Lemma \ref{lemma:EC_ContinuationValueFunction_quarterly}.}}\mbox{}
\label{proof_lemma:EC_ContinuationValueFunction_quarterly}
By definition, the value function at time \( T_{j+1}^- \), which is immediately before the \((j+1)^{th}\) view, is given by
\[
   V\bigl(T_{j+1}^-, z, \bar{x}, \bar{y}\bigr) 
   \;=\; 
   \sup_{\pi \in \mathcal{A}} \,
   \mathbb{E}\Bigl[
     U\bigl(Z(T)\bigr)
     \, | \,
     \bar{X}^y\bigl(T_{j+1}^-\bigr)=\bar{x},\;
     Z(T_{j+1}^-)=z, I(T_{j+1}^-) = \bar{y}
   \Bigr],
\]
it follows from the Principle of Optimality that the value function satisfies
   \begin{equation}
   \label{eq:EC_short_term_PoO}
       \begin{split}
           V\bigl(T_{j+1}^-, z, \bar{x}, \bar{y}\bigr)
   &=
   \mathbb{E}_{\,\bar{Y}^{j+1}}\Bigl[
      V\bigl(T_{j+1}, z, 0, \bar{Y}^{j+1}\bigr)
      \, | \,
      \bar{X}^y\bigl(T_{j+1}^-\bigr)=\bar{x},\;
      I(T_{j+1}^-)=\bar{y}
   \Bigr]\\
   &= \mathbb{E}_{\,\bar{Y}^{j+1}}\Bigl[
      V\bigl(T_{j+1}, z, 0, \bar{Y}^{j+1}\bigr)\Bigr],
       \end{split}
   \end{equation}
where the second equality follows from the fact that the view \(\bar{Y}^{j+1}(T_{j+1}, T_{j+2})\) is independent of the view \(\bar{Y}^{j}(T_j, T_{j+1})\)  and log-returns \(\bar{X}(T_{j+1}^-)\) over the previous horizon \([T_j, T_{j+1})\).

Furthermore, we can easily verify using \eqref{eq:Y-bar} that the random variable $\bar{Y}^{j+1}(T_{j+1}, T_{j+2})$ is Gaussian with mean and variance 
\begin{equation*}
    \begin{split}
        &\bar{\alpha}^{j+1} = (T_{j+2}-T_{j+1}) P \mu^x,\\
        &\bar{\Omega}^{j+1} =(T_{j+2} - T_{j+1}) P \Sigma P^\top + \Omega^{j+1, 0}.
    \end{split}
\end{equation*}
Now we derive the explicit characterization of the expectation in \eqref{eq:EC_short_term_PoO}. Recall that
\begin{equation}
\label{eq:EC_V(t,z,0)_quarterly}
\begin{split}
        V(T_{j+1},z,0,\bar{y}) &= U(z) \exp\left(g^j(T_{j+1},0, \bar{y})\right)\\
        &= U(z) \exp\left( -\frac{1}{2}\bar{y}^{ \top}C^j(T_{j+1}) \bar{y} + \bar{y}^{ \top} \hat{c}^j(T_{j+1}) + \bar{c}^j(T_{j+1}) \right).
\end{split}
\end{equation}
It follows that 
\begin{eqnarray*}
        \lefteqn{V(T_{j+1}^-, z, \bar{x}, \bar{y}) =\mathbb{E}_{\,\bar{Y}^{j+1}}\Bigl[
      V\bigl(T_{j+1}, z, 0, \bar{Y}^{j+1}\bigr)
   \Bigr]} \\
   &= & \int (2\pi)^{-K/2} \det(\bar{\Omega}^{j+1})^{-1/2} \exp\left(-\frac{1}{2}(\bar{y}^{j+1} -\bar{\mu}^{j+1})^\top (\bar{\Omega}^{j+1})^{-1}(\bar{y}^{j+1} -\bar{\mu}^{j+1}) \right) V(T_{j+1},z,0,\bar{y}^{j+1}) d\bar{y}^{j+1}\\
& = &U(z) \int (2\pi)^{-K/2} \det(\bar{\Omega}^{j+1})^{-1/2} \exp\Bigr(
-\frac{1}{2}(\bar{y}^{j+1} -\bar{\mu}^{j+1})^\top (\bar{\Omega}^{j+1})^{-1}(\bar{y}^{j+1} -\bar{\mu}^{j+1})\\
& &-\frac{1}{2}{(\bar{y}^{j+1})}^\top C^j(T_{j+1}) \bar{y}^{j+1} + {(\bar{y}^{j+1})}^\top \hat{c}^j(T_{j+1}) + \bar{c}^j(T_{j+1})\Bigr) d\bar{y}^{j+1}
\end{eqnarray*}
where the last equality follows from \eqref{eq:EC_V(t,z,0)_quarterly}. By rearranging the terms in the exponential, we can write the value function
\begin{eqnarray}
\label{eq:EC_short_Term_int}
        \lefteqn{V(T_{j+1}^-, z, \bar{x}, \bar{y}) = U(z) \det\left(I_N  + \bar{\Omega}^{j+1} C^{j+1}(T_{j+1})\right)^{-1/2} \exp\left(\epsilon_0\right)} \\ [5pt]
        &  &\times \int (2\pi)^{-K/2} \det(\hat{\Omega}^{j+1})^{-1/2} \exp\left(-\frac{1}{2}(\bar{y}^{j+1} -\hat{\mu}^{j+1})^\top (\hat{\Omega}^{j+1})^{-1}(\bar{y}^{j+1} -\hat{\mu}^{j+1}) \right) d\bar{y}^{j+1} \nonumber
\end{eqnarray}
where 
\begin{equation*}
    \begin{split}
                \epsilon_0 = &\frac{1}{2} \left(\hat{c}^{j+1}(T_{j+1}) - C^{j+1}(T_{j+1})\bar{\alpha}_0^{j+1}\right)^\top \left(C^{j+1}(T_{j+1})^{-1} +  \bar{\Omega}^{j+1} \ \right)^{-1}  \left(\hat{c}^{j+1}(T_{j+1}) - C^{j+1}(T_{j+1})\bar{\alpha}_0^{j+1}\right)^\top\\
                &+ \frac{1}{2} \bar{\alpha}_0^{j+1\top} C^{j+1}(T_{j+1}) \bar{\alpha}_0^{j+1} + \bar{\alpha}_0^{j+1\top} \left(\hat{c}^{j+1}(T_{j+1}) - C^{j+1}(T_{j+1})\bar{\alpha}_0^{j+1}\right) + \bar{c}^{j+1}(T_{j+1})
    \end{split}
    \end{equation*}
    is a constant scalar,
    and
\begin{equation*}
    \begin{split}
        &\hat{\Omega}^{j+1} = \left((\bar{\Omega}^{j+1})^{-1} + C^{j+1}(T_{j+1})\right)^{-1} \in \mathbb{R}^{K \times K},\\
        &\hat{\mu}^{j+1} = \hat{\Omega}^{j+1} \left(\hat{c}^{j+1}(T_{j+1}) - C^{j+1}(T_{j+1}) \alpha_0^j \right) + \bar{\mu}^{j+1} \in \mathbb{R}^K. \\
    \end{split}
\end{equation*}
Notice that the term inside the integral in \eqref{eq:EC_short_Term_int} is the density function of a multivariate Gaussian random variable with mean $\hat{\mu}^{j+1}$ and covariance matrix $\hat{\Omega}^{j+1}$, so it integrates to \(1\). It follows that
\begin{equation*}
\begin{split}
    V(T_{j+1}^-, z, \bar{x},\bar{y})&= U(z) \exp\left(g^j(T_{j+1},\bar{x},\bar{y}) \right)\\&= U(z) \det\left(I_N  + \bar{\Omega}^{j+1} C^{j+1}(T_{j+1})\right)^{-1/2} \exp\left(\epsilon_0\right)
    \end{split}
\end{equation*}
which gives 
\begin{equation}
\label{eq:EC_conditions_boundaries_quarterly_1}
    \frac{1}{2} \bar{x}^\top A^j(T_{j+1}) \bar{x} + \bar{x}^\top \left(B^j(T_{j+1}) \bar{y} + \bar{b}^j(T_{j+1})\right) -\frac{1}{2}\bar{y}^{\top}C^j(T_{j+1}) \bar{y} + \bar{y}^{ \top} \hat{c}^j(T_{j+1}) = 0,
\end{equation}
and 
\begin{equation}
\label{eq:EC_conditions_boundaries_quarterly_2}
\exp\left(\bar{c}^j(T_{j+1}) \right)  = \det\left(I_N  + \bar{\Omega}^{j+1} C^{j+1}(T_{j+1})\right)^{-1/2} \exp\left(\epsilon_0 \right).
\end{equation}
From \eqref{eq:EC_conditions_boundaries_quarterly_1}, it follows by separation of variables that
\begin{eqnarray*}A^j(T_{j+1}) = \mathbf{0}_{N \times N}, \,
         B^j(T_{j+1}) = \mathbf{0}_{N \times K},\,
         C^j(T_{j+1}) = \mathbf{0}_{K \times K}, \,
         \bar{b}^j(T_{j+1}) = \mathbf{0}_{N},\,
         \hat{c}^j(T_{j+1}) = \mathbf{0}_{K},\end{eqnarray*} 
 and from \eqref{eq:EC_conditions_boundaries_quarterly_2}, we directly get
\begin{equation*}
    \begin{split}
        \bar{c}^j(T_{j+1}) & =  \ln\left(\det(I_K + \bar{\Omega}^{j+1} C^{j+1}(T_{j+1}))^{-\frac{1}{2}}\right)\\ 
            & \quad + \frac{1}{2} \left(\hat{c}^{j+1}(T_{j+1}) - C^{j+1}(T_{j+1})\bar{\alpha}_0^{j+1}\right)^\top \left(C^{j+1}(T_{j+1})^{-1} +  \bar{\Omega}^{j+1} \ \right)^{-1}  \left(\hat{c}^{j+1}(T_{j+1}) - C^{j+1}(T_{j+1})\bar{\alpha}_0^{j+1}\right)^\top\\
                &\quad + \frac{1}{2} \bar{\alpha}_0^{j+1\top} C^{j+1}(T_{j+1}) \bar{\alpha}_0^{j+1} + \bar{\alpha}_0^{j+1\top} \left(\hat{c}^{j+1}(T_{j+1}) - C^{j+1}(T_{j+1})\bar{\alpha}_0^{j+1}\right) + \bar{c}^{j+1}(T_{j+1}).
    \end{split}
\end{equation*}

This concludes the proof. \hfill $\square$
    \paragraph{\textbf{Proof of Lemma \ref{Lemma:EC_Solutions_quarterly}.}}\mbox{}  
    \label{proof_lemma:EC_Solutions_quarterly}

From Lemma \ref{Lemma:EC_Solutions_quarterly}, \(C^j(t)\) satisfies
\[
    \begin{cases}
        C^{j \prime}(t) - \dfrac{1 - \gamma}{\gamma} \bar{\eta}_t^j P \Sigma P^\top \bar{\eta}_t^j + 
        \dfrac{1}{\gamma} \left( C^j(t) P \Sigma P^\top \bar{\eta}_t^j + \bar{\eta}_t^j P \Sigma P^\top C^j(t) \right) \\
        \qquad - \dfrac{1}{\gamma} C^j(t) P \Sigma P^\top C^j(t) = 0, \\
        C^j(T_{j+1}) = 0,
    \end{cases}
\]

and \(\hat{c}^j(t) \in \mathbb{R}^K\) satisfies
\[
    \begin{cases}
        \begin{aligned}
            \hat{c}^{j \prime}(t) & - \frac{1 - \gamma}{\gamma} \bar{\eta}_t^j P \big( \alpha_0^j(t) - r_f \mathbf{1}_N \big) 
            + \frac{1}{\gamma} \bar{\eta}_t^j P \Sigma P^\top \hat{c}^j(t) 
            - \frac{1}{\gamma} C^j(t) P \Sigma P^\top \hat{c}^j(t) \\
            & + \frac{1 - \gamma}{\gamma} C^j(t) P \big( \alpha_0^j(t) - r_f \mathbf{1}_N \big) 
            + C^j(t) P \left( r_f - \frac{1}{2} \diag(\Sigma) \right) = 0,
        \end{aligned} \\
        \hat{c}^j(T_{j+1}) = 0,
    \end{cases}
\]
where  
\begin{align*}
        \alpha_0^j(t) &= \mu + (T_{j+1} - t) \Sigma P^\top \bar{\eta}_t^j P \mu^x\\
    \bar{\eta}^j_t &= - \left((T_{j+1} - t) P \Sigma P^\top + \Omega^{j,0}\right)^{-1}.
\end{align*}

To avoid redundancy, note that this setting is a special case of Lemma \ref{Lemma:EC_Solutions_revision}. Thus, the solutions to the above ODEs follow directly from Lemma \ref{Lemma:EC_Solutions_revision} by replacing \( T \) with \( T_{j+1} \), \( \Omega^M \) with \( \Omega^{j,0} \), and \( y^M \) with \( \bar{y}^j \). This completes the proof. \hfill \(\square\)

\newpage

\section{Extension -- Views over Different Time Horizons}
\label{Sec:EC_ViewsDifferentHorizons}
In this part, we explore a scenario where the expert gives forward-looking views with varying horizons. For instance, for different portfolios or assets, the expert might provide forecasts over different time intervals.
\subsection{Views Model}
At time $t=0$, the investor receives a vector of $K$ views on future log-returns, where each view $j \in [K]$ is a noisy observation of the log-returns vector $X(t)$ at time $t = T_j$. Specifically, conditional on $X(T_j)$, we assume
\begin{equation}
\label{eq:ViewsDifferentHorizons}
 Y_j(0,T_j) \,|\, X(T_j) = p_j^\top X(T_j) + \sqrt{T_j} \epsilon_j \sim \mathcal{N}\big(p_j^\top X(T_j), T_j \omega_{jj}\big), \; j \in [K],
\end{equation}
where $p_j \in \mathbb{R}^N$ is a linear mapping from log-returns to the $j$-th view, and $\omega_{jj} > 0$ is the view's noise variance. 

\paragraph{Non-expired views.}
Without loss of generality we  order the views by their respective time horizons  ($T_i \leq T_j \, \, \, \text{for} \, \, i \leq j$), and denote $T_0 = 0$. Thus, on the \(j\)-th interval $[T_{j-1},\,T_j]$, we consider the non-expired views $j,j+1,\dots,K$.  Let
\begin{eqnarray*}
    Y^j(0,T_{j:K}) = \big(Y_{j}(0,T_{j}), \dots, Y_K(0,T_K)\big)^\top
\end{eqnarray*}
be the vector of these non-expired views during the \(j\)-th interval \([T_{j-1}, T_j]\), with $T_{j:K} := (T_j,\dots,T_K)^\top\in\mathbb{R}^{K-j+1}$ is the vector of their respective horizons.

Similarly, we collect the associated noise terms into 
\begin{equation*}
    \epsilon^j = \begin{pmatrix}
        \epsilon_{j} \\ \vdots \\ \epsilon_K
    \end{pmatrix}
    \sim \mathcal{N}\big(0,\Omega^j\big),
\end{equation*}
where $\Omega^j \in \mathbb{R}^{(K-j+1)\times(K-j+1)}$ is positive-definite.  Its entries $\omega_{ik}$, for $i,k \in \{j,\dots,K\}$, represent the covariances among the noise terms $\epsilon_i$ and $\epsilon_k$. Consequently, on each interval $[T_{j-1}, T_j]$, the random vector $\epsilon^j$ represents the noise of the remaining (non-expired) views $Y^j(0,T^j)$.

\begin{figure}[!ht]
\centering
\begin{tikzpicture}
[
roundnode/.style={circle, draw=black!80, fill=gray!5, very thick, minimum size=7mm},
squarednode/.style={rectangle, draw=black!80, fill=gray!5, very thick, minimum size=5mm},
]
\node[squarednode]      (observation)                              {$Y_1(0,T_1)$};
\node[roundnode]        (rT)       [above=of observation] { $X(T_1)$};
\node[]        (dots2)       [left=of rT] {\dots};
\node[squarednode]        (r0)       [left=of dots2] { $X(0)$};
\node[]        (dots3)       [right=of rT] {\dots};
\node[roundnode]        (rT2)       [right=of dots3] { $X(T_2)$};
\node[]        (dots4)       [right=of rT2] {\dots};
\node[roundnode]        (rT3)       [right=of dots4] { $X(T_K)$};
\node[squarednode]      (observation2)        [below=of rT2]                      {$Y_2(0,T_2)$};
\node[squarednode]      (observation3)        [below=of rT3]                      {$Y_K(0,T_K)$};

\draw[->] (rT.south) -- (observation.north);
\draw[->] (rT2.south) -- (observation2.north);
\draw[->] (rT3.south) -- (observation3.north);
\draw[<-] (rT.west) -- (dots2.east);
\draw[<-] (dots2.west) -- (r0.east);
\draw[<-] (dots3.west) -- (rT.east);
\draw[<-] (dots4.west) -- (rT2.east);
\draw[->] (dots3.east) -- (rT2.west);
\draw[->] (dots4.east) -- (rT3.west);

\end{tikzpicture}
\caption{Bayesian network of the Dynamic Black-Litterman model with Different Views Horizons. \textmd{The figure shows a discrete time version of the problem where $t = 0, \dots, T_K$, and the noisy views $Y(0,T_j)$ of the log-return $X(T_j)$ for $j \in [K]$ are revealed at $t = 0$}}\label{fig:figure6}
\end{figure}

Figure \ref{fig:figure6} illustrates a Bayesian network for the discrete-time version of our setting, in which the views\(\{Y(0,T_j)\}_{j=1,\dots,K}\) are revealed at time \(t=0\).
From the graph, notice that \(Y_2(0,T_2)\) is not only an observation of the log-return vector \(X(T_2)\) but also provides information about \(X(T_1)\).  The reason is that \(X(T_2)\) itself can be regarded as a ``noisy observation'' of \(X(T_1)\).  More generally, extending this argument shows that each view \(Y_j(0,T_j)\) can be viewed as a noisy measurement of \(X(T_1)\).

Consequently, if we represent each vector \( Y^j(0,T_{j:K})  = \bigl(Y_j(0,T_j),\dots,Y_K(0,T_K)\bigr)^\top\)
as observations tied to \(X(T_j)\), then Proposition~\ref{proposition1} applies directly to derive the posterior dynamics of the asset returns.

\subsection{Market Dynamics Under Expert Views}
The following Proposition shows how views from different time horizons can be transformed into views regarding the same time horizon.

\begin{proposition}
\label{proposition_DifferentHorizons}
    Consider a stock price following equation \eqref{eq8}, and expert views following \eqref{eq:ViewsDifferentHorizons}. For $j \in [K]$, define the transformation
    \begin{equation}
    \label{eq:barY_ViewsDifferentHorizons}
        \bar{Y}^j(0,T_j) = Y^j(0,T_{j:K}) - \bar{\mu}^j(T_j,T_{j:K})
    \end{equation} 
    where $\bar{\mu}^j(T_j,T_{j:K}) \in \mathbb{R}^{K-j+1}$ with elements
    \begin{equation*}
        \bar{\mu}_i^j(T_j,T_{j:K}) = (T_i - T_j) p_i^\top \mu^x, \, \, \, \text{for} \, \, i \in \{j, \dots, K\}.
    \end{equation*}
    Then, conditional on the true realization of $X(T_j)$, the views vector $\bar{Y}^j(0,T_j)$ is Gaussian with
    \begin{equation*}
        \bar{Y}^j(0,T_j)|X(T_j) = P_{j:K} X(T_j) + \bar{\epsilon}^j \sim \mathcal{N}\big(P_{j:K}X(T_j) , T_j \bar{\Omega}^j\big),
    \end{equation*}
    where \[P_{j:K} = (p_j, \dots, p_K)^\top \in \mathbb{R}^{(K - j + 1) \times N},\] and $\bar{\Omega}^j$ is positive definite with 
    \begin{equation*}
        \bar{\Omega}^j = P_{j:K}\bar{\Omega}^{jW}(P_{j:K})^{\top} + \bar{\Omega}^{jV}
    \end{equation*}
    where 
    \begin{equation*}
        \begin{cases}
            \bar{\Omega}^{jV}_{ik} = \dfrac{\sqrt{T_iT_k}}{T_j}\Omega_{ik} \\ 
            \bar{\Omega}^{jW}_{ik} = \dfrac{1}{T_j}\min\{T_i - T_j, T_k - T_j\}\Sigma_{ik}.
        \end{cases}
    \end{equation*}
    for 
    $i ,k \in \{1, \dots, K-j+1\}$.
    
Furthermore
\begin{equation*}
    X(t) \, | \, Y^j(0,T_{j:K}) = X(t) \, | \, \bar{Y}^j(0,T_j), \quad \text{for} \, \, \, t \in [T_{j-1},T_j], \, \, j \in [K].
\end{equation*}

\end{proposition}

Proposition \ref{proposition_DifferentHorizons}  shows how a vector of views, each with distinct time horizons, can be transformed into a vector of views with a single time horizon $T_j$ by leveraging the graphical structure of the model. Once transformed, Proposition \ref{proposition1} can be applied to derive the conditional dynamics.

\begin{corollary}
\label{corollary:EC1}
Suppose that the price process satisfies \eqref{eq8} and expert views $Y(0,T)$ satisfy \eqref{eq:ViewsDifferentHorizons}. Assume that $PL_j \neq 0$ for $j \in [N]$. Conditional on $Y(0,T) = y$, the log-returns $X(t)$ satisfy
\begin{equation}
\label{eq:dX_ViewsWithDifferentHorizons}
\begin{split}
        dX^y(t) =&   \bigg(\mu^x + \beta_1^j \big(\bar{y}^j - T_j P_{j:K} \mu^x \big) + \beta_2^j(t) \big(\mathbb{E}[X^y(t)] - X^y(t) \big)\bigg) dt + dW^y(t)
\end{split}
\end{equation}
where $\bar{y}^j$ is sampled from the transformed views vector \eqref{eq:barY_ViewsDifferentHorizons}, with
\begin{eqnarray*}
    \beta_1^j & = & \frac{1}{T_j}\Sigma (P_{j:K})^{\top} (P_{j:K} \Sigma (P_{j:K})^{\top}+ \bar{\Omega}^j)^{-1} \in  \mathbb{R}^{N \times K},\\ [8pt]
    \beta_2^j(t) & = & \Sigma (P_{j:K})^{\top} \left( (T_j-t) P_{j:K} \Sigma (P_{j:K})^{\top} + \bar{\Omega}^j\right)^{-1} P_{j:K} \in \mathbb{R}^{N \times N},
\end{eqnarray*}
and
\begin{eqnarray*}
   \mathbb{E}[X^y(t)]  &=& t \left(\mu^x + \beta_1^j \big(\bar{y}^j -T_j P_{j:K} \mu^x \big)\right)
\end{eqnarray*}
is the expected log-return over the horizon $[0,t]$ given $Y(0,T) = y$.
$W^y(t) \sim \mathcal{N}\big(0,t\Sigma\big)$ is a $N-$dimensional Brownian motion adapted to the filtration $\mathcal{F}_t^Y$.
\end{corollary}

Corollary \ref{corollary:EC1} shows that having access to views with varying time horizons affect the conditional dynamics solely through the covariance matrix $\Omega$. This further supports our results in Section \ref{sec4}, where we show that the hitting times of the Brownian bridge associated with the log-returns are increasing in the covariance of the views.

\subsection{Optimal Policy}
Let $T = T_K$ be the investor's investment horizon, and 
\begin{equation*}
    V(t,z,x) = \max_{\pi \in \mathcal{A}} \mathbb{E}\big[ U(Z(T)) | X(t) = x, \; Z(t) = z, \; Y_1(0,T_1) = y_1, \; \dots, \; Y_K(0,T_K) = y_k  \big]
\end{equation*}
her value function at time \(t \in [T_{j-1}, T_j)\), where the wealth process $Z(t)$ satisfies 
\begin{equation*}
    dZ(t) = Z(t) \bigg( r_f dt + \pi(t)^\top \big(\tilde{\mu}(t,X^y(t), \bar{y}^j) - r_f \mathbf{1}_N\big) dt + \pi(t)^\top dW^y(t) \bigg),
\end{equation*} 
with
\begin{equation*}
    \tilde{\mu}^j(t,x, \bar{y}^j) = \mu + \beta_1^j (\bar{y}^j - T_j P_{j:K} \mu^x) + \beta_2^j(t) (\mathbb{E}[X^y(t)] - x), \; t \in [T_{j-1}, T_{j}).
\end{equation*}
The log-returns $X^y(t)$ satisfy \eqref{eq:dX_ViewsWithDifferentHorizons}.
The following Proposition gives the corresponding optimal investment strategy.

\begin{proposition}
\label{prop:ViewsWithDifferentHorizons}
    Suppose $\gamma > 1$. For $j \in [K]$ and $t \in [T_{j-1}, T_j)$, the solution to the HJB equation is
    \begin{equation*}
        V(t,z,x) = \frac{z^{1 - \gamma}}{1 - \gamma} \exp(g^j(t,x)), \; t \in [T_{j-1},T_j)
    \end{equation*}
    where 
    \begin{equation*}
        g^j(t,x) = \frac{1}{2}x^\top A^j(t)x + x^\top b^j(t) + c^j(t), \; t \in [T_{j-1},T_j).
    \end{equation*}
    The matrix $A^j(t)$ is symmetric negative semi-definite for $t \in [T_{j-1},T_j)$ and satisfies a Riccati equation
    \begin{equation}
    \begin{cases}
         A^{j'}(t) + \dfrac{1 - \gamma}{\gamma} \eta_t^j \Sigma \eta_t^j + \dfrac{1}{\gamma} (A(t) \Sigma \eta_t^j + \eta_t^j \Sigma A(t)) + \dfrac{1}{\gamma} A^j(t) \Sigma A^j(t) = 0 ,  \\
         A^{j}(T_j) = A^{j+1}(T_j),
    \end{cases}
    \label{eq:Riccati_extension3}
    \end{equation}
    with terminal condition \(A^K(T) = 0\),
    and
    \begin{equation}
             \eta_t^j = -(P_{j:K})^{\top} ((T_j - t) P_{j:K} \Sigma (P_{j:K})^{\top} + T_j \bar{\Omega}^j)^{-1} P_{j:K},  \; t \in [T_{j-1},T_j).
             \label{eq:eta_j}
    \end{equation}
    $b^j(t)$ solves a system of linear ODEs
    \begin{equation}
    \begin{cases}
                b^{j'}(t) + \dfrac{1}{\gamma}\big(\eta_t^j + A^j(t)  \big) \Sigma b^j(t) + \dfrac{1 - \gamma}{\gamma}\big( \eta_t^j +A^j(t) \big)(\alpha_t^j - r_f \mathbf{1}_N) + A^j(t) \big(\alpha_t^j - \dfrac{1}{2}\diag(\Sigma) \big) = 0,\\
                b^j(T_j) = b^{j+1}(T_j),
    \end{cases}
        \label{eq:ODEsys_extension3}
    \end{equation}
    with terminal condition \(b^K(T) = 0\), and
    \begin{equation*}
            \alpha_t^j = \mu + \dfrac{1}{T_j}\beta_1^j (\bar{y}^j - T_jP_{j:K} \mu^x) - \Sigma \eta_t^j \mathbb{E}[X^y(t)],
    \end{equation*}
    and $c^j(t)$ is the solution of
    \begin{equation}
    \begin{cases}
        \begin{split}
         c^{j'}(t) &+ (1 - \gamma) r_f + \frac{1}{2} \Tr\big(A^j(t) \Sigma \big) + \frac{1 - \gamma}{2 \gamma}\big(\alpha_t^j - r_f \mathbf{1}_N \big)^\top \Sigma^{-1} \big(\alpha_t^j - r_f \mathbf{1}_N\big) + \big(\alpha_t^j - \frac{1}{2}\diag(\Sigma)\big)^\top b^j(t) \\
         &+ \frac{1 - \gamma}{\gamma} \big(\alpha_t^j - r_f \mathbf{1}_N\big)^\top b^j(t) + \frac{1}{2\gamma} b^{j\top} (t) \Sigma b^j(t) = 0,
    \end{split}\\
    c^j(T_j) = c^{j+1}(T_j).
    \end{cases}
        \label{eq:c^j_extension3}
    \end{equation}
    with \(c^K(T) = 0\).
    
    There exists a unique optimal allocation policy
        \begin{equation}
        \pi^{j*}(t,x,\bar{y}^j) = \frac{1}{\gamma} \Sigma^{-1} \big(\tilde{\mu}^j(t,x,\bar{y}^j) - r_f \mathbf{1}_N \big) +\frac{1}{\gamma}\frac{\partial g^j}{\partial x}(t,x,\bar{y}^j), \; t \in [T_{j-1},T_j)
    \end{equation}
    where
\begin{equation*}
    \tilde{\mu}^j(t, x, \bar{y}^j) =  \mu  +  \beta_1^j \big(\bar{y}^j - T_j P_{j:K} \mu^x\big) + \beta_2^j(t) \big(\mathbb{E}[X^y(t)] -x \big), \; t \in [T_{j-1},T_j)
\end{equation*}    
    and
    \begin{equation*}
    \frac{\partial g^j}{\partial x}(t,x,\bar{y}^j) = A^j(t) x + b^j(t), \; t \in [T_{j-1},T_j).
\end{equation*}
\end{proposition}

\subsection{Proofs of the results}
\paragraph{\textbf{Proof of Proposition \ref{proposition_DifferentHorizons}.}}\mbox{}
As views are ordered according to their horizon, let  $k, \;j \in [K]$ such that $k \geq j$. We can write
\begin{equation*}
    \begin{split}
        X(T_k) &= T_k \mu^x + W(T_k)\\
        &= T_j \mu^x + W(T_j) + (T_k - T_j)\mu^x + W(T_k) - W(T_j)\\
        &= X(T_j) + \underbrace{(T_k - T_j)\mu^x + W(T_k) - W(T_j)}_{\text{Adjustments}},
    \end{split}
\end{equation*}
thus, the log-returns at time $T_k \geq T_j$ can be expressed in terms of the log-returns at time $T_j$, with some adjustments. It follows that for $k \in [K]$, the view $Y_k(0,T_k)$ can be transformed to a view about the log-return realization at time $T_j \leq T_k$
\begin{equation*}
    \begin{split}
        Y_k(0,T_k) &= p_k^\top X(T_k) + \sqrt{T_k}\epsilon_k\\
        &= p_k^\top (X(T_j) + (T_k - T_j)\mu^x + W(T_k) - W(T_j)) + \epsilon_k\\
        &= p_k^\top X(T_j) + \bar{\mu}_k^j(T_j, T_{j:K}) + \bar{\epsilon}_k^j,
    \end{split}
\end{equation*}
with
\begin{equation*}
    \begin{split}
        \bar{\mu}_k^j(T_j, T_{j:K}) = (T_k - T_j) p_k^\top \mu^x,\\
    \end{split}
\end{equation*}
is the additional bias due to the adjustment in the time horizon of the the view, and
\begin{equation*}
           \bar{\epsilon}_k^j = p_k^\top(W(T_k) - W(T_j)) + \sqrt{T_k}\epsilon_k
\end{equation*}
is a zero mean noise term which represents the uncertainty in the view. Notice that the noise term can be split into two part: The first term, \(p_k^\top(W(T_k) - W(T_j))\), captures the uncertainty related to the structure of the log-returns, while the second term, \(\sqrt{T_k}\epsilon_k\), accounts for the uncertainty of the experts.

Now given the views horizons vector
\begin{equation*}
    T_{j:K} = \left(T_j ,\dots, T_K\right)^\top \in \mathbb{R}^{K-j+1},
\end{equation*}
and the linear mapping
\begin{equation*}
    P_{j:K} = (p_j, \dots, p_K)^\top \in \mathbb{R}^{(K-j+1) \times N},
\end{equation*}
we can write the forward-looking views vector $Y^j(0,T_{j:K}) \in \mathbb{R}^{K-j+1}$ available to the investor at time $t \in [T_{j-1}, T_j)$ as
\begin{equation*}
    Y^j(0,T_{j:K}) = P_{j:K} X(T_j) + \bar{\mu}^j(T_j,T_{j:K}) + \bar{\epsilon}^j,
\end{equation*}
where $\bar{\epsilon}^j \in \mathbb{R}^{K-j+1 \times K - j+1}$ is Gaussian with
\begin{equation*}
    \bar{\epsilon}^j \sim \mathcal{N}\big(0, T_j \bar{\Omega}^j\big),
\end{equation*}
and the covariance matrix $\bar{\Omega}^j$ takes the form
\begin{equation*}
    \bar{\Omega}^j =  \bar{\Omega}^{Vj} + P_{j:K}\bar{\Omega}^{Wj}(P_{j:K})^{\top},
\end{equation*}
with
    \begin{equation*}
            \bar{\Omega}^{Vj}_{ik} = \dfrac{\sqrt{T_iT_k}}{Tj}\Omega_{ik}, \, \, \, \text{for} \,\, i,k \in \{1, \dots, K-j+1\}\\ 
    \end{equation*}
    is the covariance related to pushing the view horizon from times $T_k \geq T_j$ to $T_j$, and
    \begin{equation*}
    \bar{\Omega}^{Wj}_{ik} = \dfrac{1}{T_j}\min\{T_i - T_j, T_k - T_j\}\Sigma_{ik}, \, \, \, \text{for} \,\, i,k \in \{1, \dots, K-j+1\}
    \end{equation*}
    is the covariance related to using the structure of the log-returns to predict the realization of $X(T_j)$ from that of $X(T_k)$. It follows that for
    \begin{equation*}
        \bar{Y}^j(0,T_j) = Y^j(0,T_{j:K}) - \bar{\mu}^j(T_j,T_{j:K}),
    \end{equation*}
   the view $\bar{Y}^j(0,T_{j})$ satisfies
\begin{equation*}
    \bar{Y}^j(0,T_j) \,|\, X(T_j) = P_{j:K}X(T_j) + \bar{\epsilon}^j \sim \mathcal{N}\big(P_{j:K}X(T_j), T_j \bar{\Omega}^j\big).
\end{equation*}
 Furthermore, since $\bar{Y}^j(0,T_j)$ contain the same information as $Y^j(0,T_{j:K})$, it follows that
 \begin{equation*}
X(t) \, | \, Y^j(0,T_{j:K}) = X(t) \,|\, \bar{Y}^j(0,T_{j}), \, \, \, \text{for} \, \, \, t \in [T_{j-1},T_j], 
 \end{equation*}
 which complete the proof. \hfill $\square$

 \paragraph{\textbf{Proof of Corollary  \ref{corollary:EC1}}}\mbox{} Since we showed that for $j \in [K]$ and $t \in [T_{j-1}, T_{j})$, the conditional log-returns process can be written as
\begin{equation*}
    \begin{split}
        X^y(t) &= X(t) \,|\, (\bar{Y}^j(0,T_{j}) = \bar{y}^j),
    \end{split}
\end{equation*}
where
\begin{equation*}
    \bar{Y}^j(0,T_{j:K}) \,|\, X(T_j) \sim \mathcal{N}\big(P_{j:K} X(T_j), T_j \bar{\Omega}^j\big).
\end{equation*}
It follows from Proposition \ref{proposition1} that 
 \begin{equation*}
\begin{split}
        dX^y(t) =&   \bigg(\mu^x + \beta_1^j \big(\bar{y}^j - T_j P_{j:K} \mu^x \big) + \beta_2^j(t) \big(\mathbb{E}[X^y(t)] - X^y(t) \big)\bigg) dt + dW^y(t)
\end{split}
\end{equation*}
where
\begin{eqnarray*}
    \beta_1^j & = & \frac{1}{T_j}\Sigma P_{j:K}^{\top} (P_{j:K} \Sigma (P_{j:K})^{\top}+ \bar{\Omega}^j)^{-1} \in  \mathbb{R}^{N \times K},\\ [8pt]
    \beta_2^j(t) & = & \Sigma P_{j:K}^{\top} \left( (T_j-t) P_{j:K} \Sigma (P_{j:K})^{\top} + \bar{\Omega}^j\right)^{-1} P_{j:K} \in \mathbb{R}^{N \times N},
\end{eqnarray*}
and
\begin{eqnarray*}
   \mathbb{E}[X^y(t)]  &=& t \left(\mu^x + \beta_1^j \big(\bar{y}^j -T_j P_{j:K} \mu^x \big)\right)
\end{eqnarray*}
This concludes the proof. \hfill $\square$

\paragraph{\textbf{Proof of Proposition \ref{prop:ViewsWithDifferentHorizons}.}}\mbox{}
The proof of this proposition follows directly by applying the results in Section \ref{sec5} and replacing the horizon  $T$  with  $T_j$ and the view \(y\) with \(\bar{y}^j\)  for each  $j \in [K]$. The derivation of the ODEs \eqref{eq:Riccati_extension3} -- \eqref{eq:c^j_extension3} will follow directly.

For the boundary conditions, observe that during the last interval \([T_{K-1},T]\), the value terminal condition of the value function is
\begin{equation*}
    V(T,z,x) = 0,
\end{equation*}
it follows that 
\begin{equation*}
    \begin{split}
        A^K(T) &= 0,\\
        b^K(T) &= 0,\\
        c^K(T) &= 0.
    \end{split}
\end{equation*}
For the boundary conditions of the coefficients  $A^j$, $b^j$,  and  $c^j$ at time period \(\{T_j, j \in \{1,K-1\}\}\). 
Note that by Principle of Optimality we must have
\begin{equation*}
    V\bigl(T_{j}^-, z, x\bigr)
   =
   \mathbb{E}\Bigl[
      V\bigl(T_{j}, z, x\bigr)
      \, | \,
      X^y\bigl(T_{j}^-\bigr)=x,\;
     Z(t) = z \Bigr]
\end{equation*}
and since the views are fixed and are given at the beginning of the horizon \(t = 0\), it follows by continuity of the value function that
\begin{equation*}
    V(T_j^-,z,x) = V(T_j,z,x)
\end{equation*}
which translates to
\begin{equation*}
\begin{split}
A^j(T_j) &= A^{j+1}(T_j), \\
b^j(T_j) &= b^{j+1}(T_j), \\
c^j(T_j) &= c^{j+1}(T_j).
\end{split}
\end{equation*}
This concludes the proof. \hfill $\square$

\end{document}